\documentclass[12pt]{iopart}

\usepackage{graphicx}
\usepackage{xspace}
\usepackage[usenames,dvipsnames]{color}
\usepackage{amssymb}

\usepackage{mathrsfs}
\usepackage{bm}
\usepackage[normalem]{ulem} 

\usepackage{cite}

\usepackage[utf8]{inputenc}

\usepackage[colorlinks=true,citecolor=blue,urlcolor=blue]{hyperref}

\newcommand{\eqref}[1]{(\ref{#1})}
\newcommand{\spec}{\texttt{SpEC}}

\begin{document}

\title[Fundamental frequencies and resonances from eccentric and precessing BBH]{
Fundamental frequencies and resonances from eccentric and precessing binary black hole inspirals
}

\author{Adam G.~M.~Lewis, Aaron Zimmerman, Harald P.~Pfeiffer} 
\address{Canadian Institute for Theoretical
    Astrophysics, 60 St.~George Street, University of Toronto,
    Toronto, ON M5S 3H8, Canada}
\ead{alewis@physics.utoronto.ca}

\date{\today}

\begin{abstract}
  Binary black holes which are both eccentric and undergo precession
  remain unexplored in numerical simulations.  We present simulations of such
  systems which cover about 50 orbits at comparatively high mass ratios 5 and 7.  The configurations correspond to the generic motion
  of a nonspinning body in a Kerr spacetime, and are chosen to study
  the transition from finite mass-ratio inspirals to point particle
  motion in Kerr.  We develop techniques to extract analogs of the
  three fundamental frequencies of Kerr geodesics, compare our
  frequencies to those of Kerr, and show that the differences are
  consistent with self-force corrections entering at first order in
  mass ratio.  This analysis also locates orbital
  resonances where the ratios of our frequencies take
  rational values.  At the considered mass ratios, the binaries pass
  through resonances in one to two resonant cycles, and we find no
  discernible effects on the orbital evolution.  We also compute the
  decay of eccentricity during the inspiral and find good agreement
  with the leading order post-Newtonian prediction.

\end{abstract}


\maketitle

\section{Introduction}

The recent landmark detections of gravitational waves by the Advanced LIGO 
interferometers gave definitive proof that binary black hole (BBH) systems 
exist and merge in nature~\cite{LIGOVirgo2016a,TheLIGOScientific:2016qqj,Abbott:2016nmj}. 
Detection and characterization of the resulting signals 
relies on knowledge of the expected gravitational waveforms for tasks as varied as 
detection, parameter estimation, and tests of 
general relativity~\cite{LIGO-DataAnalysis-Whitepaper:2015}.
The use of theoretical waveforms enhances gravitational wave (GW) astronomy's scientific potential where waveform models are available, as 
demonstrated by the GW detections and analyses in 
Advanced LIGO's first observing run~\cite{TheLIGOScientific:2016pea}.  Conversely, the absence of 
accurate and reliable waveform models may limit detection sensitivity, and 
hinders source parameter estimation and tests of gravitational theories.

Due to both physical motivations and computational complexity, 
waveform modeling for BBH systems has focused on quasi-circular binaries. 
  Orbital eccentricity is damped away quickly during a GW-driven
  inspiral of two compact objects~\cite{PetersMathews1963} because GW
  emission peaks at periastron.  Thus ``field binaries'' formed from
  the respective collapse of both partners in a high-mass stellar
  binary are expected to be almost exactly circular by the time they
  enter the sensitive band of ground-based GW detectors.
  Quasi-circularity removes the two degrees of freedom related to
  orbital eccentricity, and thus reduces the dimensionality of the BBH
  parameter space that needs to be modelled.  Mature waveform models
  exist for quasi-circular, aligned-spin BBH systems~\cite{Khan:2015jqa,Taracchini:2013rva}, 
  while models of quasi-circular BBH systems with generic spin (i.e.~precessing binaries)   
  are maturing quickly
  (e.g.~\cite{Hannam:2013oca,Pan:2013rra,Babak:2016tgq}).  These
  waveform models are based on direct numerical solutions of Einstein's
  equations (e.g.~\cite{baumgarteShapiroBook}) that provide the 
  gravitational waveforms for the late inspiral and merger of the two
  black holes; such numerical simulations, too, are most mature for
  quasi-circular systems
  (e.g.~\cite{Ajith:2012az,Hinder:2013oqa,Mroue:2013PRL,Chu:2015kft,Jani:2016wkt,Husa:2015iqa}).

In contrast, very few numerical simulations have been performed for 
eccentric BBH systems~\cite{SperhakeEtAl:2008, Hinder2008, Hinder:2008kv, Mroue2010, East:2015yea, Gold:2012tk}, and a systematic numerical exploration of the eccentric parameter space has not even started.  
Complete inspiral-merger-ringdown waveform models for eccentric BBH systems 
are also in a nascent state, with current models~\cite{Huerta:2016rwp} neither 
reaching the accuracy of quasi-circular models, nor accounting for black hole 
spins (not even aligned-spin).

Several astrophysical scenarios have been explored in recent years
which can lead to nonzero eccentricity late in the inspiral: Within
dense stellar environments such as globular clusters (GCs) and
galactic cores, binaries can form dynamically
(e.g.~\cite{Morscher:2014doa, 2008ApJ...676.1162S,Rodriguez:2015oxa,
  Rodriguez:2016}), and there can be significant residual eccentricity
from direct dynamical capture \cite{ Hopman:2006, OLeary2009,
  Kocsis:2012, Tsang:2013}.  The scattering of single stars off
binaries \cite{SamsingEtAl:2014} can also lead to high eccentricities,
although these events are likely rare \cite{AntoniniEtAl:2016}.  Even
more promisingly, eccentricity can be generated by the Kozai-Lidov
mechanism \cite{Lidov:1962, Kozai:1962,Ford-Kozinsky-Rasio2000,
  Ford-Kozinsky-Rasio2004, FabryckyAndTremaine:2007, NaozEtAl:2013,
  MerritBook:2013} in three body systems with moderate separations
\cite{AntoniniAndPerets:2012, KatzAndDong:2012, AntoniniEtAl:2014,
  BodeAndWegg:2014, Seto:2013wwa}.  Such triples can occur in GCs
\cite{Wen2003, AntoniniEtAl:2016} or galactic nuclei
\cite{AntoniniAndPerets:2012}, and may provide a source for eccentric
compact binaries in advanced detectors \cite{Stephan:2016,
  Antognini:2014, AntoniniEtAl:2016} (in particular reference
\cite{AntoniniEtAl:2016} estimates $e>0.1$ should occur at rate of
about 0.2 yr$^{-1}$, but see also~\cite{AmaroSeoane2016}).

Ultimately gravitational wave observations will be the primary tool to 
measure eccentricity of compact object inspirals and to
constrain the rate of eccentric compact mergers.  Unfortunately, the
present detection template bank used by LIGO is entirely
circular~\cite{TheLIGOScientific:2016pea,Abbott:2016ymx}, hindering
detection of systems with in-band eccentricities above about
$ e \sim 0.1$ \cite{BrownZimmerman2009,Huerta:2013qb,Huerta:2016rwp}.  
Due to this same absence of eccentric waveform models, parameter estimation must 
likewise assume quasi-circularity.

 Looking ahead, the ``extreme mass-ratio inspirals" (EMRIs) visible to space-based detectors such as LISA~\cite{AmaroSeoane:2012je,Seoane:2013qna} may be detected at quite large eccentricities (e.g.~\cite{AmaroSeoane:2009ui}). 
While eccentric waveforms are being developed~\cite{Huerta:2016rwp,Tanay:2016zog}, further numerical and analytical modeling of eccentric systems are required to constrain eccentricity and prepare for future ground- and space-based detections of eccentric systems.

Eccentric systems also yield
additional insight into strong-field dynamics, especially at high mass
ratios~\cite{AckayEtAl:2015, Tiec:2015cxa,LoutrelEtAl:2016}.
While circular orbits must remain
  outside the innermost stable circular orbit (ISCO) until the final plunge, 
  eccentric orbits can reach as deep as the innermost bound
  orbit, thus probing stronger gravitational
  fields~\cite{1972ApJ...178..347B}. Furthermore, binaries which are both eccentric and precessing open up the
possibility of qualitatively new phenomena when orbits pass through
resonances and radiation reaction effects accumulate secularly
\cite{Flanagan:2010cd, Flanagan:2012kg}. 

This paper presents a survey of 12 numerical simulations over a range of initial eccentricities as high as $e = 0.2$, mass ratios $q=5$ and $7$, and initial inclinations between the spin and the orbital plane between $0^\circ$ and $80^\circ$, performed with the Spectral Einstein Code (\spec)~\cite{SXSWebsite}.
We focus on higher mass ratios in order to make contact with the extreme mass ratio limit, where the BBHs can be modelled to leading order by bounded motion of a test particle in a Kerr spacetime.
Beyond the test particle limit, the motion is corrected by higher order, self-force (SF) effects due to the spacetime perturbations sourced by the particle \cite{Poisson2011}.
In this case the ratio of the particle mass $\mu$ to that of the Kerr black hole $M_{\rm Kerr}$ serves as a small expansion parameter. 
While SF results for Kerr are still in development \cite{Shah:2012gu,vandeMeent:2015lxa,vandeMeent:2016pee,Merlin:2016boc}, mounting evidence suggests that SF results may be extended much closer to the nearly equal mass cases than expected \cite{LeTiec-Mroue:2011,LeTiec:2011dp,Tiec:2013twa}, which is the regime appropriate for our our high (but not extreme) mass ratio binaries.
Meanwhile, the lowest order approximation to the SF-expansion, namely the test particle limit, incorporates the effects of strongly curved spacetime, highly relativistic motion, and is fully understood.
At the same time, post-Newtonian approximations have difficulty describing higher mass ratio binaries, where the constituents remain at close separations for many orbits.
In addition, higher mass ratio binaries can potentially remain at orbital resonances for multiple cycles of the resonance, and our simulations allow for the first investigation of the role of passage through resonances in BBHs.

In order to compare to the test particle limit and investigate these SF corrections, our primary interest is the computation of the characteristic frequencies in the azimuthal ($\Omega^\phi$), radial ($\Omega^r$), and polar ($\Omega^\theta$) directions. 
Except very close to ISCO these frequencies furnish a one-to-one map with (self-force corrected) Kerr geodesics \cite{Warburton:2013yj} and thus provide a natural point of comparison with the latter.  
In the Kerr spacetime the effect of timelike coordinate transformations that do not involve time-dependent
rotations is to multiply all of the frequencies by the \emph{same} factor \cite{Schmidt:2002qk}. 
Thus their ratios $K^{ab} \equiv \Omega^a/\Omega^b$ are insensitive to such transformations. 
In the particular case of nearly circular orbits, the periastron precession rate of BBHs simulations as encoded in $K^{r\phi}$ was investigated \cite{LeTiec-Mroue:2011,Tiec:2013twa}, showing that analytic approximations provide highly accurate descriptions of the simulations.
Additionally, the precession rates $K^{ab}$ are of interest because the points in the orbit when these frequency ratios are rational are precisely the moments of orbital resonance. 
Their extraction can be used to characterize and explore such resonances in our simulations.

Extracting the characteristic frequencies accurately from an
eccentric, precessing numerical relativity simulation turns out to be
challenging due to both strong dissipation as well as modulations arising from interactions of radial and
polar motion. Fourier-based methods of frequency extraction require several
  orbits to achieve percent level accuracy, resulting in unacceptable dissipative contamination.
We therefore rely upon time-domain methods based on 
 intervals between successive periastron passages, which achieve better accuracy with only a single orbit.

This paper is structured as follows. In section~\ref{sec:KerrMotion} we introduce the basic context of our analysis by discussing generic test orbits in the Kerr spacetime. In section~\ref{sec:Simulations} we review the simulations we have performed in detail and describe the association we make between simulation trajectories and Kerr geodesics. In section~\ref{sec:EquatorialFrequencies} we detail our frequency extraction methodology and its results for our equatorial runs. Section~\ref{sec:Inclined} provides the same analysis for the inclined runs, and showcases our ability to detect resonances in these simulations. We discuss our results and future directions in section~\ref{sec:Conclusions}.

\section{Motion in Kerr spacetimes}
\label{sec:KerrMotion}

We interpret our high mass ratio, eccentric, and precessing simulations in terms of motion in the Kerr spacetime.
From the perspective of dynamical systems, bound test orbits in Kerr form an integrable Hamiltonian system. Such systems admit action-angle coordinates, in which the generalized momenta are the conserved actions $J^a$. The conjugate positions are circulating angles  
which evolve at fixed frequencies.
The motion of the particle is multiperiodic, and can be expanded in a Fourier series in the frequencies.
In Hamiltonian perturbation theory the action-action angle formalism elucidates the perturbed motion, the loss of integrability, and the onset of chaos \cite{LichtenbergLieberman}.

For Kerr geodesics, the transformation to action-angle variables was
first discussed by Schmidt~\cite{Schmidt:2002qk}.~The resulting $J^a$
are geometric invariants, and the associated proper time frequencies
describe the motion in a coordinate-independent manner. The frequencies measured by alternative observers are related to the proper time frequencies through multiplication by a Lorentz-like factor. 
In particular, distant inertial observers measure frequencies $\Omega^a$.

Self-force corrections to geodesic motion decompose usefully into conservative corrections to the orbital dynamics and dissipative effects which drive inspiral \cite{Mino:2003yg}.
The small size of the dissipation means that this system is amenable to evolution using a two-timescale approach \cite{Hinderer:2008dm,Pound:2015wva}.
We work in the paradigm where (accounting for the slow dissipation) the fundamental frequencies are perturbed by the conservative SF effects.
This perturbed Hamiltonian perspective on the SF problem has been used to compute invariant quantities in Kerr spacetimes \cite{Isoyama:2014mja}\footnote{In fact, it is not a priori guaranteed that Kerr orbits perturbed by the conservative SF are Hamiltonian, although there are strong indications that they are \cite{Vines:2015efa}. 
Therefore, it is not clear that this approach is formally valid. Nevertheless, since conservative SF effects are multiperiodic in the underlying orbits, it is reasonable to expect that the techniques of perturbed Hamiltonian systems apply. }.
Our goal in sections~\ref{sec:EquatorialFrequencies} and \ref{sec:Inclined} is to extract the fundamental frequencies from our simulations, which have physical meaning in terms of precession rates of the binary as viewed by distant, inertial observers. The differences between these frequencies
and those of Kerr can be viewed as (possibly high-order) self-force corrections; thus our work may be useful for comparison with future self-force results.

\subsection{Geodesic orbits in Kerr}

A test particle on a bound orbit in the Kerr spacetime has four conserved quantities: 
the specific energy $\mathcal E$, the specific angular momentum along the symmetry axis 
$\mathcal L_z$, the specific Carter constant $\mathcal Q$, and the Hamiltonian constraint 
$\mathcal H = (1/2) g^{\mu \nu}p_\mu p_\nu = - \mu^2/2$, where $\mu$ denotes the mass of the test particle.
It is convenient to use Boyer-Lindquist coordinates $x^\mu = ( t , r, \theta, \phi)$ (see \ref{sec:KerrDetails})
and to define the Carter-Mino time $\lambda$ through \cite{Carter:1968rr,Mino:2003yg}
\begin{equation}
\frac{d\tau}{d\lambda} = \rho^2\,, \qquad \rho^2 = r^2 + a^2 \cos^2 \theta\,,
\end{equation}
where $\tau$ is the proper time of the trajectory.
Parameterized by $\lambda$ rather than $\tau$, the equations for the radial and polar motions of the particle separate,
\begin{equation}
\label{eq:RadialEoM}
\left(\frac{dr}{d\lambda}\right)^2 =   R(r) \,, \qquad \left(\frac{d\theta}{d\lambda}\right)^2  =  \Theta(\theta) \,,
\end{equation}
where the potentials $R$ and $\Theta$ are given in \eqref{eq:RPotential} and \eqref{eq:ThetaPotential}.
This results in independent cyclic motions in the $r$ and $\theta$ directions. 

Unfortunately, the equations of motion for $t(\lambda)$ and $\phi(\lambda)$ depend on both the radial and polar motions,
\begin{eqnarray}\label{eq:mino_time}
\frac{dt}{d\lambda} & = & T_r(r) + T_\theta(\theta) + a \mathcal L_z \,, \\
\label{eq:PhiEoM}
\frac{d\phi}{d\lambda} & = & \Phi_r(r) + \Phi_\theta(\theta) - a \mathcal E \,,
\end{eqnarray}
with the potentials on the right hand side given by \eqref{eq:TPotential} and \eqref{eq:PhiPotential}.
This means that $r(t)$, $\theta(t)$, and $\phi(t)$ are multiperiodic, complicating the extraction of the fundamental frequencies of motion. 
For geodesics, this problem can be solved by moving to action-angle coordinates \cite{Schmidt:2002qk}.

When studying bound orbits it is useful to introduce a Keplerian parametrization of the orbit, which represents the radial motion as a precessing eccentric orbit.
The radius evolves as 
\begin{equation}
\label{eq:RadEllipse}
r = \frac{p M}{1 + e \cos \chi^r}\,,
\end{equation}
where $p$ and $e$ respectively denote the semilatus rectum and eccentricity of the orbit.  
The phase $\chi^r(\lambda)$ represents the position of the particle along the precessing ellipse. 
The particle oscillates between the periastron $r_p = p/(1+e)$ and apastron $r_a = p/(1-e)$.
Similarly, one can express the polar motion as an oscillation between symmetric turning points above and below the equatorial plane, by writing \cite{Drasco:2003ky,Fujita:2009bp}
\begin{equation}
\label{eq:theta_phase}
\cos\theta \equiv \cos \theta_{\rm min} \cos \chi^\theta\,,
\end{equation}
where $\theta_{\rm min}$ is the smallest polar angle the particle reaches at the height of its vertical motion. We define the inclination of the orbit by $i = \pi/2 - \theta_{\rm min}$ as the inclination angle of the orbit above the equatorial plane. Note that our use of $i$ to define the inclination angle of our orbit differs from the commonly used (e.g.~\cite{Hughes:2001jr}) inclination angle $\iota$ defined through the constants of motion by $ \cos \iota = \mathcal L_z/(\mathcal Q+ \mathcal L_z^2)^{1/2}.$

Together, $(p, e, i)$ characterize bound orbits.
Given these orbital parameters, we can compute the corresponding constants of motion $(\mathcal E, \mathcal L_z, \mathcal Q)$ \cite{Schmidt:2002qk} and solve the equations of motion, see e.g.~\cite{Fujita:2009bp}.

\subsection{Fundamental frequencies of Kerr}

In the case of equatorial, eccentric orbits, the fundamental frequencies are straightforward to define.
The particle oscillates between periastron and apastron, while advancing in the azimuthal angle $\phi$.
The two relevant frequencies are the frequency between successive radial passages, $\Omega^r$, and the azimuthal frequency averaged over successive passages, $\Omega^\phi$.

The frequency ratio $K^{r\phi} = \Omega^r/\Omega^\phi$ measures the periastron precession rate.
It is a well-defined observable that can be measured by distant inertial observers, for whom the particular coordinate time that the frequencies reference (whether it be $t$, $\tau$, or $\lambda$) divides out.

For generic, non-equatorial orbits the situation is more complicated. 
In terms of $\lambda$ the radial and polar motions decouple. 
However, azimuthal motion is modulated by both the radial and polar motions.
Similarly, the Boyer-Lindquist time $t(\lambda)$ along the world line is modulated by the radial and polar motions.
This results in multiperiodic behaviour in the coordinate graphs: the time between successive coordinate extrema is variable, and only the long-term average time between extrema approaches the fundamental periods $T^a = 2\pi/\Omega^a$.

Thus the fundamental frequencies $\Omega^r$, $\Omega^\theta$, $\Omega^\phi$ are infinite time averages.
Once again, the ratios of frequencies $K^{r\phi}$, $K^{r\theta} = \Omega^r/\Omega^\theta$, and $K^{\theta \phi} = \Omega^\theta/\Omega^\phi$ eliminate the dependence on the particular choice of time coordinate, and are related to the precession rates of the periastron and of the instantaneous orbital plane.

We use the method of Schmidt~\cite{Schmidt:2002qk} to compute the fundamental 
frequencies in terms of our chosen time parameter. Fujita and Hikida~\cite{Fujita:2009bp} 
provide analytic solutions for the bound orbits and their fundamental frequencies
in terms of elliptic integrals. We do not exploit these, instead numerically 
integrating the geodesic equations to validate our frequency extraction methods
discussed in section \ref{sec:Simulations}. We choose to use frequencies 
$\Omega^a$ in terms of Boyer-Lindquist time $t$. The precise choice of time-coordinate
will cancel in the frequency ratios, so long as the coordinates under study
do not differ by time-dependent rotations.

\subsection{Orbital resonances}

Unlike the familiar quasi-circular inspirals, eccentric and especially
eccentric, precessing systems can access the qualitatively new dynamical
effect of orbital resonances. These occur when the
fundamental frequencies $\Omega^a$ are in (or near) rational ratio. In
that case the trajectory is exactly (or approximately) closed in phase space and
perturbations can accumulate secularly. During resonance the perturbed 
system's evolution can deviate dramatically from its unperturbed counterpart.
In the generic case of a Hamiltonian system this leads to
``islands" of chaotic motion in the phase space. When dissipation is
included perturbed systems pass through resonances. 
This may lead to distinct changes in the evolution of the frequency ratios (see, e.g.~\cite{LukesGerakopoulos:2010rc}), the resonant ``kicks'' discussed below, or even capture into the resonance \cite{Henrard1982,vandeMeent:2013sza}.
If the passage is fast enough, however, the system may not display chaotic behaviour or other signatures of resonant passage.
For late-inspiral BBH systems no evidence for such chaotic motion has been
observed, presumably since the dissipative timescales are too short
for any significantly ergodic motion to manifest.

Resonances can nevertheless have an important effect on the motion of high mass-ratio systems, due to concordant asymmetries in the gravitational radiation reaction. The emission of energy, momentum, and angular momentum to infinity is controlled by the Weyl scalar $\Psi_4$, which when sourced by bound orbits in Kerr can be Fourier expanded as \cite{Drasco:2003ky}
\begin{equation}
\Psi_4 = \frac{1}{(r- i a \cos \theta)^{-4}} \sum_{lmkn} R_{lmkn}(r) S_{lmkn}(\theta) e^{i m \phi - \Omega_{mkn} t}  \,,
\end{equation}
where $R_{lmkn}(r)$ is the radial wave function which solves the sourced radial Teukolsky equation \cite{Teukolsky}, $S_{lmkn}(\theta)$ are the spin-weight $s = -2$ spheroidal harmonics, and 
\begin{equation}
\Omega_{mkn} = m \Omega^\phi + n \Omega^\theta + k \Omega^r \,.
\end{equation}
Fluxes to infinity are built from the time integrals of $|\Psi_4|^2$, followed by angular integrals over the sphere at infinity, possibly weighted by additional angular terms.
The result is interference between harmonics $(m,n,k)$ and $(m',n',k')$. Typically 
the interference oscillates rapidly and does not contribute to the time integral. 
For example, for the flux of energy only terms where $(m,n,k)=(m',n',k')$ contribute.

At resonances, however, the generically-unimportant terms accumulate, leading to resonantly enhanced or diminished fluxes \cite{Flanagan:2012kg}.
Resonances can also lead to gravitational wave beaming: during resonance the BBH trajectory (or its projection into the $r$-$\theta$ plane) closes on itself, taking the form of a Lissajous figure. 
That figure need not be symmetric in space, and gravitational wave emission need not be symmetric either. 
These resonant kicks have been studied for circular, precessing orbits at $\theta$-$\phi$ resonances~\cite{Hirata:2010xn}, and for equatorial, eccentric orbits at $r$-$\phi$ resonances~\cite{vandeMeent:2014raa}.
In these studies, the resonances required for symmetry breaking of the orbit are high order and occur only for very close orbits.
For example, the $r$-$\phi$ kicks require $K^{r\phi} = 1/p$ with $p$ an integer and $p \geq 2$;
such orbits are zoom-whirl orbits.
Initial investigation in \cite{vandeMeent:2014raa} indicated that $r$-$\phi$ kicks are relatively strong and may be effective for inspirals with mass ratios $q = m_1/m_2 \gtrsim 5$. However, these kicks require that a finite mass ratio inspiral achieve these extreme orbits before plunge and merger.

Resonant kicks do not lead to chaotic motion since their influence is self-limiting: the dissipation will inevitably push the inspiral off the resonant trajectory. At this point the system will resume its approximately adiabatic motion. Nevertheless the precise kick dynamics depend sensitively on the orbital phase at which the binary enters resonance. This would challenge a
description of systems undergoing resonant kicks using e.g.~a template bank of gravitational
waveforms.

Of particular interest are the resonances between radial and polar motion described in \cite{Flanagan:2012kg,Berry:2016bit}. 
These resonances cause secular accumulation in SF effects and drive a rapid change in the conserved quantities, especially the Carter constant $\mathcal Q$. In the extreme mass-ratio limit $\mu/M \to 0 $ an $O(1)$ change to the phase
prior to resonance leads to an $O(\sqrt{M/\mu})$ correction to the phase post-resonance. Importantly, these resonances are effective at many more resonant frequencies than the kicks. Any rational ratio for $\Omega^r$ and $\Omega^\theta$ will do, although lower order resonances like $2$:$3$ or $3$:$4$ are expected to have a larger effect than higher orders. Systems passing through successive $r$-$\theta$ resonances could accumulate a sensitive dependence on initial conditions. 

While resonances are a high mass-ratio effect, the precise value of $q$ at which they may
become important is as yet unknown, partly due to the current absence of any systematic
numerical studies of eccentric inclined black hole binaries. 
It is indeed possible that resonances may become relevant at the mass ratios accessible, or nearly accessible, to numerical relativity. This would potentially seriously frustrate attempts by LIGO to measure moderate mass-ratio eccentric inspirals using matched filtering.
In sections \ref{sec:EquatorialFrequencies} and \ref{sec:Inclined} we test to see whether the high mass ratio, eccentric and generic inspirals we discuss in section \ref{sec:Simulations} display interesting resonant behaviour.

\section{Numerical relativity simulations}
\label{sec:Simulations}

\subsection{Simulations of eccentric, precessing black hole binaries}

We perform simulations with the Spectral Einstein Code
\cite{SpECwebsite}, which uses a generalized harmonic formulation
\cite{Friedrich1985, Garfinkle2002, Pretorius2005c, Lindblom2006} to
integrate the Einstein equations in damped harmonic gauge
\cite{Lindblom2009c, Choptuik:2009ww, Szilagyi:2009qz}. The code uses
an adaptively refined grid \cite{Lovelace:2011nu} between two sets of
boundaries. Lying within the holes, the ``excision boundaries" are
chosen to conform to the shapes of the apparent horizons
\cite{Szilagyi:2009qz, Scheel2009, Hemberger:2013hsa,
  Ossokine:2013zga}. After merger there is only one excision boundary
\cite{Scheel2009, Hemberger:2013hsa}. Being inside the holes, the
excision boundaries are pure-outflow and no boundary conditions are
required. The grid extends from the excision boundaries to an
artificial outer boundary endowed with constraint-preserving boundary
conditions \cite{Lindblom2006, Rinne2006, Rinne2007}. The evolution
proceeds from the construction \cite{Pfeiffer2003} of
quasi-equilibrium \cite{Caudill-etal:2006, Lovelace2008} initial
conditions satisfying the Einstein constraint equations
\cite{York1999}.

We run each of our simulations at three resolutions, which we label L1
(lowest resolution), L2, and L3 (highest resolution). Each choice of
resolution provides different tolerances on the adaptive mesh
refinement. Each resolution thus has a different
refinement history. While this prevents us from showing strict
convergence of quantities extracted from the simulations, the range
quantities take over the three resolutions is a measure of our
numerical error.  When it is practical we plot all three resolutions,
and when we report a single result from a given simulation it is from
the highest resolution, L3.  All of our results are consistent across
resolutions.

We label the masses of our black holes $m_1$ and $m_2$, using the
convention that $m_1 > m_2$.  The total mass of the black holes
$M = m_1 + m_2$ sets all scales in our simulations.  Our black holes
have angular momenta ${\bm S}_1$ and ${\bm S}_2$, 
computed using quasi-local angular momentum
diagnostics~\cite{Lovelace2008,OwenThesis},
and we define dimensionless
spin vectors through ${\bm \chi}_i = {\bm S}_i/m_i^2$, 
where here $i=1,2$ labels the black holes.  
We focus on two sequences of simulations. One sequence has
mass ratio $q = m_1/m_2=5$ and $\chi_1 = |{\bm \chi}_1| = 0.6$,\footnote{Note that for
  coordinate simulation quantities such as ${\bm \chi}$ we use a flat
  Euclidean norm.  } and the other sequence has $q= 7$ with
$\chi_1 = 0.8$.  In both cases, we set ${\bm \chi}_2 = 0$.  These
parameters were chosen so that the binary orbits might be modelled by
motion in Kerr spacetime with mass $m_1$ and spin parameter $\chi_1$,
together with radiation reaction effects and finite mass ratio
corrections to the motion.  By selecting two choices of BH parameters
we have some freedom to investigate the effect of varying those
parameters while keeping computational expense manageable.

For our two choices of $(q,\chi_1)$ we ran two sets of simulations,
using ``low'' and ``high'' initial eccentricities.  We targeted the
initial eccentricities by using simple Keplerian relations between the
initial orbital separation and angular velocity of the binary at the
moment of apastron passage, where we began our simulations.
Specifically, the initial data solver takes as input an initial
expansion factor $\dot{a}_0$, an initial orbital frequency $\Omega_0$,
and an initial coordinate separation distance $D_0$. We set
$\dot{a}_0$ to zero in all cases, to fix the orbit at apastron.  For a
given initial distance $D_0$, we target a Newtonian eccentricity
$e_N$, using the so-called ``vis-viva" equation (which expresses
energy conservation for the orbit) at apastron,
\begin{equation}
\Omega_0^2 = (m_1 + m_2)\left(\frac{1-e_N}{D_0^3} \right) \,,
\end{equation}
to solve for the appropriate $\Omega_0$.
We chose $e_N = 0.2$ for our ``low eccentricity'' and $e_N = 0.3$ for our ``high'' eccentricity runs. 

The initial separation $D_0$, together with the achieved eccentricity
$e$ and the mass ratio $q$, controls the length of the inspiral.  In
order to facilitate accurate frequency extraction we set the
simulations to be quite long, using initial distances $D_0$ of $19.5M$
and $21.125M$. This gave the same initial Newtonian semi-major axis of
$D_0/(1+e_N) = 16.25M$ for all of our simulations, and they proceeded
through $\approx30 - 60$ radial oscillations before merger.

\begin{figure}
  \vspace*{-.2cm}
  \parbox{0.5\textwidth}{$ $\\[-.5em]

    \includegraphics[scale=0.86,trim=60 -25 40 0,clip=true]{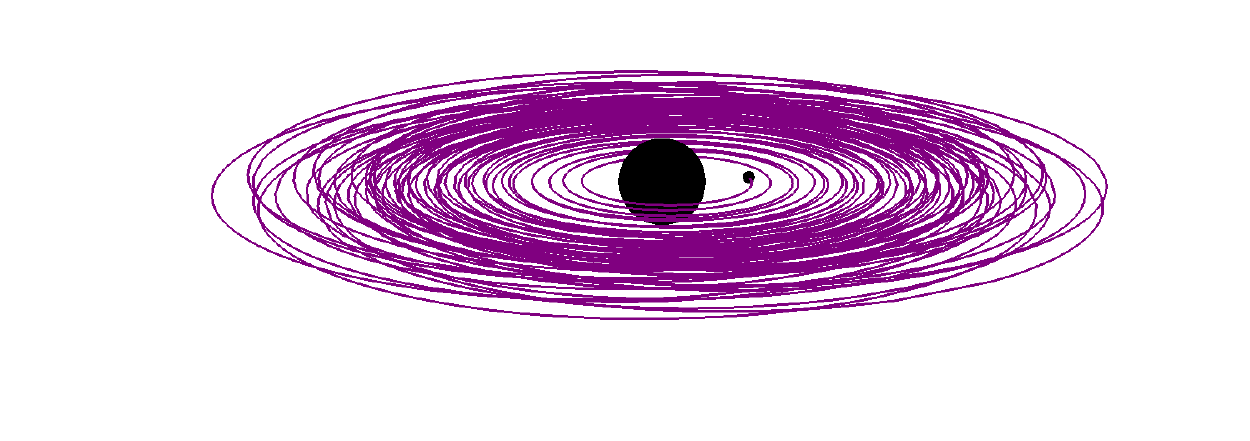}}
  \parbox{0.5\textwidth}{\includegraphics[scale=0.68,trim=43 30 40 30,clip=true,angle=161]{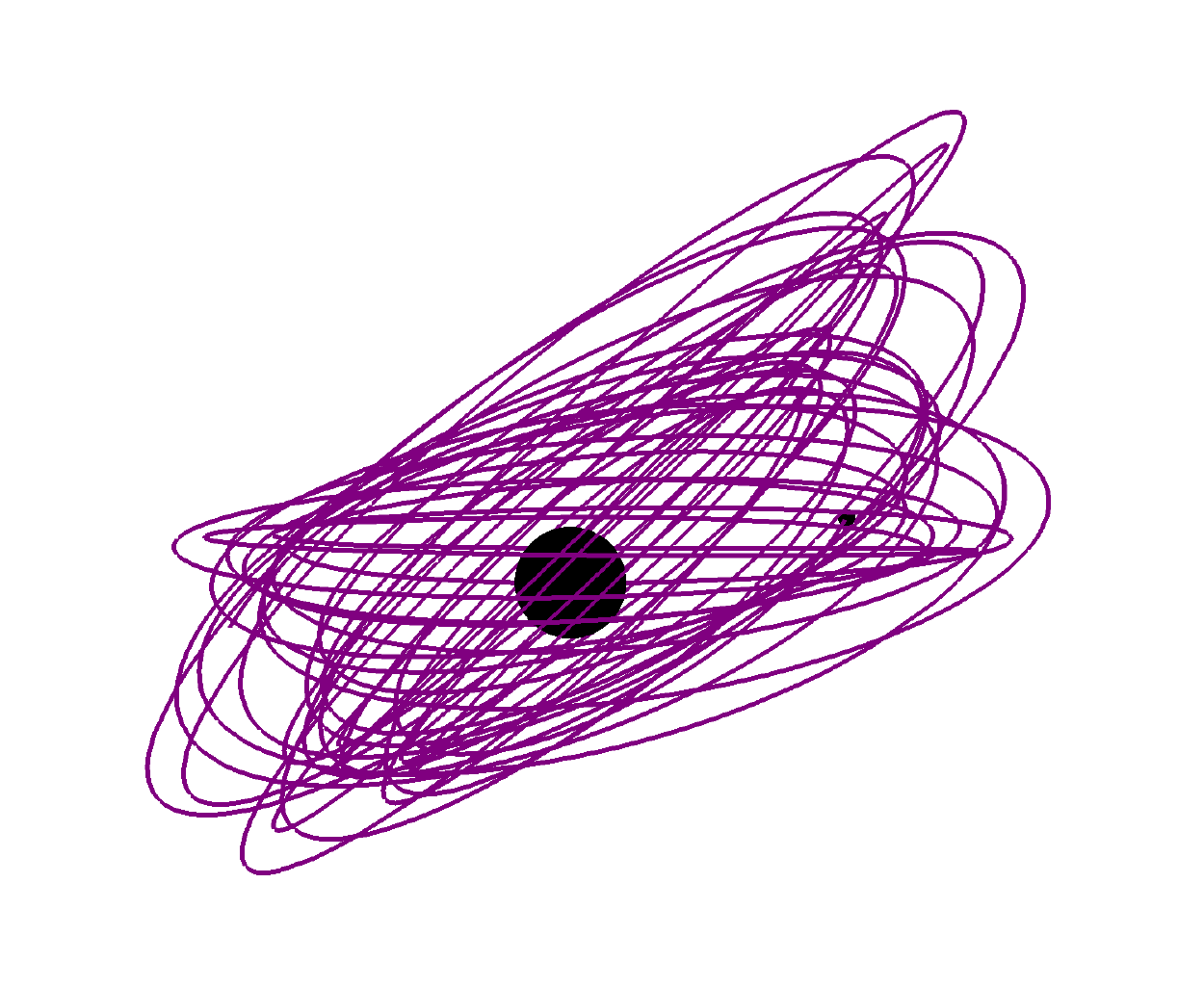}}
  \vspace*{-1.5cm}
  \caption{\label{fig:3Dtrajectories} Coordinate trajectories for two $q=7$, $\chi_1 = 0.8$, high
eccentricity simulations. Left: An equatorial inspiral, depicted up to the final orbit before merger. Right: About $28$ periastron passages from the middle of an inspiral with a $40^\circ$ inclination.
Both panels show a 3-dimensional perspective view.
}
\end{figure}

To investigate the effects of precession we ran each simulations for
each of the above choices of $(q,\chi_1)$ and $e_N$ three times,
initializing the spin vector ${\bm \chi}_1$ to form angles of
$0^\circ$, $10^\circ$, and $20^\circ$ with the computational $z$-axis
in the $q = 5, \chi_1 = 0.6$ cases and $0^\circ$, $40^\circ$, and
$80^\circ$ in the $q=7$, $\chi_1 = 0.8$ cases. The $z$-axis is normal
to the initial orbital plane. These choices of initial inclinations of
the orbital plane to the spin of the more massive hole spans the full
range of orbits from the perspective of motion in Kerr, from
equatorial to near-equatorial to nearly polar orbits.  Figure
  \ref{fig:3Dtrajectories} plots the trajectories for portions of
  two of our $q = 7$, high eccentricity orbits.

In table \ref{tab:SimParams} we present the full range of simulation
parameters. The number of radial oscillations before merger
$\mathcal{N}_r$ is estimated by counting the number of maxima in the
coordinate separation $r$ of the centers of the black holes, following
the initial junk phase. 
The number of orbits $\mathcal{N}_\phi$ is estimated by integrating
the orbital frequency $\Omega(t)$ defined in \eqref{eq:OmegaOrb} below
over the entire inspiral.  For our precessing runs, this
$\mathcal{N}_\phi$ does not correspond to the number of azimuthal
cycles in the fixed simulation coordinates, since $\Omega(t)$ is the
angular velocity in the instantaneous, precessing orbital plane.
Computation of the simulation eccentricity $e$
[and thus its initial eccentricity $e_0 = e(t=0)$] is described below
in section \ref{sec:SimulationQuantities}; the expression for $e(t)$
is given by \eqref{eq:eccentricity}. The target eccentricity $e_N$
turns out to produce an overestimate  of
the initial eccentricity $e_0$, but one which is stable across our
simulations.

\begin{table}
\centering
\caption{\label{tab:SimParams} Simulations used in this study. Tabulated here are mass-ratio $q$, dimensionless spin ${\bm \chi}_1$ of the larger black hole, initial separation $D_0$, initial orbital frequency $\Omega_0$, the Newtonian eccentricity $e_{\rm N}$ used to choose $\Omega_0$, the actual eccentricity of the simulation at its start $e_0$, and the initial inclination $i$.  The last two columns report the number of azimuthal and radial cycles during the inspiral.}
\footnotesize
\begin{tabular}{ l *{9}{c} }
\br
Run & $q$ &  ${\bm \chi}_1$ & $D_0$ & $\Omega_0 \times 10^2 $ & 
$e_{\rm N}$ & $e_0$ & $i$ & $\mathcal N_\phi $  & $\mathcal N_r$  \\
\mr
q5\_i00\_low-e & 5 & $(0,0,0.6)$ & 19.5 & 1.0387 & 0.2 & 0.07689 & $0^\circ$ & 55 & 41\\
q5\_i10\_low-e & 5 & $(-0.10419, 0, 0.59089)$ & 19.5 & 1.0387 & 0.2 & 0.07700 & $10^\circ$ & 55 & 42 \\
q5\_i20\_low-e & 5 & $(-0.20521,0,0.56382)$ & 19.5 & 1.0387 & 0.2 & 0.07749 & $20^\circ$ & 54 & 41\\
 \mr
q5\_i00\_high-e & 5 & $(0,0,0.6)$ & 21.125 & 0.8617 & 0.3 & 0.2126 & $0^\circ$ & 42 & 31\\
q5\_i10\_high-e& 5 & $(-0.10419, 0, 0.59089)$ & 21.125 & 0.8617 & 0.3 & 0.2128 & $10^\circ$ & 42 & 31\\
q5\_i20\_high-e & 5 & $(-0.20521,0,0.56382)$ &  21.125 & 0.8617 & 0.3 & 0.2134 & $20^\circ$ & 41 & 31\\
 \mr
 q7\_i00\_low-e & 7 & $(0,0,0.8)$ &19.5 & 1.0387 & 0.2 & 0.06657 & $0^\circ$ & 74 & 59\\
 q7\_i40\_low-e & 7 & $ (-0.51423,0,0.612836) $ & 19.5 & 1.0387 & 0.2 & 0.06942 & $40^\circ$ & 72 & 58\\
 q7\_i80\_low-e & 7 & $ (-0.787846,0,0.138919) $ & 19.5 & 1.0387 & 0.2 & 0.07954 & $80^\circ$ & 60 & 46\\
  \mr
q7\_i00\_high-e   & 7 & $(0,0,0.8)$ & 21.125 & 0.8617 & 0.3 & 0.2032 & $0^\circ$ & 60 & 45\\
q7\_i40\_high-e   & 7 & $ (-0.51423,0,0.612836) $ & 21.125 & 0.8617 & 0.3 & 0.2069 & $40^\circ$ & 55 & 42\\
q7\_i80\_high-e   & 7 & $ (-0.787846,0,0.138919) $ & 21.125 & 0.8617 & 0.3 & 0.2195 & $80^\circ$ & 44 & 32\\
\br
\end{tabular}\\
\end{table}

\subsection{Dynamics of simulated binaries}
\label{sec:SimulationQuantities}

In order to connect our simulations with analytic theory we need to
first construct quantities that can be used to extract fundamental
frequencies.  Numerical simulations yield the position vectors
${\bm x}_1(t)$ and ${\bm x}_2(t)$ of the Cartesian
coordinate-centres of the holes, and a spin-vector ${\bm \chi}_1(t)$
describing the spin angular momentum of the more massive hole.

\begin{figure}
\includegraphics[scale=0.4]{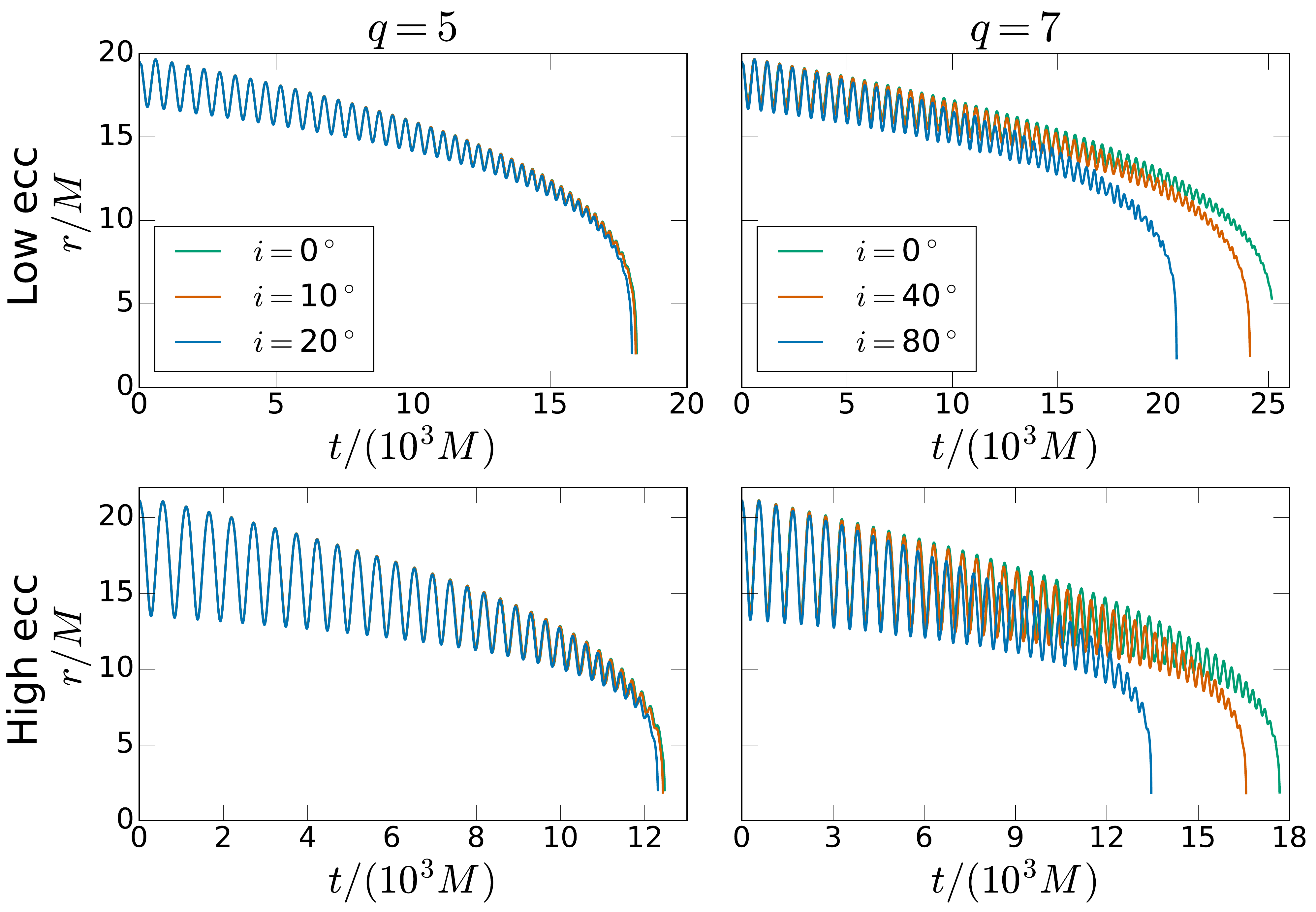}
\caption{\label{fig:BBHsepvst} Coordinate separations of our BBH simulations as a function of simulation time. The simulations are grouped into
    four panels according to mass-ratio (left, $q=5$; right, $q=7$) and
    eccentricity (top, low eccentricity; bottom, high eccentricity).  Each panel shows runs at the three different inclinations $i$.}
\end{figure}

From the position vectors we compute a radial separation vector
${\bm r}(t) \equiv {\bm x}_1(t) - {\bm x}_2(t)$ along with its
magnitude $r \equiv |{\bm r}|$.  The separation $r$ has no invariant
meaning, but it illustrates the dynamics of the inspiral; we thus plot
it in figure~\ref{fig:BBHsepvst}.  The number of extrema of $r$
indicate the number of apastron and periastron passages, and the
decline in eccentricity during the inspiral is evident in the
decreasing amplitude of the oscillations of $r$. 

\begin{figure}
\includegraphics[scale=0.4]{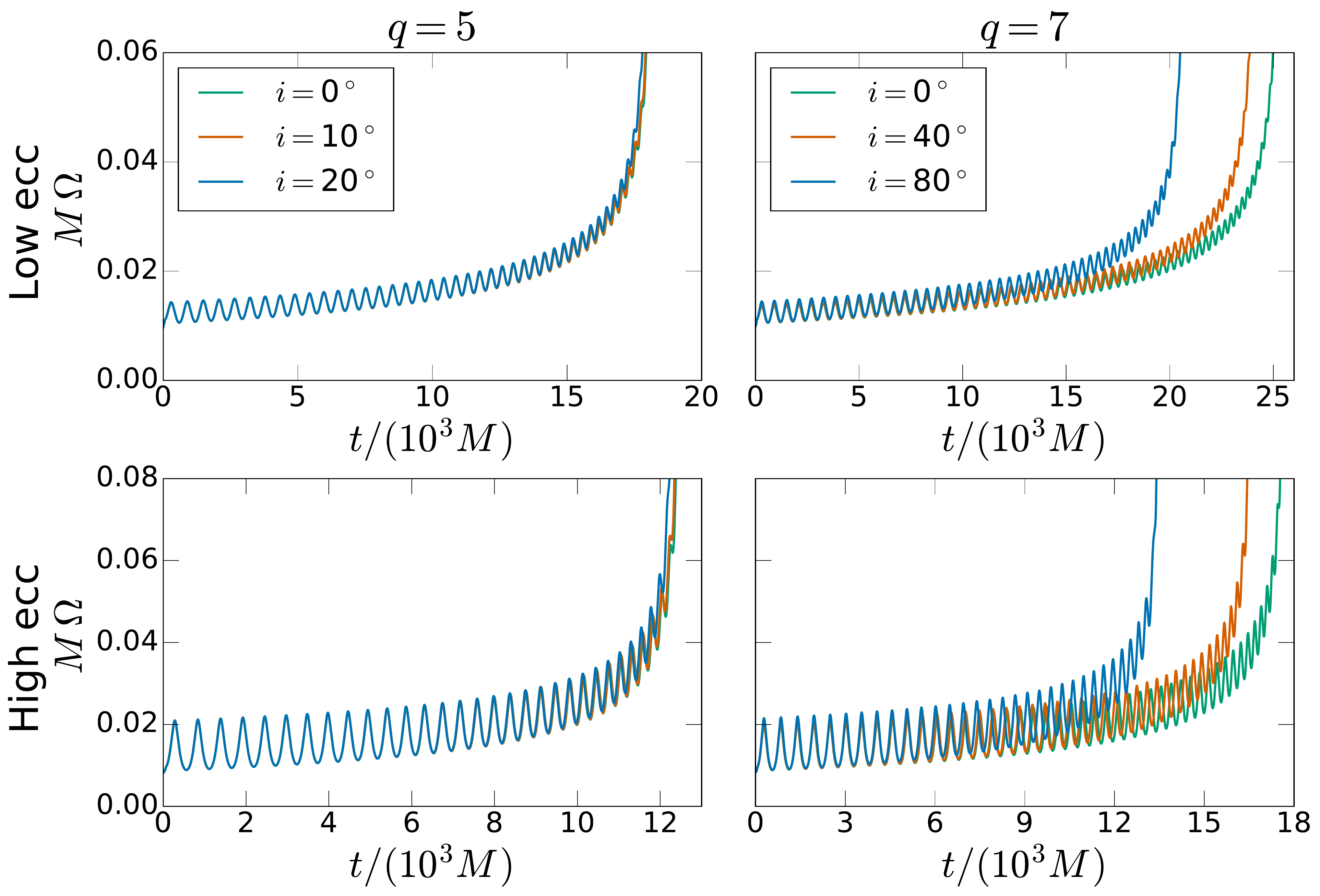}
\caption{\label{fig:BBHomvst} Orbital frequency $\Omega$ from \eqref{eq:OmegaOrb} as a function of simulation time, organized in the same way as in figure \ref{fig:BBHsepvst}.}
\end{figure}

For the equatorial runs we do not work with ${\bm r}$ directly. Instead we compute from it an orbital frequency~\cite{Buonanno:2010yk}
\begin{equation}
\label{eq:OmegaOrb}
\Omega(t) \equiv \frac{|{\bm r} \times \dot{\bm r}|}{r^2} \,,
\end{equation}
from which we are able to extract the fundamental frequencies $\Omega^\phi(t)$ and
$\Omega^r(t)$.
In practice, we compute $\dot{\bm r} = d {\bm r}/dt$ with a 3rd order Savitzky-Golay filter
windowed over groups of seven points
\cite{SavitzkyGolay1964}.
Our time coordinate
is asymptotically inertial \cite{Boyle2007}, justifying our
identification of simulation coordinate frequencies with those
measured by asymptotic, inertial observers.

In figure~\ref{fig:BBHomvst} we plot $\Omega(t)$ for our simulations, 
which clearly shows 
modulations due to the eccentricity. The modulation amplitude
decreases throughout the inspiral as eccentricity is radiated
away. The modulations are not simple sinusoidal oscillations about a
chirping mean, but instead exhibit sharper peaks at periastron, an
effect more pronounced at higher eccentricities.  

Our analysis of eccentric motion is based on the
  oscillatory features of $\Omega(t)$.  For non-dissipative orbits,
  one would simply utilize the extrema of $\Omega(t)$.  However, as shown in
  figure~\ref{fig:BBHomvst}, GW-driven inspiral adds an overall
  monotonic trend to the radial motion of the binary, so that extrema
  of $\Omega(t)$ do not correspond precisely to the reversals of the
  underlying oscillatory behavior.  Should the growth rate of the
  monotonic trend ever exceed the range of the growth rate of the
  oscillatory one, as can easily happen at late inspiral, or even
  early inspiral when eccentricity is low, $\Omega(t)$ will no longer
  display extrema at all. 

\begin{figure}
\centering
\centering{\includegraphics[scale=0.33]{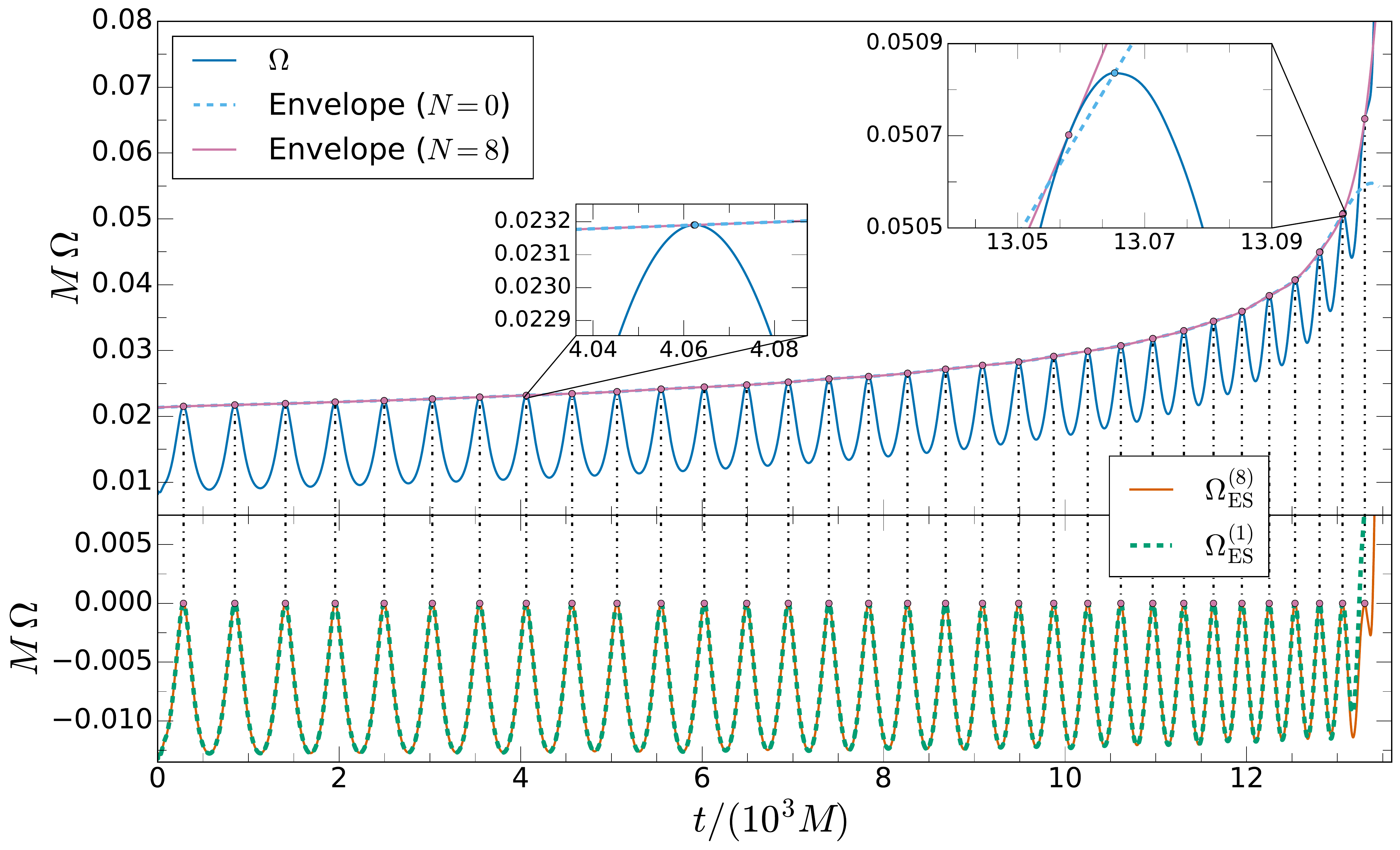}}
\caption{\label{fig:refinement} Illustration of refinement procedure applied to $\Omega$ from \eqref{eq:OmegaOrb} 
  computed from our q7\_i80\_high-e run. The top panel shows
  $\Omega$
  itself, while the bottom shows the same after $N=1$ and $N=8$ iterations of envelope subtraction, when 
  the location of the maxima are stable up to machine precision. The
  refined curve is computed by subtracting a 4th order spline
  interpolation of the maxima (the ``envelope'') from the
  original curve; each iteration denotes a subsequent envelope
  subtraction from the result of this procedure. This results in an
  envelope which, rather than passing through the maxima of the
  original curve, lies tangent to it, as seen in the
    inserts. The difference between the maxima of each
    iteration is due to dissipation and thus becomes more pronounced
  further into the inspiral. 
  The refinement procedure also sometimes finds
  additional peaks which were saddle points in the original curve
  (note the differences between the envelopes as well as the
  $\Omega_{\rm ES}^{(N)}$ curves close to the end of the inspiral).} 
\end{figure}

We begin by computing the maxima and minima of
  $\Omega(t)$\footnote{Extrema are found in practice by a
      quartic spline fit to the discretely sampled $\Omega$ near each
      extremum, and computing the extremum of the spline.  
      which we denote by $\Omega_i^+$ and $\Omega_i^-$.}. We denote
  the times where $\Omega(t)$ takes its maximum and minimum values as
  $t_i^+$ and $t_i^-$, respectively, where $i$ labels successive maxima/minima.  
    For a conservative
orbit these extrema are precisely the points of physical interest and
the procedure is complete.  

For inspiraling orbits, we apply a refinement procedure
based on the subtraction of the envelope of the upper extrema. This procedure
is motivated by analyzing a simple model in \ref{sec:EnvSubtract} and is
illustrated in figure~\ref{fig:refinement}. We compute a spline-fit to the extrema $\{(t_i^+, \Omega(t_i^+))\}$, and subtract this fit from $\Omega(t)$.
  This results in the first-stage ($N\!=\!1$)
``envelope subtracted" frequency $\Omega_{\rm ES}^{(1)}(t)$, cf.~the
  lower panel of figure~\ref{fig:refinement}.  We now iterate this
  procedure: Find the abscissas of the maxima of
  $\Omega_{\rm ES}^{(1)}(t)$; compute a spline-fit to the respective
  points in $\Omega(t)$; subtract to generate the $N\!=\!2$ envelope
  subtraction, $\Omega_{\rm ES}^{(2)}(t)$.  We iterate the procedure
  $N_f$-times until no more change occurs 
    to within machine precision (typically $N_f\!=\!8$). 
    We redefine the extremal times $t^+_i$,
$t^-_i$ as the maxima/minima of $\Omega_{\rm ES}^{(N_f)}(t)$. 
We finally define $\Omega^\pm(t)$ as the spline-interpolant to $\{(t_i^\pm, \Omega(t_i^\pm) ) \}$.  (In practice, we are most interested in $\Omega^+$ when computing radial frequencies, since
the sharper peaking of $\Omega(t)$ at periastron allows the former to
be located more accurately.)
Envelope-subtraction generates a spline-interpolant $\Omega^+(t)$ which appears 
tangent to the original $\Omega(t)$, whereas the original envelope
passing through the bare peaks would cross it
(cf. the upper right inset of figure~\ref{fig:refinement}). 
During our analysis of inclined runs in section~\ref{sec:Inclined} we
use an envelope-subtracted separation
\begin{equation}\label{eq:rES}
r_{\rm ES}(t) \equiv r(t) - r^+(t),
\end{equation}
which is generated with the identical procedure, substituting $r(t)$ for $\Omega(t)$ above.

\begin{figure}
\centering
\includegraphics[scale=0.41]{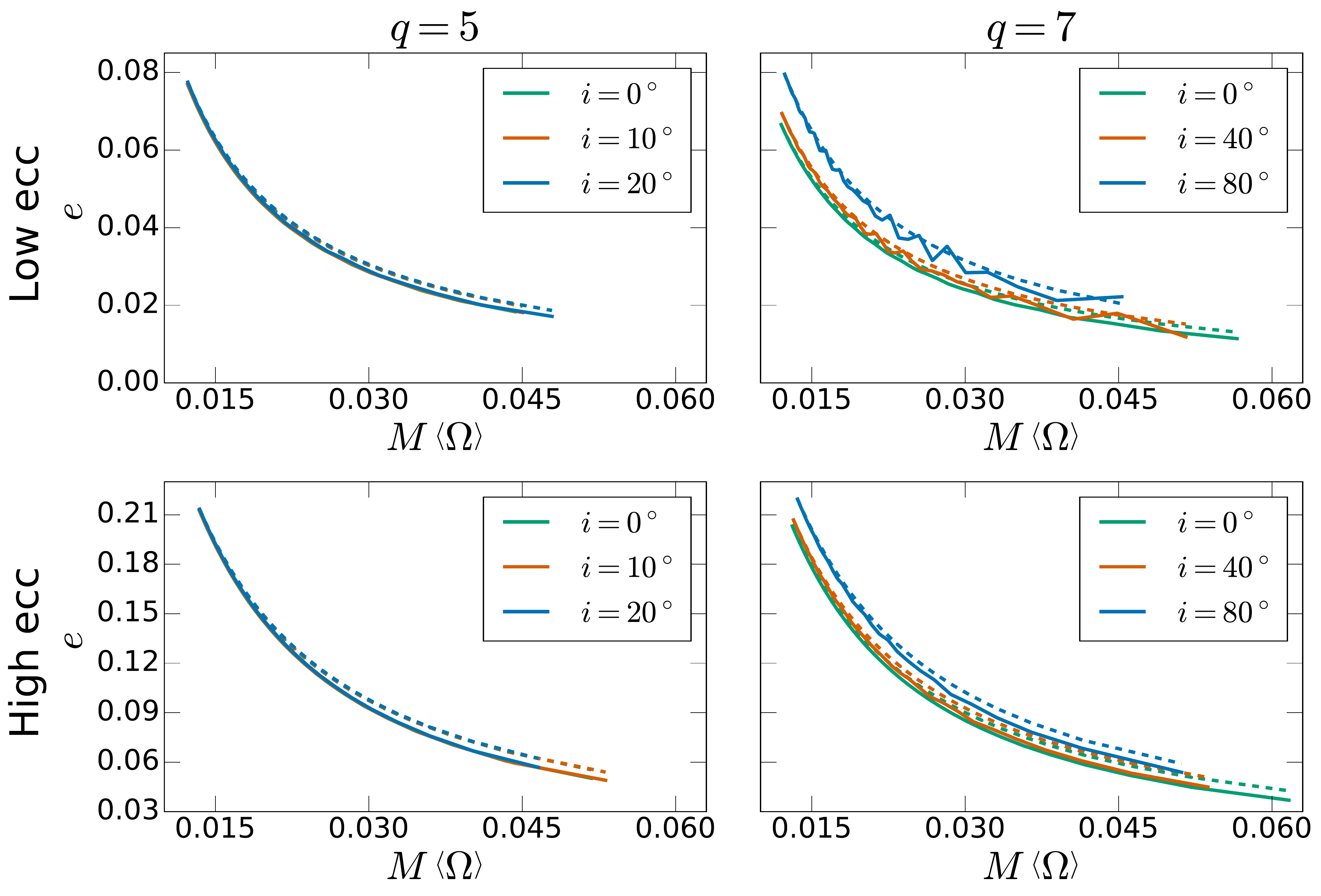}
\caption{\label{fig:eccentricities} 
Eccentricity vs. orbital frequency for our numerical simulations (solid lines).
The simulations are grouped into four panels according to mass-ratio (left, $q=5$; right, $q=7$) and initial eccentricity (top, low eccentricity; bottom, high eccentricity).  Dashed lines show the
eccentricity decay predicted by leading-order post-Newtonian formulae~\cite{PetersMathews1963,Peters1964}.
}
\end{figure}

With the output quantities of envelope-subtraction, we can now define several more useful quantities.  
First, the average orbital frequency in the $i$-th radial oscillation period is given by
\begin{equation}
\label{eq:omphiequatorial}
\langle \Omega \rangle_i \equiv \frac{1}{t^+_{i+1} - t^+_i} \int_{t^+_i}^{t^+_{i+1}}\Omega(t) \mathrm{d}t,
\end{equation}
and we assign the values $\langle \Omega \rangle_i $ to the midpoint times
\begin{equation}
\label{eq:MidpointTimes}
\tilde t_i^+\equiv \frac{t^+_i +t^+_{i+1}} {2} \,. 
\end{equation}
Furthermore,  we define eccentricity using the Keplerian formula
\begin{equation}
\label{eq:eccentricity}
e(t) \equiv \frac{\sqrt{\Omega^+(t)} - \sqrt{\Omega^-(t)}}{\sqrt{\Omega^+(t)} + \sqrt{\Omega^-(t)}} \,.
\end{equation}
We define $e(t)$ based on orbital frequency rather
  than the separation
since the extrema of the former can be located with marginally
better accuracy.  

For the inclined runs, we also need analogs for the Boyer-Lindquist
coordinate $\theta$.  We take $\theta$ to be
the angle between ${\bm \chi}_1$ and ${\bm r}$; thus
\begin{equation}
  \cos \theta \equiv \frac{{\bm \chi}_1 \cdot {\bm r}}{\chi_1 r}.
\end{equation}
We view the ``equatorial plane" as that normal to ${\bm \chi}_1$.
The coordinate $\theta$, or alternatively $\cos \theta$,
  oscillates between maximum and minumum values during the orbit, and
  this motion is the basis of our definition of the polar frequency of
  the inclined binaries discussed in section~\ref{sec:Inclined}.

\subsection{Eccentricity decay}

As a first result of our definitions in
  section~\ref{sec:SimulationQuantities}, we consider the decay of
  eccentricity for equatorial binaries.
  Figure~\ref{fig:eccentricities} plots $e( \tilde t_i^+)$ at the
  midpoint times $\tilde t_i^+$ versus
  $\langle\Omega\rangle_i$.  This figure illustrates the fast decay of
  eccentricity in a GW-driven inspiral, as first computed
  by~\cite{PetersMathews1963,Peters1964}.
  Figure~\ref{fig:eccentricities} also includes a direct comparison
  with the leading-order post-Newtonian results
  of~\cite{PetersMathews1963,Peters1964}, through a parametric plot
  of $\Omega$ as a function of eccentricity,
  cf. equation~\eqref{eq:Omega_vs_e} derived in~\ref{sec:PetersEvo}.
  The agreement for eccentricity decay is quite remarkable,
  generalizing results in~\cite{Mroue2010} to the larger
  eccentricities considered here.


\section{Equatorial binaries}
\label{sec:EquatorialFrequencies}

In this section we extract the two fundamental frequencies $\Omega^r$
and $\Omega^\phi$ for our eccentric, equatorial simulations.  For
these binaries, the frequencies have simple definitions in terms of
finite-time averages over radial passages and can be extracted cleanly
once we account for dissipation.

\subsection{Frequencies for equatorial orbits}
\label{sec:Extrema}

Fourier-based methods of frequency extraction are inaccurate over the short timescales forced upon us by  by gravitational dissipation (see \ref{sec:FreqValidation}).
We therefore rely on time-averages or period measurements between extrema 
of the orbital frequency $\Omega(t)$, cf.~\eqref{eq:OmegaOrb}. 

For circular, equatorial, conservative orbits  $\Omega(t)$ is
constant and equals the frequency $\Omega^\phi$ of the motion in the orbital plane. Adding dissipative 
inspiral will cause $\Omega(t)$ to increase monotonically with time. 

Including eccentricity but not dissipation yields a periodic
$\Omega(t)$ whose time-average is the constant $\Omega^\phi$ of the
analogous circular, conservative case. These periodic oscillations
track the radial motion of the orbit, with successive maxima
$\Omega^+_i$ at periastron and minima $\Omega^-_i$ at apastron.  These
oscillations will not be symmetric about their mean, instead peaking 
near periastron. 
This foils attempts at frequency extraction using rolling fits~\cite{Mroue2010,LeTiec-Mroue:2011,Tiec:2013twa}
which have been used in the past to extract orbital frequencies in low-eccentricity BBH simulations.
It is difficult to find suitable fitting-functions at higher eccentricities without
introducing unacceptable model-dependence to the procedure.

Based on the envelope-subtracted maxima of the orbital
  frequency $\{(t^+_i, \Omega^+_i)\}$
  (cf.~section~\ref{sec:SimulationQuantities}), 
we define the radial frequency $\Omega^r$ through
the period between successive maxima,
\begin{equation}
\label{eq:omrequatorial}
\Omega^r_i \equiv \frac{2\pi}{t^+_{i+1} - t^+_i} \,.
\end{equation}
Further, we define the azimuthal frequency as the average orbital frequency \eqref{eq:omphiequatorial}, 
\begin{equation}
\Omega_i^\phi\equiv \langle\Omega\rangle_i,
\end{equation}
which holds exactly for equatorial Kerr orbits.
For each cycle we assign the values of $\Omega^r_i$ and $\Omega^\phi_i$ to the orbital midpoint times $\tilde t^+_i$, equation \eqref{eq:MidpointTimes}.
The $r$-$\phi$ precession rate is then
\begin{equation}
K^{r\phi}_i \equiv \frac{\Omega^r_i}{\Omega^\phi_i},
\end{equation}
which we parametrize as a function of $\Omega^\phi_i$. 
Our choice to define frequencies over \emph{one} radial
oscillation period is ideal for equatorial inspirals, as it minimizes
biasing by dissipation.

In order to compare the precession rate $K^{r\phi}$ to the precession
rate of an eccentric orbit in Kerr, we must identify both a particular
Kerr spacetime and a particular geodesic orbit to compare to.  In the
self-force formalism, the spacetime is expanded around a Kerr solution
with mass $M_{\rm Kerr}$ equal to that of the larger black hole, and
not the total mass of the system.  As such, we set the Kerr mass equal
to the mass of the larger black hole $M_{\rm Kerr} = m_1$ and the spin
parameter to $a_{\rm Kerr} / M_{\rm Kerr} = \chi_1$. With the  
Kerr spacetime fixed, two further parameters are needed to identify a
particular equatorial geodesic.
Since we wish to compare $K^{r \phi}$ to an equivalent test orbit, we
cannot use both $\Omega^r$ and $\Omega^\phi$ to select our reference
orbit.  We choose to identify our test orbit using $\Omega^\phi$ and
the eccentricity $e(\Omega^\phi)$ computed using \eqref{eq:eccentricity}.

To compute the geodesic precession rate, we numerically invert the
procedure for finding $\Omega^\phi (p, e ,i = 0)$ from
\cite{Schmidt:2002qk} to find $p$ as a function of $\Omega^\phi$ when
restricted to the eccentricities $e(\Omega^\phi)$. 
This allows us to obtain analytic predictions
$K^{r\phi}_\mathrm{Kerr}(m_1 \Omega^\phi_i, e_i)$ for the midpoint times $\tilde t^+_i$ 
at which we extract $K^{r \phi}$ from our simulations.
We subtract this analytic prediction from our extracted precession rates 
at each $\Omega^\phi_i$.  We can then reparameterize
these differences in terms of any other quantity defined at the same $\tilde t^+_i$.

This procedure is not ideal: the eccentricity estimator
\eqref{eq:eccentricity} is chosen to give stable results and does not
correspond to the Boyer-Lindquist coordinate eccentricity used in
Kerr.  However, the frequencies are only weakly
dependent on $e$ at small eccentricity (e.g.~\cite{Schmidt:2002qk})
and so this approximation impacts $K^{r\phi}$ also weakly. Ideally, we
would compute a third scalar quantity $f$, averaged over the orbit,
parametrize it by our measured $\Omega^r$ rather than the
eccentricity, and compare the extracted time series $f_i$ to analytic
predictions for $f(\Omega^\phi, \Omega^r)$. An example would be the
third, polar frequency $\Omega^\theta$. While $\Omega^\theta$ remains 
well-defined for equatorial orbits, we cannot in practice extract it
from simulation without measurable polar motion. With no such third quantity
available, we compare our eccentricity-dependent prediction to
simulation, in order to get a qualitative understanding of the
observed precession rate $K^{r\phi}$.

\subsection{Results}

\begin{figure}
\centering
\includegraphics[width=0.495\columnwidth]{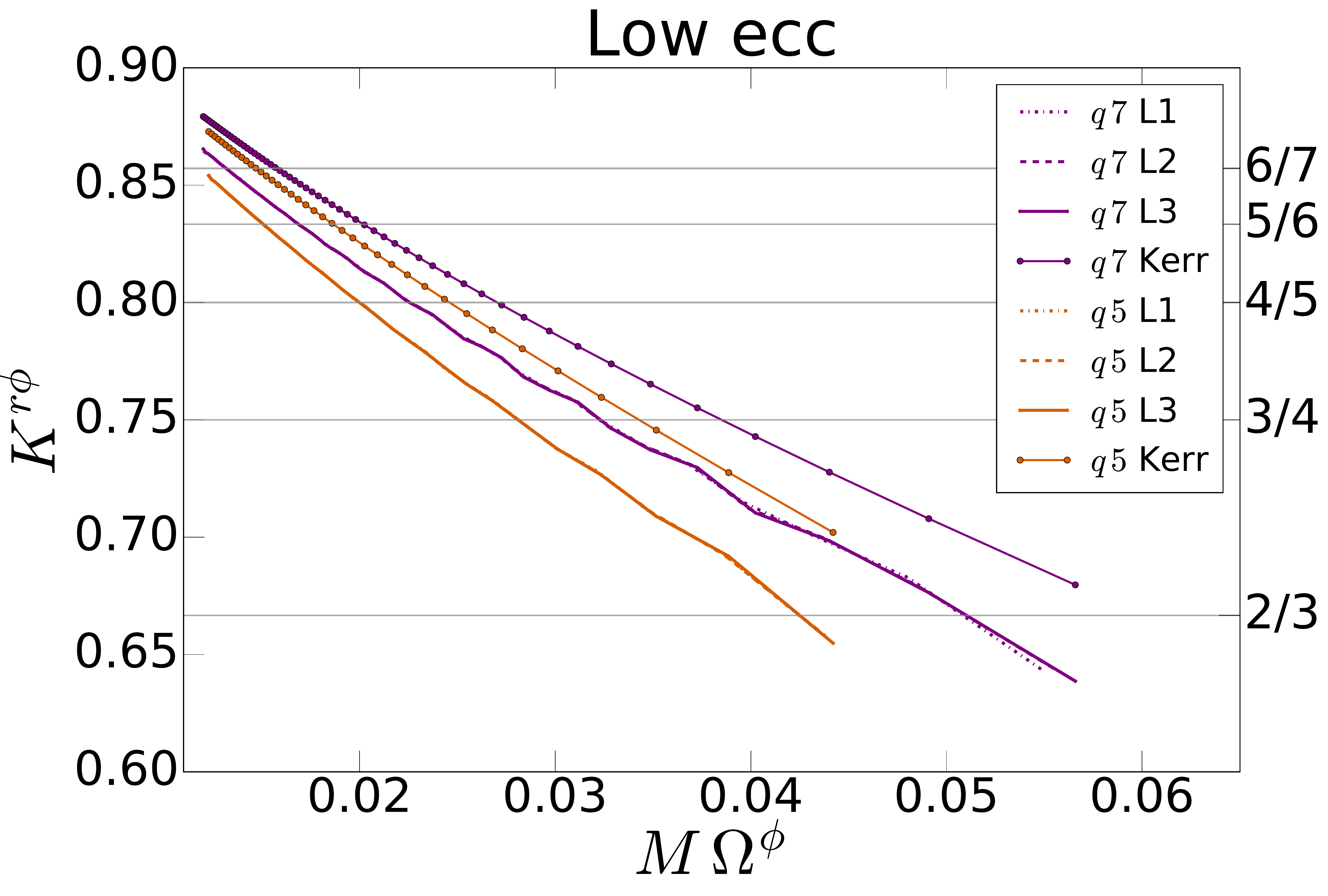}
\includegraphics[width=0.495\columnwidth]{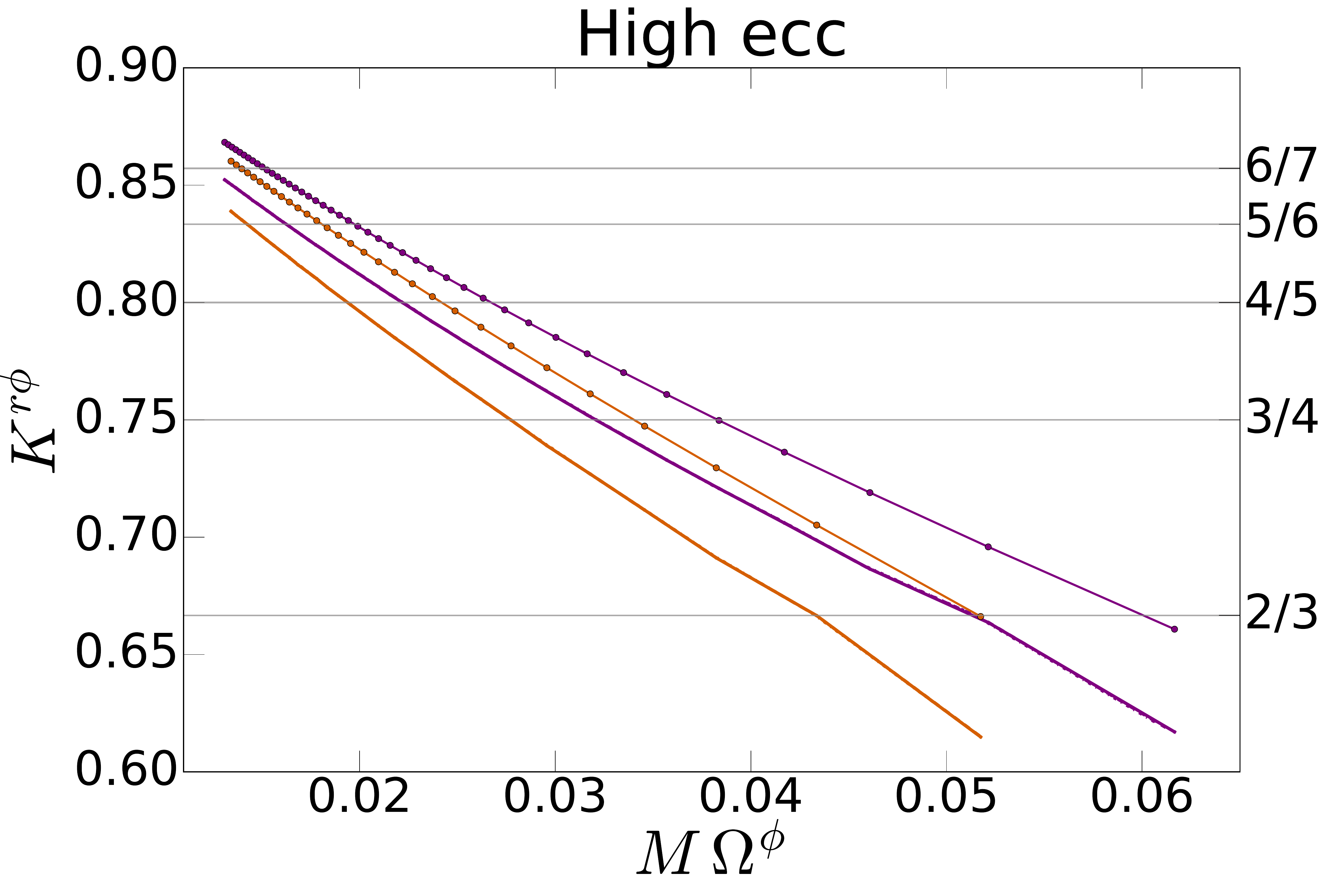}
\caption{\label{fig:ratios_equatorial} 
Precession rates $K^{r \phi}$ as a function of $M\Omega^\phi$ extracted 
from our equatorial simulations (thick solid, dashed, and dotted lines)
 and as predicted for a Kerr black hole of mass $M_{\rm Kerr} = m_1$ and
 the same eccentricity evolution (thin lines with symbols). We plot multiple 
resolutions for our simulations as indicated. {\it Left}: Precession of 
our low-eccentricity simulations. {\it Right}: Precession of our
 high-eccentricity simulations.}
\end{figure}

We plot our extracted precession rates $K^{r\phi}_i$ along with the geodesic predictions 
as functions of azimuthal frequency $M \Omega^\phi_i$ in figure \ref{fig:ratios_equatorial}.
Several numerical resolutions are plotted, and we find agreement
in our precession rates across resolutions.  In addition, it is
apparent that the finite mass corrections increase the magnitude of
general relativistic precession, by decreasing $\Omega^r$ as compared
to $\Omega^\phi$.  Higher eccentricities also enhance the precession
rate.   Figure~\ref{fig:ratios_equatorial} enables identification of $r$-$\phi$
resonances, which are marked with horizontal lines.
Our binaries plunge before the high-order $r$-$\phi$ resonances which
are expected to generate resonant kicks.  Thus, although the analytical results
of~\cite{vandeMeent:2014raa} indicate these kinds of resonant kicks
could be promising at mass ratios comparable to ours, we see that in reality the rate of inspiral is too high and significantly
higher mass ratios are required to access the relevant resonances.

\begin{figure}
\centering
\includegraphics[width=0.495\columnwidth]{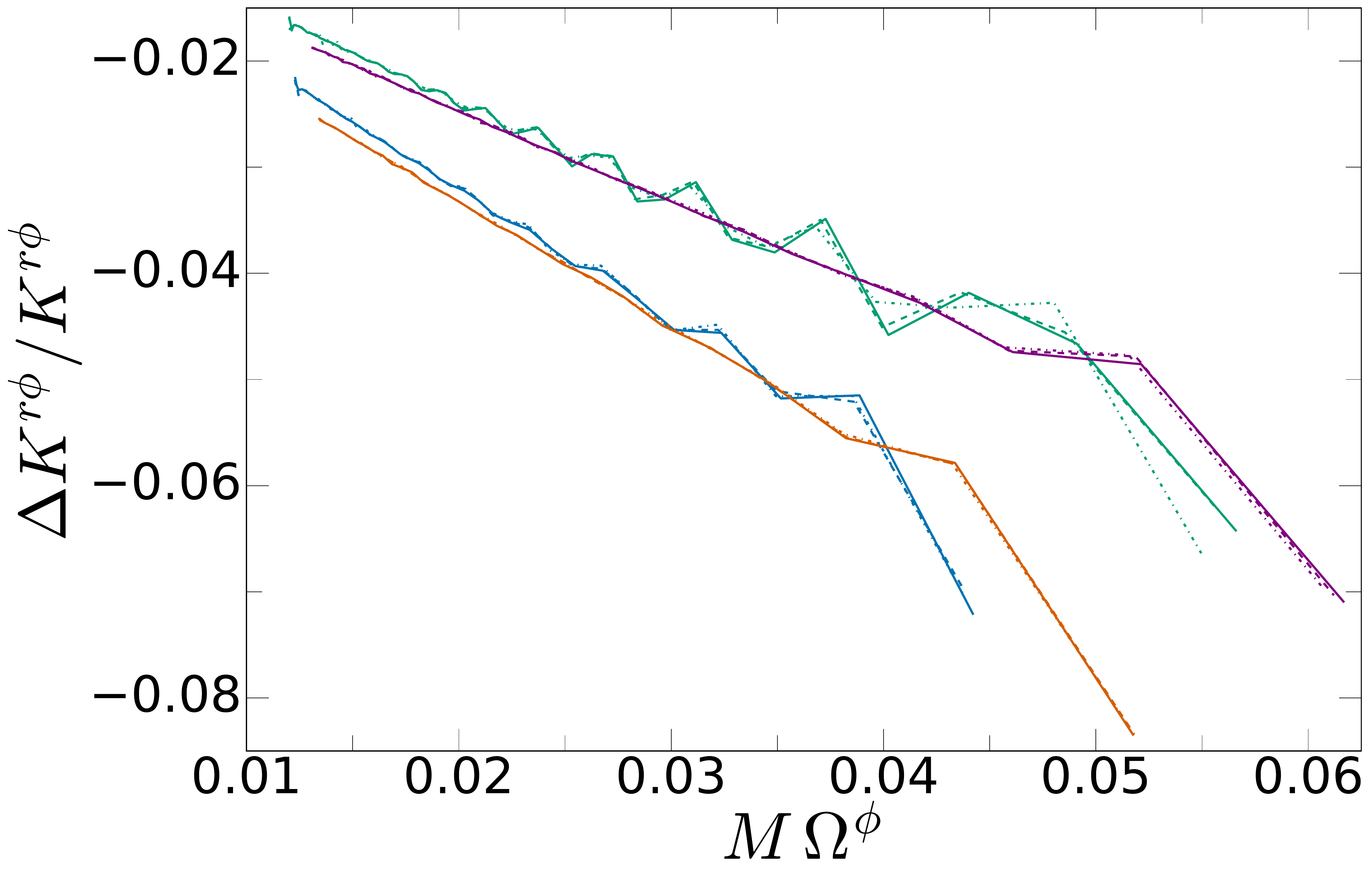}
\includegraphics[width=0.495\columnwidth]{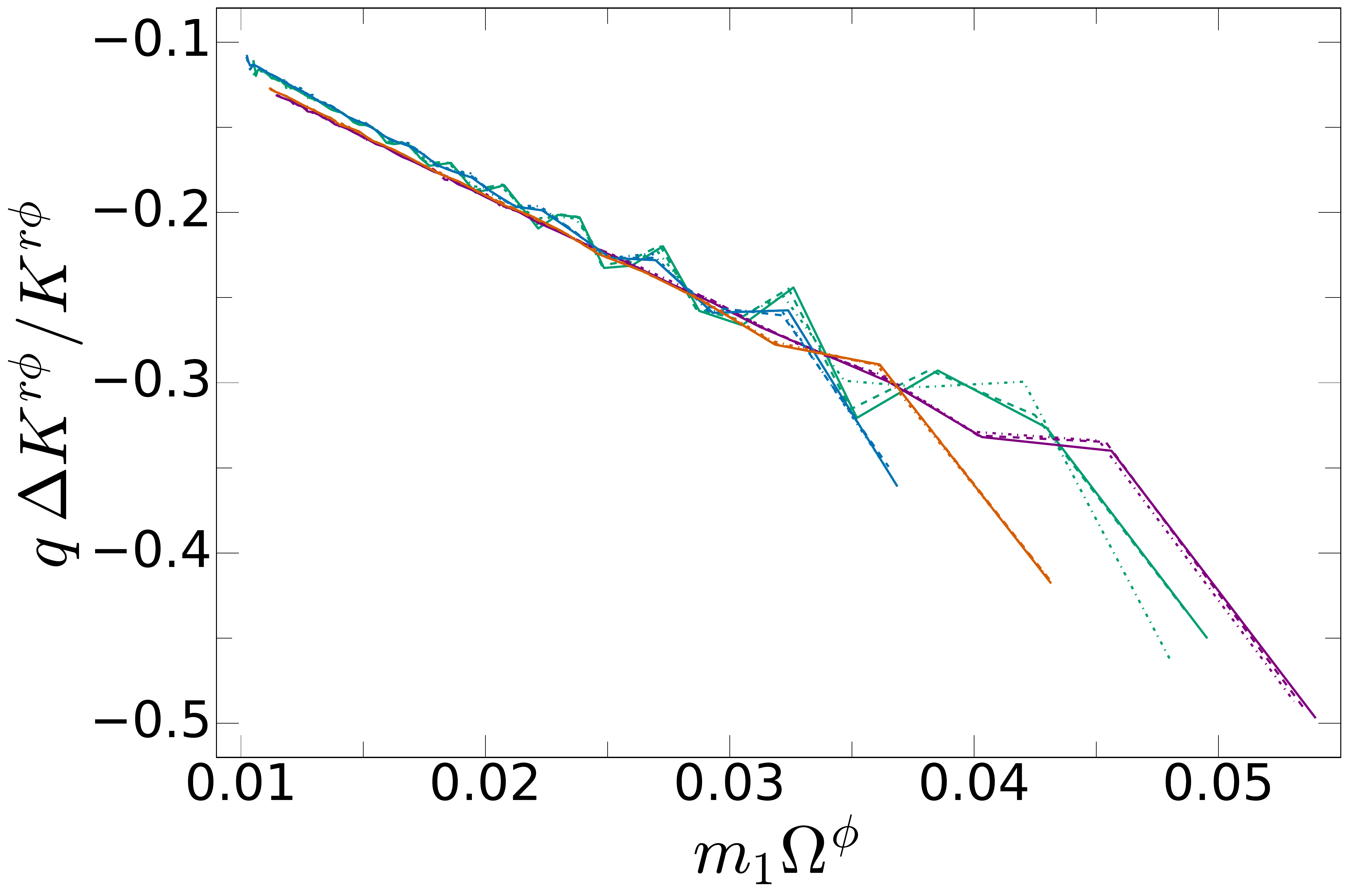}
\caption{\label{fig:rEquatorialCompare} 
Comparison between the precession rates for our equatorial binary
inspirals and Kerr geodesic theory. Specifically, we plot the
differences $\Delta K^{r\phi}$ between the numerical extraction and
geodesic prediction for the $r$-$\phi$ frequency ratio, divided by the
numerical $K^{r\phi}$.}
\end{figure}

In the left panel of figure \ref{fig:rEquatorialCompare} we plot the
relative difference between the numerical precession rates and
geodesic predictions,
$\Delta K^{r\phi}_i = K^{r\phi}_{{\rm Numeric,} i} - K^{r\phi}_{{\rm Kerr,}i}$,
normalized by the numerical $K^{r\phi}_i$. 
We see that the geodesic prediction captures the
precession rate well for our binaries, up to corrections of 
less than ten percent.  As expected, the $q=7$ simulations are closer
to the geodesic predictions than the $q=5$ simulations.  Inspecting
each mass ratio separately, we see that doubling the
eccentricity has less than a percent impact on the relative
differences.

Finally, we use our simulations to extract the $O(1/q)$
corrections to the geodesic precession rate, i.e.~the 
leading conservative SF correction to $K^{r\phi}$.  In the right
panel of figure \ref{fig:rEquatorialCompare}, we rescale the relative
differences by $q$, and plot our results against $m_1 \Omega^\phi_i$.
This guarantees that we are expanding our results around the same test
mass limit in all cases.  Remarkably, we see that after this rescaling
the $q=5$ and $q=7$ curves are nearly identical in both the
high-eccentricity and low-eccentricity cases.  This implies several
things.  First, that the higher order SF effects must be quite small,
since a priori an $O(1/q^2)$ correction would generate a similar split
in the rescaled $q =5$ and $q = 7$ curves as seen in the left panel of
figure \ref{fig:rEquatorialCompare}.  Next, it must be true that the
spin-dependence of the $O(1/q)$ SF correction is almost entirely
captured by the prefactor, which is normalized out in each case.
Finally, comparing the high- and low-eccentricity cases to each other,
we see the same limited dependence on eccentricity as discussed for
the left hand panel.

All together, we conclude that the periastron precession rate takes
the form
\begin{equation}\label{eq:Krph-equatorial}
K^{r\phi}(m_1 \Omega^\phi) = K^{r\phi}_{\rm Kerr}(m_1 \Omega^\phi)\left[1 + \frac{1}{q} \Delta K^{r\phi}_{\rm SF} + O\left(\frac{1}{q^2} \right)\right] \,.
\end{equation}
where $\Delta K^{r\phi}_{\rm SF}$ is nearly independent of spin and
eccentricity at modestly large spins and low eccentricities, and the
further $O(1/q^2)$ terms are numerically small.

\section{Inclined binaries}
\label{sec:Inclined}


In this section, we extract the frequencies of motion for our eccentric, inclined simulations.
In these cases, precession of the orbital plane and the black hole spin introduces a non-trivial dependence between all three of the characteristic frequencies.
This complicates our procedure, but we are able to cleanly extract $K^{r\theta}$ and compare to Kerr geodesics.
We find that our binaries pass through low order resonances in $r$-$\theta$ motion, and we investigate the fluxes of energy and angular momentum from our simulations to search for the imprint of these resonances.

\subsection{Definition of frequencies} 

  For inclined runs, there are two periodic modulations which both
  imprint themselves on the dynamics,
  cf.~(\ref{eq:mino_time}) and~(\ref{eq:PhiEoM}).  When
  choosing averaging-periods, we cannot honour both periodicities
  simultaneously.  Empirically, we find that windows over the radial
  rather than polar periods yield smoother frequencies. 
  For instance, $\Omega$ varies
  much more strongly between periastron and apastron than for
  different values of $\theta$.  These findings depend on the
  eccentricities considered here: we expect that if eccentricity were
  reduced at fixed inclination, eventually the polar oscillations would
  become more important than the radial oscillations.  

The radial motion 
couples to the polar motion through spin-orbit coupling 
via terms proportional to $\hat{\bm \chi}_1\cdot\hat {\bm r}=\cos\theta$, which results in a
$\cos \theta$-dependent modulation to the envelopes of $\Omega(t)$ and
$r(t)$. This introduces some oscillation to both $\Omega^r$ defined in~\eqref{eq:omrequatorial} 
and especially to the eccentricity $e$~\eqref{eq:eccentricity}. 
Nevertheless these effects are
relatively small compared to those which would be introduced by choice
of a different window for the $\Omega^r$ computation, and we continue
to use the same relation as in the equatorial case: $\Omega^r$ is the
reciprocal of the periods between maxima of the orbital frequency
$\Omega(t)$, \eqref{eq:omrequatorial}.

\begin{figure}
\centering
\centering{\includegraphics[scale=0.38]{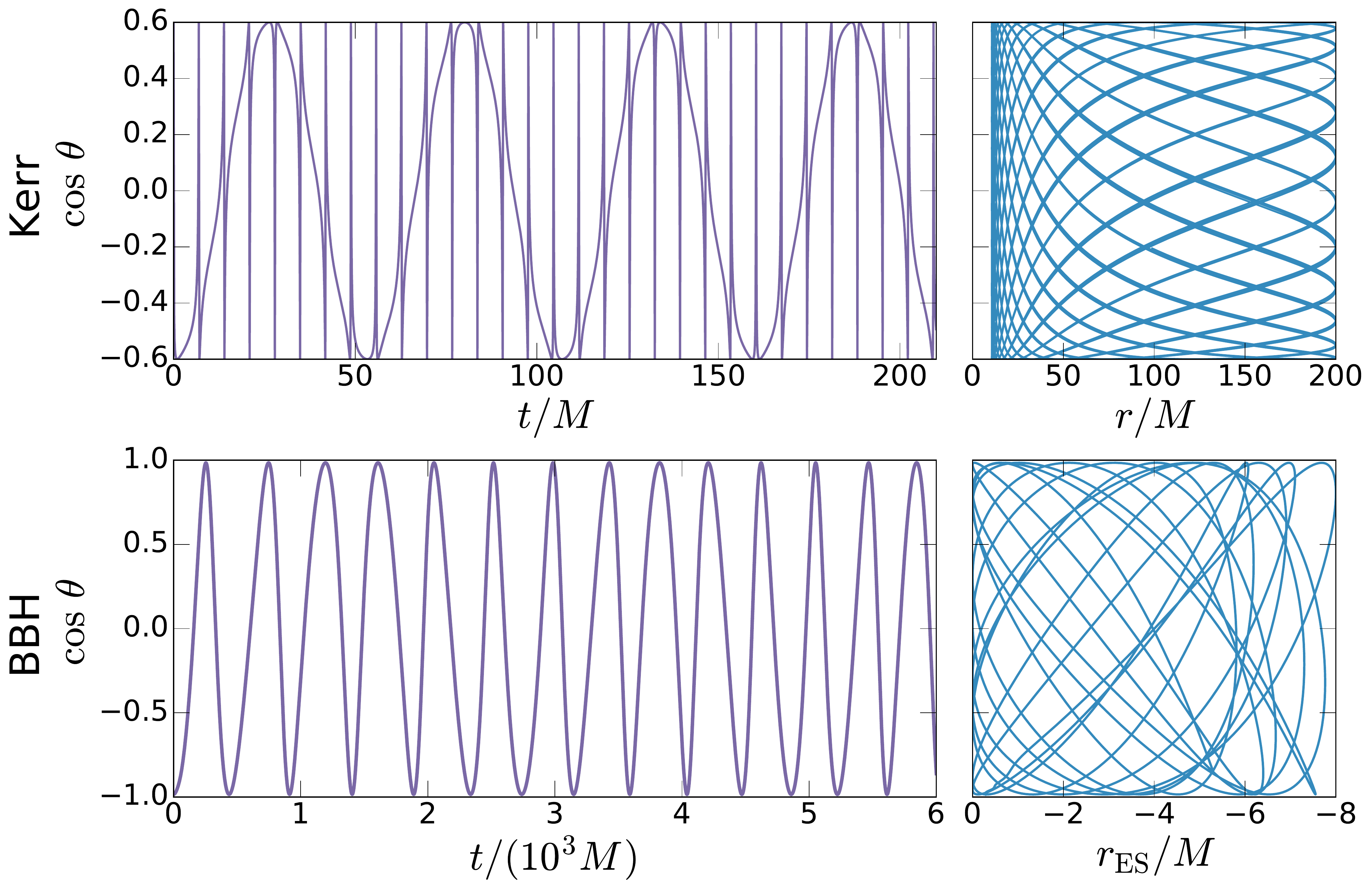}}
\caption{\label{fig:cosfig} Plot of $\cos \theta$ versus $t$ (left)
  and $r$ (right) for a Kerr geodesic with semilatus rectum $p =20M$,
  eccentricity $e=0.9$, $i = 36^\circ$, and
  spin parameter $\chi=0.9$ (top), compared with our high mass ratio
  and eccentricity $i=80^\circ$ BBH simulation (bottom). In the BBH
  case we use the envelope subtracted separation
  $r_{\rm ES}$, computed as described in section~\ref{sec:Simulations}. 
  Note in particular the strong dependence
  of $\partial_r \cos \theta$ upon $r$, most obviously visible in the
  Kerr phase plot. This strongly modulates the polar motion in time,
  resulting in the beating effect visible in both time series
  plots. This effect necessitates a more complex frequency-extraction
  strategy than taking simple peak-to-peak periods of the polar
    motion. }
\end{figure}

We now turn to the polar frequency $\Omega^\theta$, which we would
like to define in terms of the binary black hole system's polar motion 
$\theta(t)$. This angle $\theta(t)$ shows clear modulations due to the 
interdependence of the radial and polar motion: roughly, $\theta(t)$ 
varies most quickly near periastron. While the resultant modulations to
$\theta(t)$ are most pronounced
for highly eccentric orbits such as the Kerr geodesic shown in the upper 
  panel of figure~\ref{fig:cosfig}, they are still visible by eye at the
  eccentricities accessible to numerical simulations (cf.~the lower panel of 
  figure~\ref{fig:cosfig}). A straightforward 
  definition of polar frequency
  based on extrema of $\theta(t)$, while satisfactory for conservative orbits 
  once averaged over infinite polar cycles, suffers from substantial 
  interval-to-interval variations during a simulation, depending on the 
  radial phase at successive $\theta(t)$ extrema.

The most obvious way to account for the dominant dependence on separation
  is to measure the polar frequency over intervals that begin and end at
  fixed radial phase. While we do not have access to such a phase in general,
  any reasonable definition of one will attain a fixed value at periastron. 
  We therefore employ the same periastron-to-periastron intervals
  as those we used to define $\Omega^r$ \eqref{eq:omrequatorial}.
  As an additional advantage, this choice results in an $\Omega^\theta$ defined at the same
  midpoint times $\tilde t^+_i$ as the other frequencies, such that comparisons
  do not require an interpolation.

Because the radial and polar frequencies are distinct, the 
  periastron-to-periastron intervals $[t^+_i, t^+_{i+1}]$ do not correspond
  to an integer number of polar oscillations. To account for this we 
  define a polar phase $\chi^\theta(t)$ through equation~(\ref{eq:theta_phase}),
  and we define $\cos \theta_{\rm min}$ using the maximal envelope of $\cos \theta(t)$, 
  interpolated to the time of interest. 

The function $\chi^\theta(t)$ is made monotonic and continuous by suitable choice of 
quadrant when inverting
$\cos \chi^\theta$  followed by
suitable additions of multiples of $2\pi$. We then define
\begin{equation}
\label{eq:omegatheta}
\Omega^\theta_i \equiv \frac{\chi^\theta(t^+_{i+1}) - \chi^\theta(t^+_i)}{t^+_{i+1} - t^+_i}.
\end{equation}
In \ref{sec:FreqValidation} we show that this definition applied to Kerr limits to the 
exact polar frequency, with sub-percent error over single-cycle windows.

Finally, the azimuthal frequency $\Omega^\phi$ presents 
the most difficulties.
For inclined Kerr orbits, the averaged orbital frequency $\langle\Omega\rangle$  differs from the azimuthal frequency
$\Omega^\phi$.  One solution would be to
define $\Omega^\phi$ in terms of some azimuthal angle $\phi$, in a similar manner as in \eqref{eq:omegatheta}. 
Such a definition is possible, but all our attempts have resulted in an $\Omega^\phi$
which either fails to converge properly for Kerr orbits or which oscillates
unacceptably wildly for single-cycle windows.   To appreciate the difficulty,
note for example that the ``orbital plane" of the BBH simulation might reasonably be 
defined as the plane orthogonal either to the primary spin vector $\bm{\chi}_1$
or to the total angular momentum vector ${\bm J}$. Both choices recover the
desired limit for $q\to\infty$, but are different planes at finite $q$.

On the other hand $\langle\Omega\rangle$ does furnish a fairly smooth frequency
which is itself of some dynamical interest. 
 We therefore report 
$\langle\Omega\rangle$ computed via \eqref{eq:omphiequatorial} for
the inclined orbits as well.

\subsection{Precession rates}

\begin{figure}
\includegraphics[width=0.495\columnwidth]{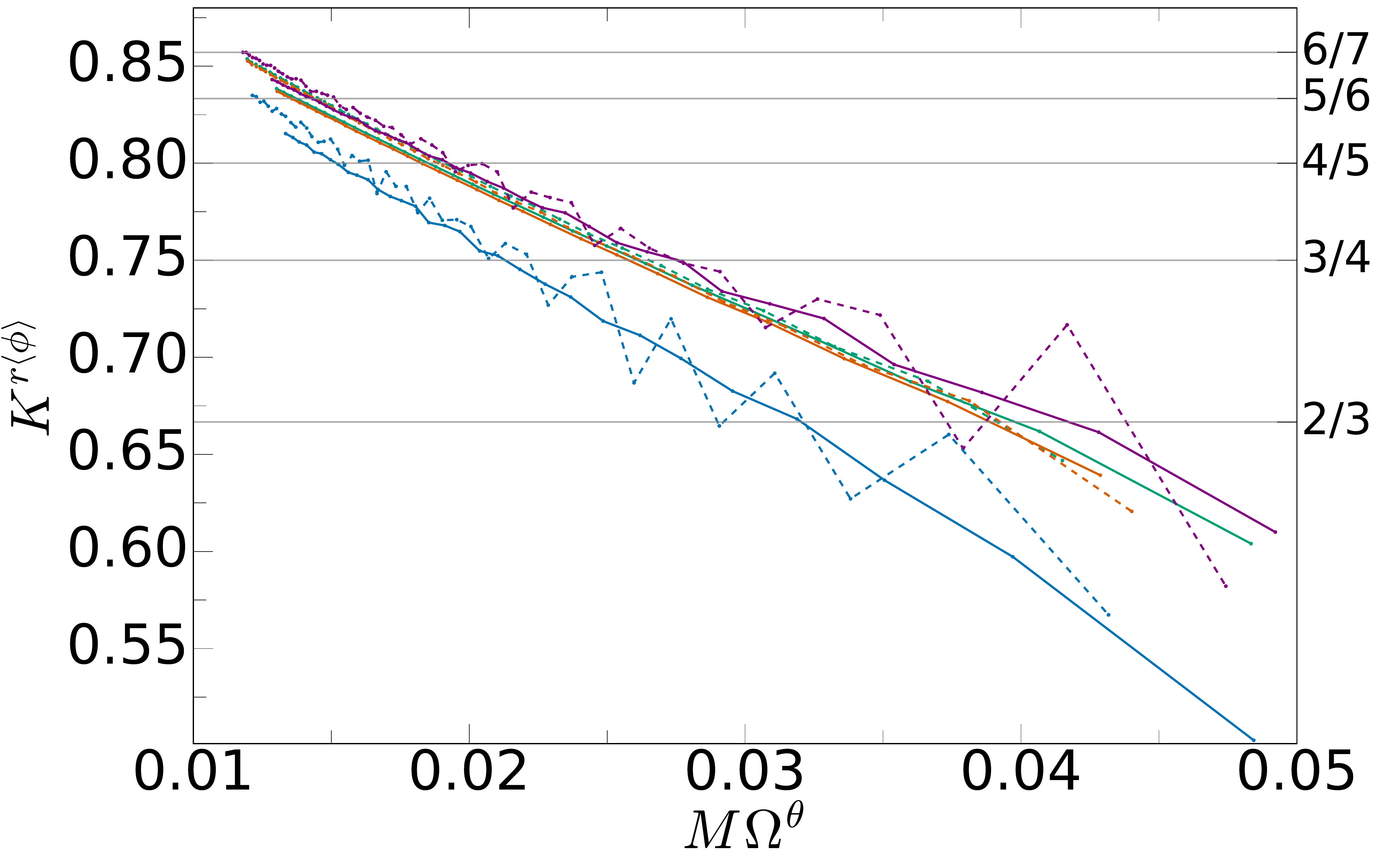}
\includegraphics[width=0.495\columnwidth]{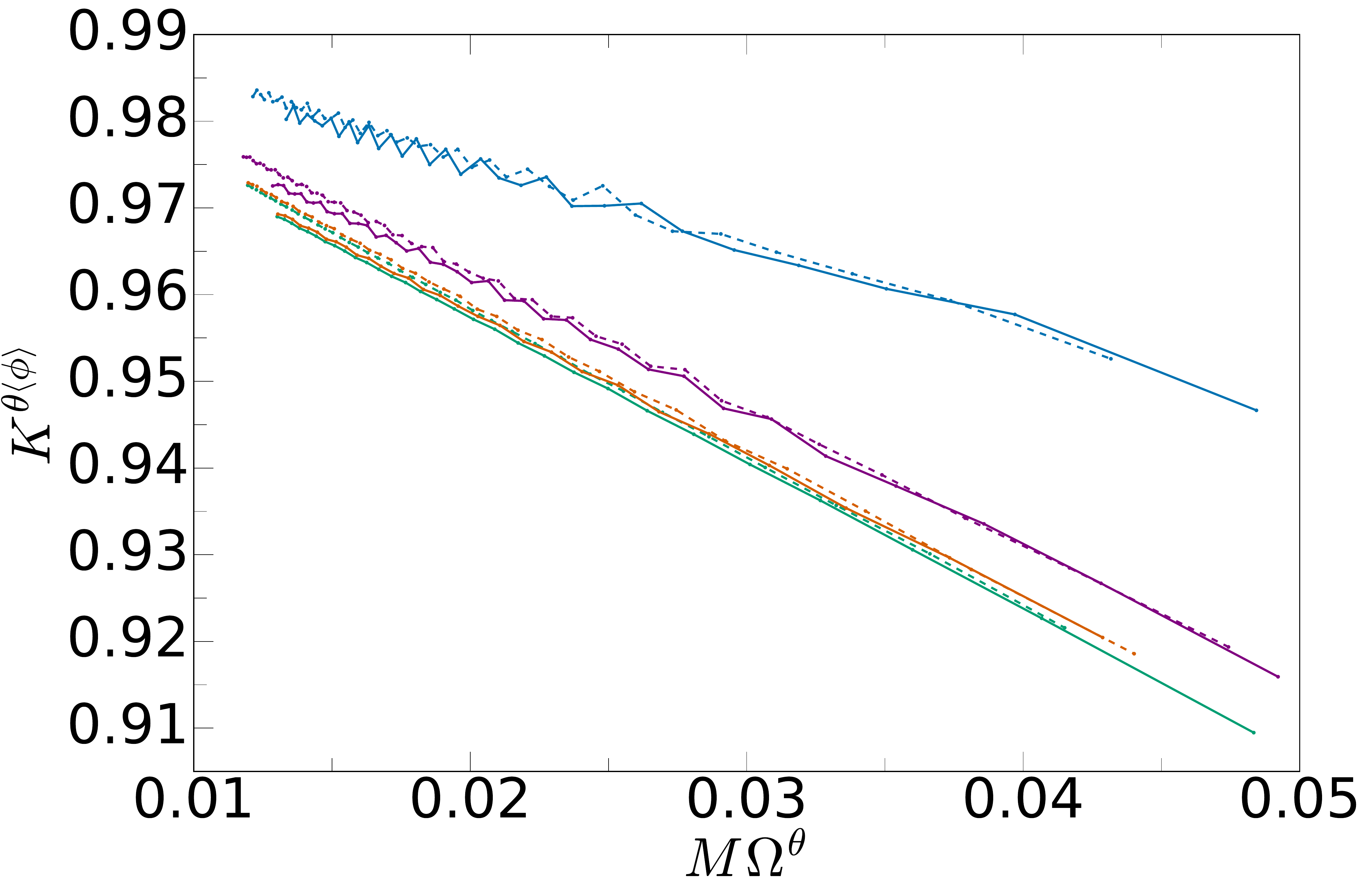} \\
\includegraphics[width=0.495\columnwidth]{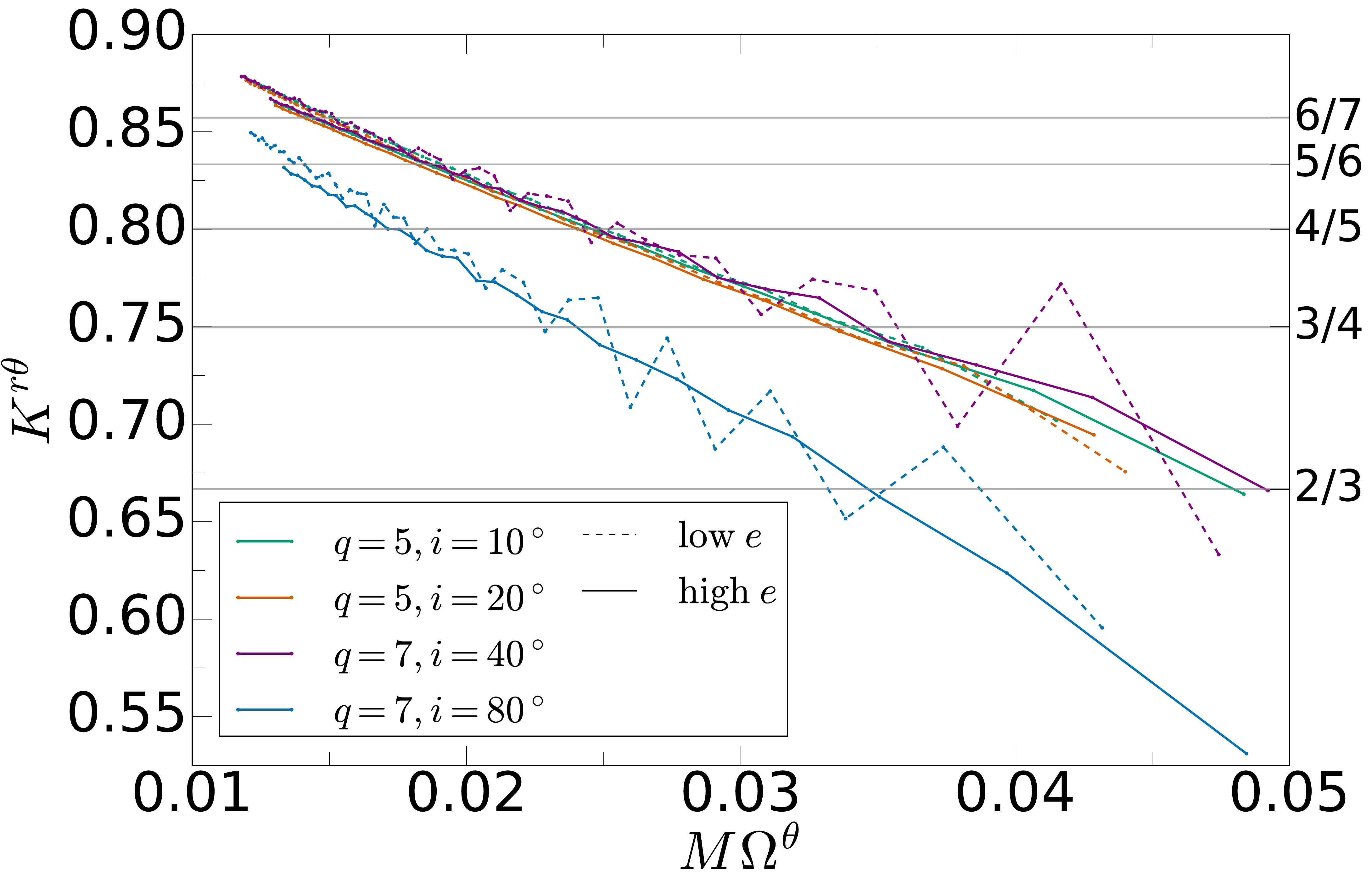}
\includegraphics[width=0.495\columnwidth]{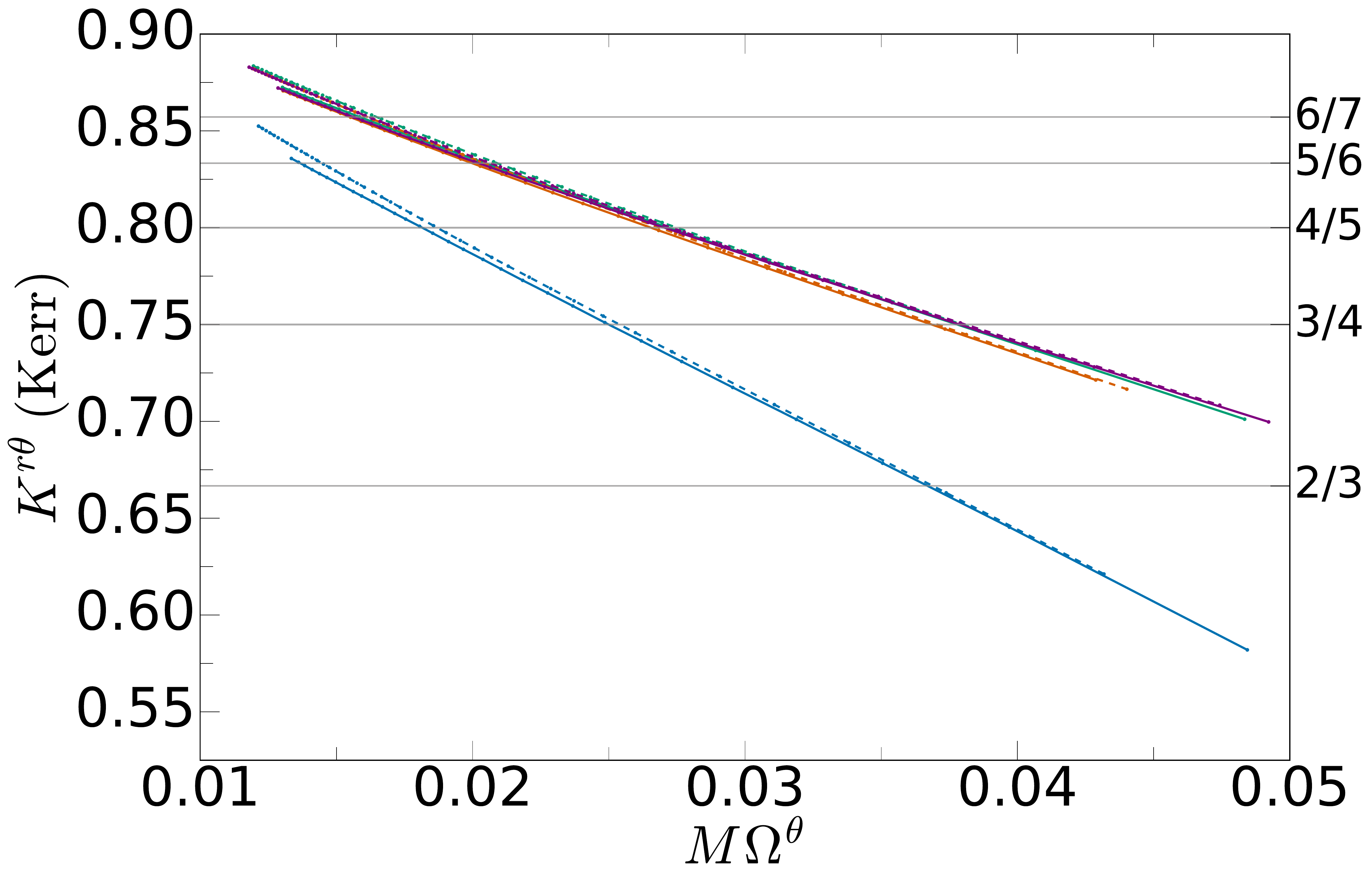}
\caption{\label{fig:kij_inc} Precession rates for our inclined runs at
  highest resolution (L3). Resonant precession rates are highlighted
  with solid horizontal bars. The bottom right plot shows the Kerr
  $K^{r\theta}$, to be compared to the BBH data at the bottom left.}
\end{figure}

From the frequencies $\Omega^r$, $\Omega^\theta$ and $\langle\Omega\rangle$ we compute
precession rates as frequency ratios $K^{ab}=\Omega^a/\Omega^b$.  
The results are shown in figure~\ref{fig:kij_inc}. As in 
section~\ref{sec:EquatorialFrequencies}, we plot against a frequency,
$M \Omega^\theta$, in order to mitigate any gauge effects.  Since we
cannot cleanly extract $\Omega^\phi$, we plot the ratios
$K^{r \langle \phi \rangle} \equiv \Omega^r/\langle \Omega
\rangle$ and $K^{\theta \langle \phi \rangle} \equiv \Omega^\theta/\langle \Omega
\rangle$ in addition to $K^{r\theta}$.  
We do not have analytic predictions for
the ratios involving $\langle \Omega\rangle$, 
but we can see some general trends. As the binary sweeps
to higher frequencies, periastron precession becomes stronger as
$\Omega^r$ lags further behind the other frequencies.  
Comparing the higher eccentricity to the lower eccentricity simulations at fixed $\Omega^\theta$, we see that $K^{ab}$ is smaller in the higher eccentricity cases. 
This is most apparent early in the inspirals (smaller $\Omega^\theta$), when there is a larger difference in $e$ between the cases; for all the $K^{ab}$, the solid lines lie at smaller values than the dashed, although the differences are small.
This is true also for the Kerr predictions (bottom right of figure~\ref{fig:kij_inc}).

There is no clear effect of inclination on the precession rates $K^{r\theta}$ and $K^{r\langle\phi\rangle}$, 
except for $i=80^\circ$, where they are systematically smaller, i.e.~periastron advance is more pronounced.
Meanwhile, the effect of increasing inclination on $K^{\theta \langle \phi \rangle}$ is
clearly visible: at higher inclinations, $\langle \Omega \rangle$ becomes more
and more nearly equal to $\Omega^\theta$, and so their ratio
approaches unity as we approach polar orbits.

Figure~\ref{fig:kij_inc} shows large oscillations in the extracted precession rates $K^{r\theta}$ and $K^{r\langle\phi\rangle}$ for the simulations {q7\_i40\_low-e} and {q7\_i80\_low-e}.  These oscillations can be traced to distinct features in the separation $r(t)$ for these two simulations.  Compared to the other simulations considered here, $r(t)$ shows extra modulations for these two cases. We discuss this in more detail in~\ref{sec:FreqValidation}.  We presently do not understand the origin of these additional features, nor the reason why they only appear in the simulations {q7\_i40\_low-e} and {q7\_i80\_low-e}.  All $q=7$ simulations were run with identical  source-code revision and configuration files.  

\begin{figure}
\includegraphics[width=0.51\columnwidth]{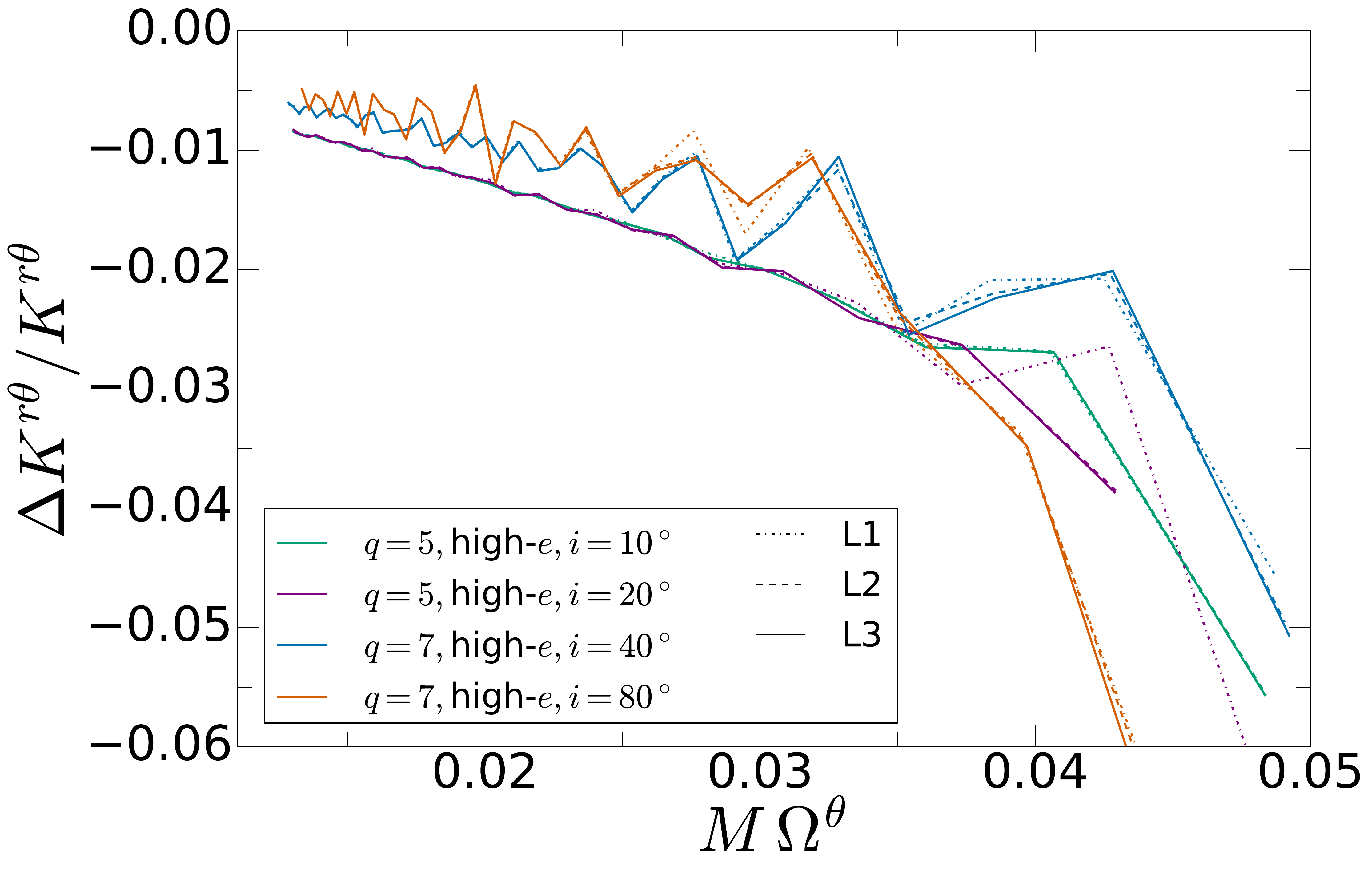}
\includegraphics[width=0.495\columnwidth]{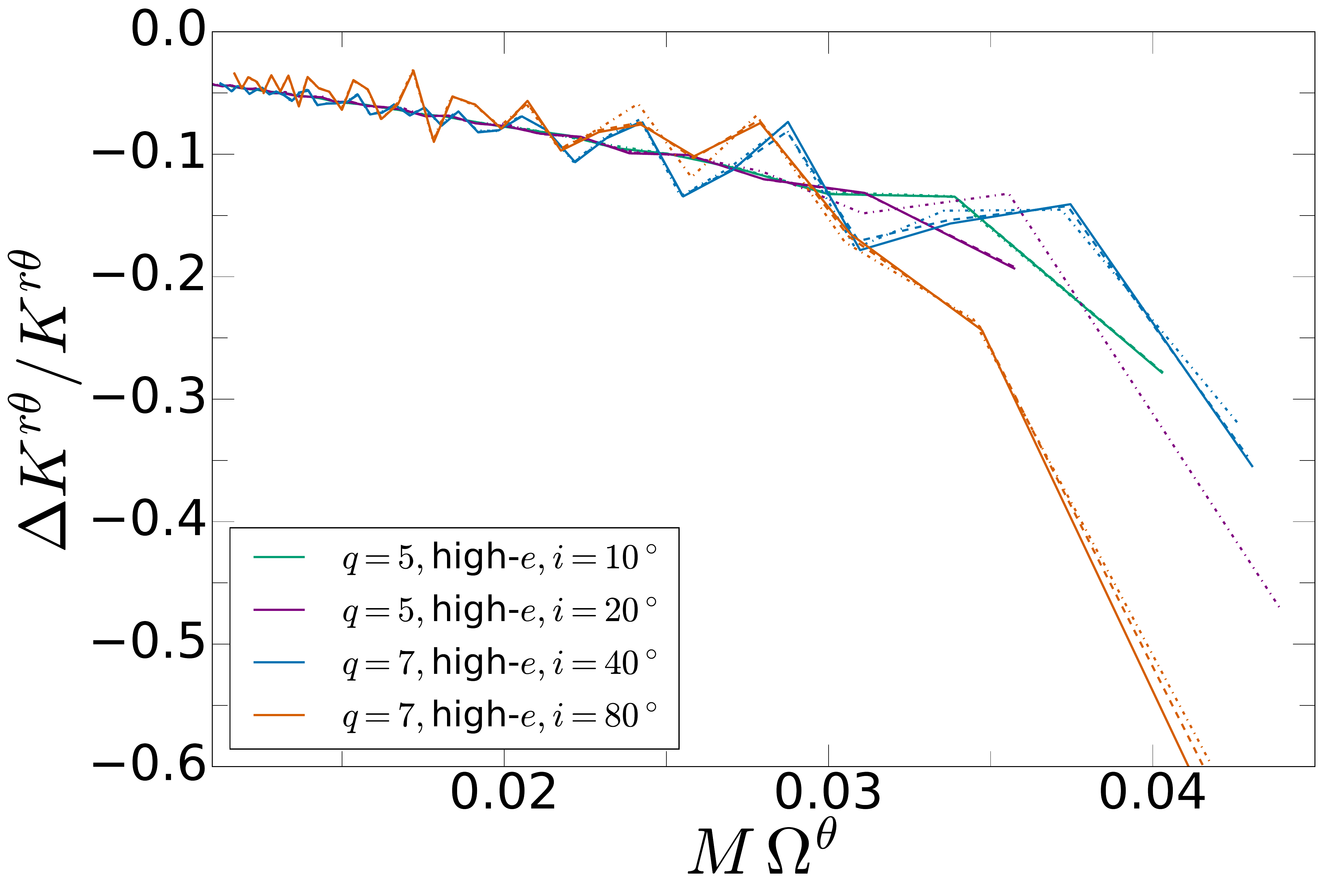} \\
\includegraphics[width=0.495\columnwidth]{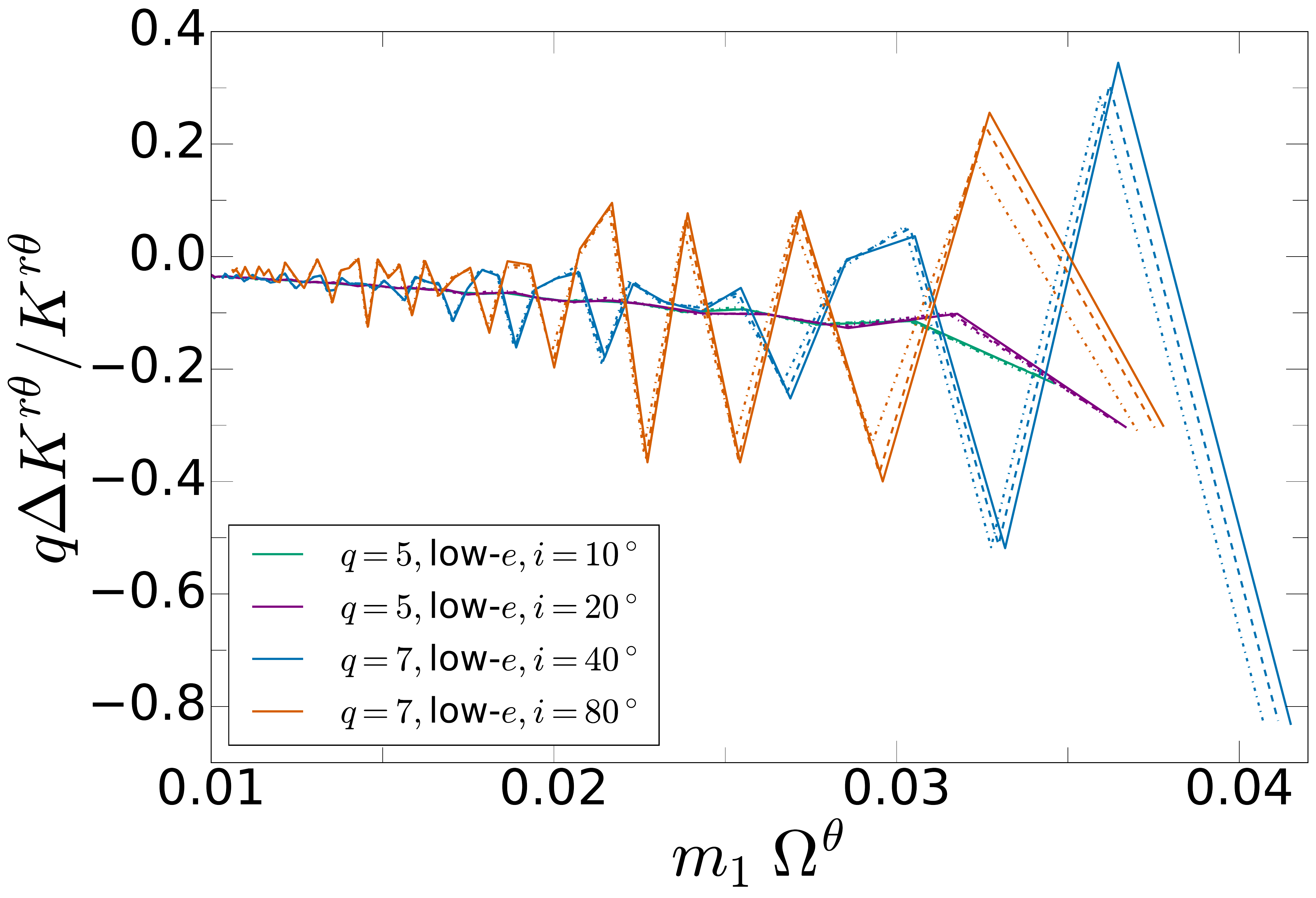}
\includegraphics[width=0.495\columnwidth]{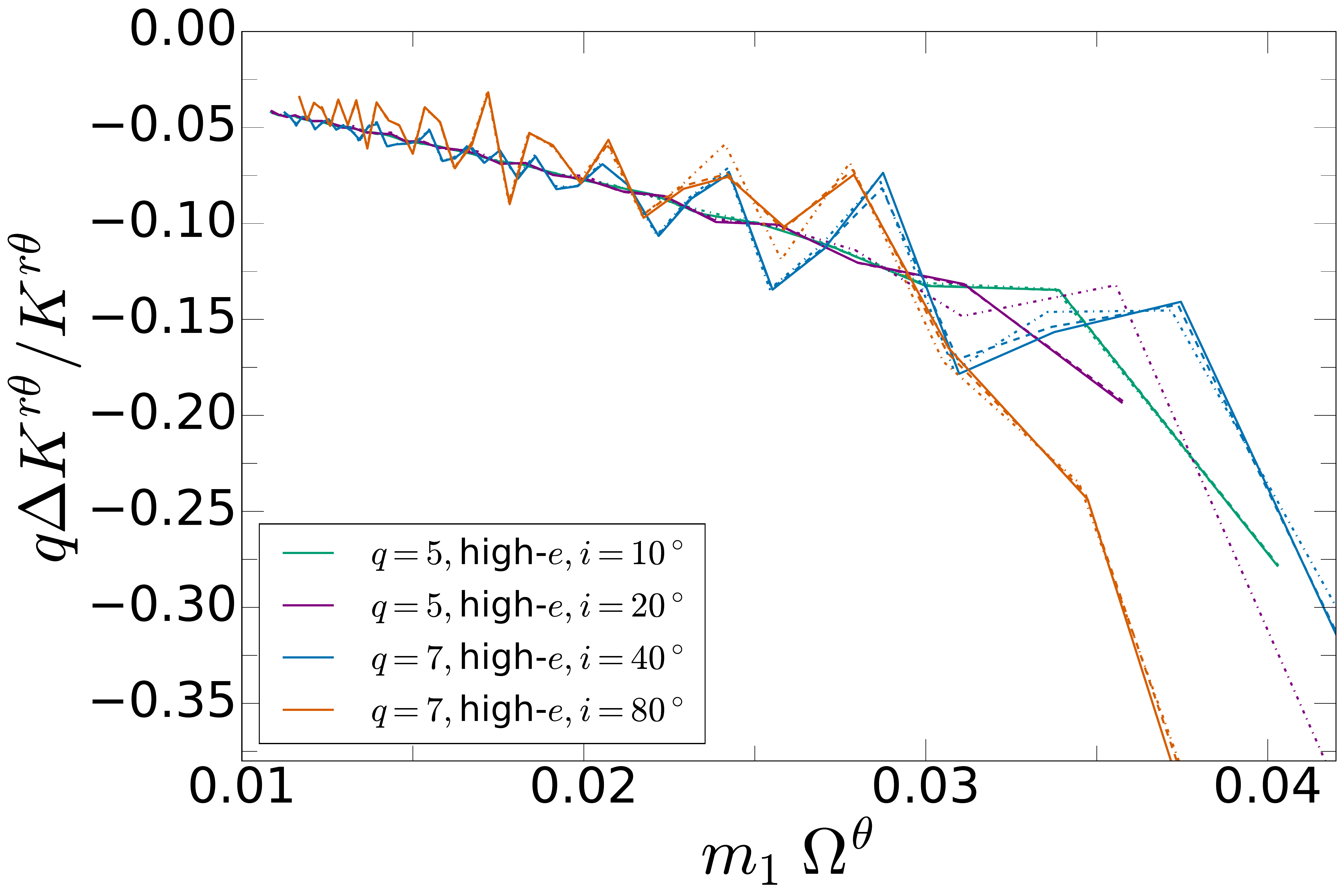}
\caption{\label{fig:krthcompare} {\it Top}: Comparison of the
    numerical extraction of $K^{r\theta}(M \Omega_\theta)$ to geodesic
    theory, for each of $q = 5$ and $q=7$ (higher eccentricity
    simulations only). The frequencies for the low eccentricity $q=7$
    simulations display large modulations, as can be seen in
    Fig.~\ref{fig:kij_inc}. {\it Bottom}: Plot of
    $\Delta K^{r\theta}/K^{r\theta}$ rescaled by $q$ and plotted
    against $m_1 \Omega_\theta$ for each simulation. The left panel
    features the lower eccentricity simulations, and the right panel
    features the higher eccentricity simulations.}
\end{figure}

Figure \ref{fig:krthcompare} illustrates the relative differences
between the geodesic predictions and our numerical precession rates.
We focus on $K^{r \theta}$, which we can compare to analytic theory.  
The top left panel of figure~\ref{fig:krthcompare} 
shows the $q=5$ data, where we find no discernible difference between the inclinations $i=10^\circ$ and $i=20^\circ$.  Further, as in the case of the equatorial inspirals, the 
difference between low eccentricity and high eccentricity is mild.
The top right panel shows the high eccentricity case for the $q=7$
inspirals, and again we see only a mild dependence on the inclination,
although it is discernible.   
The low $e$ runs for $q=7$, unfortunately, are dominated by the large modulations mentioned above and do not add any further insights.

The lower panels of figure \ref{fig:krthcompare} plot the rescaled
residual differences, with the lower eccentricity inspirals on the
left and the higher eccentricities on the right.  As seen before in
figure \ref{fig:rEquatorialCompare}, the fact that all the lines are
nearly on top of each other indicates that the $O(1/q^2)$ corrections
are unexpectedly small, cf.~equation~(\ref{eq:Krph-equatorial}).  Even in the case of the $q=7$, low
eccentricity runs, we see that the midline of the large modulations
agrees well with the $q=5$ scaled residuals.  Thus, the curves here
roughly give the leading SF conservative corrections to the
$r$-$\theta$ precession rate. Further, we see that the inclination,
spin, and eccentricity dependence of this correction is mostly
captured by scaling out the geodesic results.

\subsection{Resonances}
\label{sec:InclinedResonances}

The frequencies computed in section~\ref{sec:Inclined} can be
immediately exploited to detect resonances.
We focus on the lowest order $r$-$\theta$ resonances our simulations achieve, which are the 5:6, 4:5, 3:4 and 2:3 resonances.  
In particular, self-force calculations using PN models
\cite{Flanagan:2012kg,Berry:2016bit} have highlighted the importance of the 2:3 resonance.

\begin{figure}
\includegraphics[width=0.495\columnwidth]{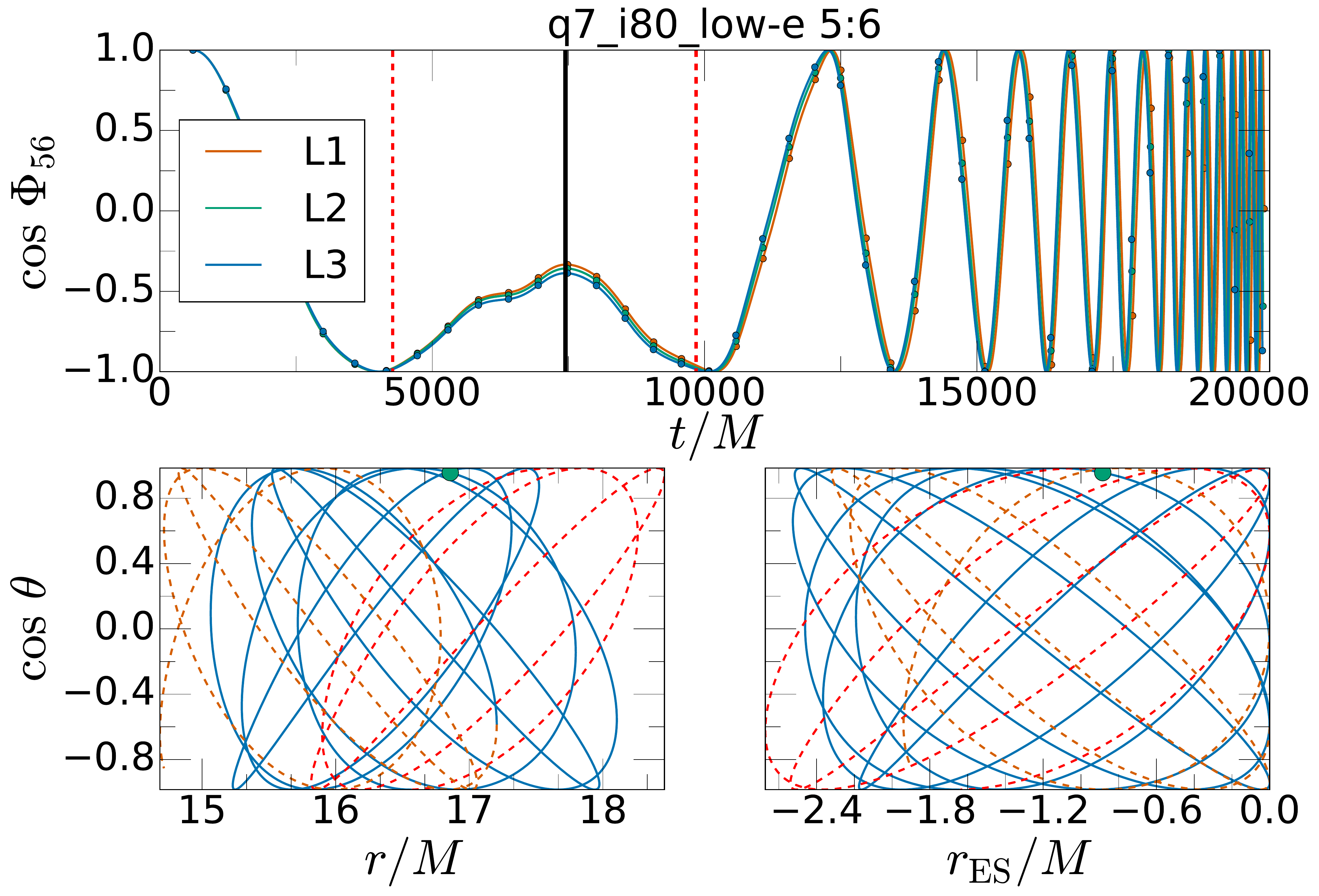}
$\quad$\includegraphics[width=0.495\columnwidth]{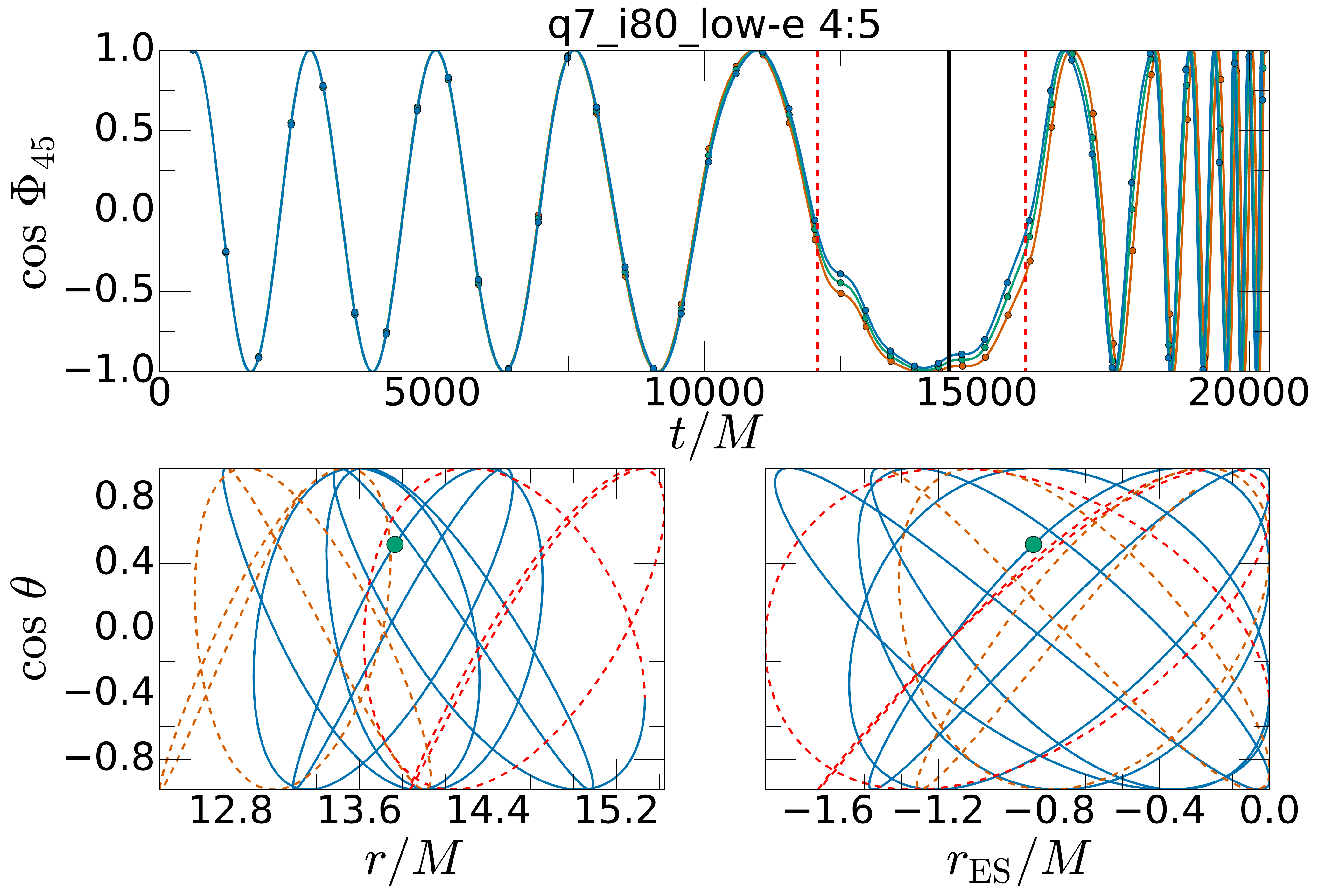}\\
\includegraphics[width=0.495\columnwidth]{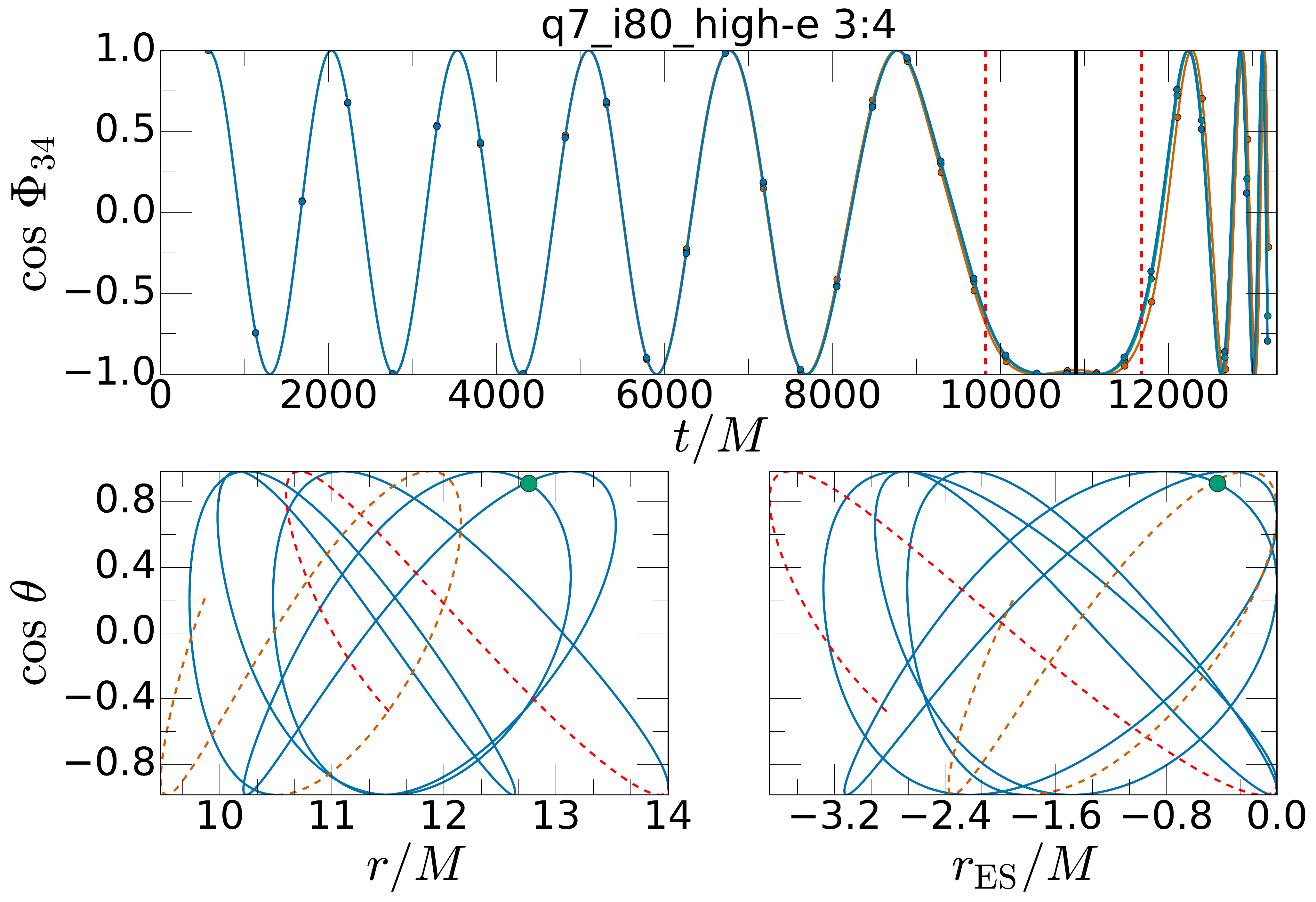}
$\quad$
\includegraphics[width=0.495\columnwidth]{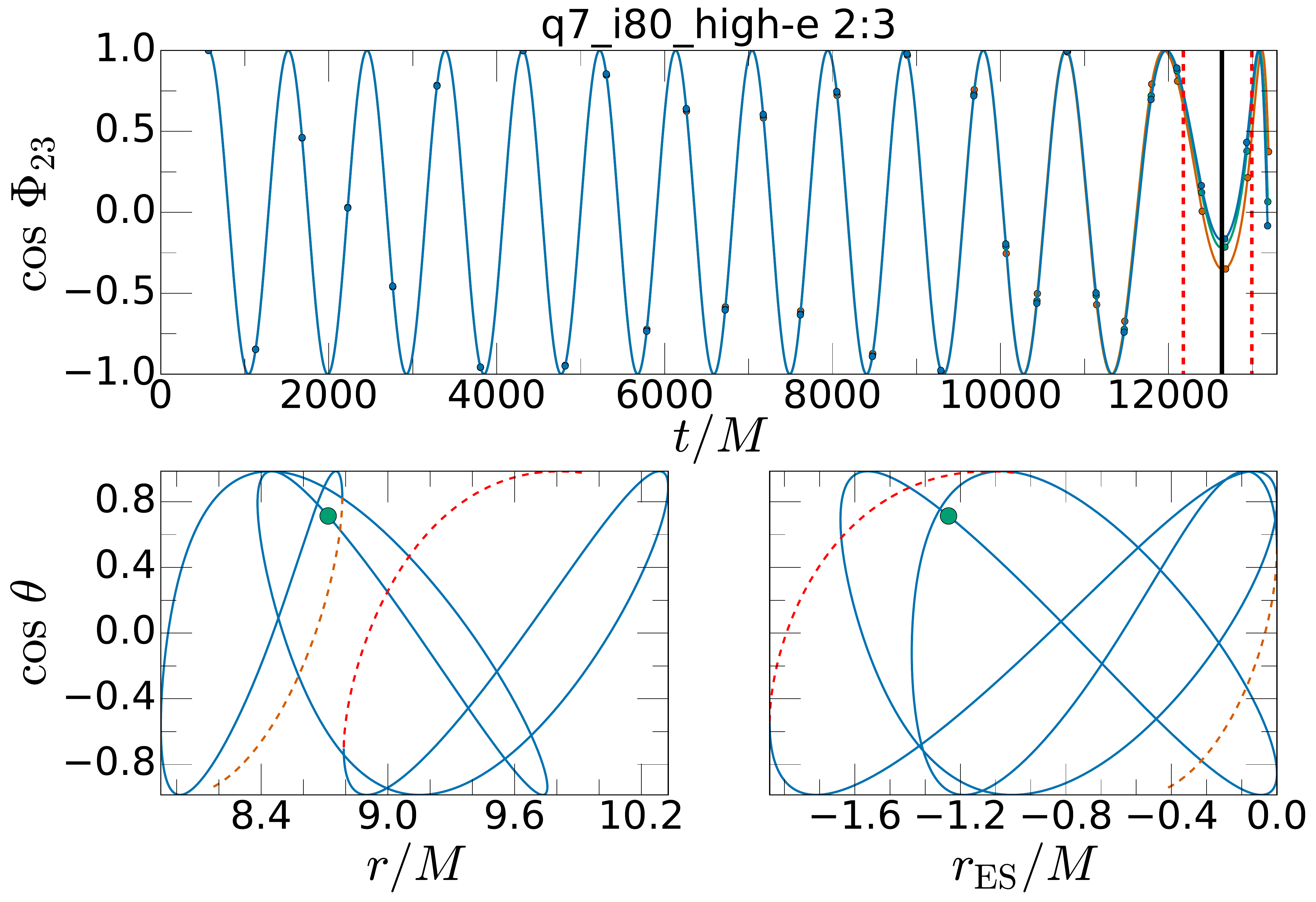}
\caption{\label{fig:resphases} Pictorial analysis of $r$-$\theta$
  resonances for four select resonances.  For each example, a block of
  three graphs is presented, titled by the simulation and the
  resonance under consideration.  Each block is organized as follows: \emph{Top graph:} Resonant phase $\Phi_{kn}$ with vertical solid lines marking the resonance, vertical dashed lines bounding one radian in resonant phase in either direction, 
  and black dots indicating midpoint times, $\tilde t^+_i$,
  cf.~(\ref{eq:MidpointTimes}).  The resonant phase $\Phi_{56}(t)$ in the
    top-left panel is plotted at three different numerical
    resolutions.  \emph{Bottom left graph:} Orbital trajectory during the resonance using two line-styles, solid
  lines represent one resonant cycle centered on the resonance
  (marked by a circle); dashed lines represent the remaining
  evolution within the resonant window.  \emph{Bottom right graph:} Trajectory with inspiral-motion removed, by
  using the envelope-subtracted radius $r_{\rm ES}$~(\ref{eq:rES}).
  }
\end{figure}

The rough locations of the coordinate resonances can be identified from figure \ref{fig:kij_inc}, where the relevant values of
$K^{ab}$ are highlighted with horizontal lines. A somewhat better
estimate can be obtained by solving the equation
$k\Omega^a(t) - n\Omega^b(t)=0$ for integers $n, k$, where $n/k \leq 1$ gives the value
of $K^{ab}$ at resonance. This avoids the noise in the
quotient. Solving this equation gives us an estimate for the
simulation time $t_{\rm res}$ at which a given resonance is
encountered. 
Interpolating $\Omega^\theta$ onto this value gives an
estimate of the corresponding polar frequency at resonance,
$\Omega^\theta_{\rm res}$.

Following \cite{Ruangsri:2013hra} we estimate the time spent on $r$-$\theta$ resonance using the following procedure. First, we construct a resonant phase $\Phi_{kn}(t)$
\begin{equation}
\label{eq:PhaseApprox}
\Phi_{kn}(t) = \int_{t_0}^t\left[k\Omega^\theta(t') - n\Omega^r(t')\right] {\mathrm d}t',
\end{equation}
where the arbitrary reference time $t_0$ is chosen after initial transients of the numerical simulation\footnote{In practice, we evaluate $k\Omega^\theta_i - n\Omega^r_i$, spline-interpolate this discrete time series, and integrate the interpolant.}.
The significance of the phases $\Phi_{kn}$ arises from black hole perturbation theory, where the
effect of the dissipative self-force upon the time derivative of some
Kerr constant of motion $\mathcal{C}$ expands into Fourier modes $k$
and $n$ of the form
\begin{equation}
\label{eq:selfforce}
\frac{d\mathcal{C}}{dt} = \dot{\mathcal{C}}_{00} + \sum_{k=-\infty}^{\infty} \sum_{n=-\infty}^{\infty} \dot{\mathcal{C}}_{kn} e^{-i\Phi_{kn}(t)} , \qquad n,k\neq 0.
\end{equation}
The exponential terms in most cases average to zero since they
oscillate rapidly compared to the overall evolution of
$\mathcal{C}$. At resonance, however, the phase is 
constant and the respective term contributes secularly to $d\mathcal{C}/dt$.

A representative resonant phase for each order $k\!:\!n$ is shown in figure~\ref{fig:resphases}. 
Still following \cite{Ruangsri:2013hra}, we say the system is near
resonance when $\Phi_{kn}$ differs from its resonant value by no more
than 1 radian.  
This dephasing is indicated in figure~\ref{fig:resphases} by the dashed vertical lines; our simulations satisfy this condition for durations of a few $1000M$.  
The duration $\Delta t_{\rm res}$ and frequency range $\Delta\Omega^\theta_{\rm res}$ during which our simulations stay on resonance can be readily determined from $\Phi_{kn}$.  
We estimate the number of radial and polar cycles
spent on resonance by counting the number of relevant coordinate peaks
within the resonance. These data are collected in table~\ref{tab:Resonances}.
Inspection of table \ref{tab:Resonances} reveals that, with the 2:3
resonance excepted, our simulations remain on resonance long enough to
trace out one or even two full resonant cycles.
(cf.~the coordinate plots in figure \ref{fig:resphases}). 
On resonance, the $r$-$\theta$ motion of a geodesic is a Lissajous figure; figure~\ref{fig:resphases} includes the corresponding plots for the full BBH simulation, highlighting that the character of the Lissajous figures are preserved. 

\begin{table}
\centering
\caption{\label{tab:Resonances} The $r$-$\theta$ resonances identified in our simulations.  Given is the order of the resonance, the polar frequency $\Omega^\theta$ at resonance, the duration in terms of time $\Delta t_{\rm res}$ and polar frequency $\Delta \Omega^\theta_{\rm res}$, and the number of radial cycles $\mathcal{N}_r$ and polar cycles $\mathcal{N}_\theta$ during the resonance. 
For many of the 2:3 and some of the 6:7 resonances, the resonances were detected, but occurred too near plunge for their width to be measured; in those cases we report the frequency at resonance and mark the other columns n/a. 
}
\footnotesize
\begin{tabular}{ l *{8}{c} }
\br
Run & Order & $M \Omega^\theta_{\rm res}$ & $\Delta t_{\rm res}/M$ & $M\Delta \Omega^\theta_{\rm res}$ & $\mathcal{N}_r$ & $\mathcal{N}_\theta$ \\
\mr
q5\_i10\_low-e & 6:7 & 0.0264 & 4540 & 0.00787 & 9 & 11 \\ 
               & 5:6 & 0.0341 & 2921 & 0.00506 & 8 & 9 \\
               & 4:5 & 0.0385 & 1737 & 0.00301 & 5 & 7 \\
               & 3:4 & 0.0408 & 920.3 & 0.00159 & 4 & 5 \\
\mr
q5\_i20\_low-e  & 6:7 & 0.0268 & 4646 & 0.00876 & 10 & 11 \\ 
                & 5:6 & 0.0353 & 3026 & 0.00571 & 8 & 9 \\
                & 4:5 & 0.0403 & 1792 & 0.00338 & 5 & 7 \\
                & 3:4 & 0.0430 & 989.0 & 0.00187 & 4 & 5 \\
\mr
q5\_i10\_high-e  & 6:7 & 0.0209 & 4713 & 0.0144 & 9 & 11 \\
                 & 5:6 & 0.0346 & 2982 & 0.00911 & 7 & 9 \\
                 & 4:5 & 0.0425 & 1757 & 0.00537 & 5 & 7\\
                 & 3:4 & 0.0466 & 935.4 & 0.00286 & 4 & 5 \\
                 & 2:3 & 0.0531 & n/a & n/a & n/a & n/a \\
\mr
q5\_i20\_high-e  & 6:7 & 0.0184 & n/a  & n/a & n/a & n/a \\ 
                 & 5:6 & 0.0301 & 3088 & 0.00814 & 7 & 9 \\
                 & 4:5 & 0.0379 & 1821 & 0.00480 & 6 & 7 \\
                 & 3:4 & 0.0416 & 964.9 & 0.00254 & 4 & 5 \\
\mr
 q7\_i40\_low-e  & 6:7 & 0.0299 & 5048 & 0.00780 & 10 & 12 \\
                 & 5:6 & 0.0385 & 3253 & 0.00502 & 9 & 10 \\
                 & 4:5 & 0.0437 & 2076 & 0.00320 & 7 & 8 \\
                 & 3:4 & 0.0469 & 279.2 & 0.000431 & 1 & 2\\
                 & 2:3 & 0.0515 & n/a & n/a & n/a & n/a \\
\mr 
 q7\_i80\_low-e & 5:6 & 0.0230 & 5569 & 0.00880 & 11 & 12\\
                 & 4:5 & 0.0341 & 3817 & 0.00603 & 9 & 11 \\
                 & 3:4 & 0.0393 & 2182 & 0.00345 & 7 & 8 \\
                 & 2:3 & 0.0447 & n/a & n/a & n/a & n/a \\
\mr
q7\_i40\_high-e & 6:7 & 0.0224 & 5317 & 0.0124 & 11 & 12 \\
                & 5:6 & 0.0360 & 3371 & 0.00784 & 8 & 10 \\
                & 4:5 & 0.0435 & 1958 & 0.00456 & 7 & 8 \\
                & 3:4 & 0.0475 & 954.9 & 0.00222 & 4 & 5\\
                & 2:3 & 0.0537 & n/a & n/a & n/a & n/a \\
\mr
q7\_i80\_high-e & 4:5 & 0.0311 & 3506 & 0.00975 & 8 & 10 \\
                & 3:4 & 0.0421 & 1858 & 0.00517 & 6 & 7 \\
                & 2:3 & 0.0469 & 815.3 & 0.00227 & 3 & 5 \\ 
\br
\end{tabular}
\end{table}

\begin{figure}
\includegraphics[width=0.33\columnwidth]{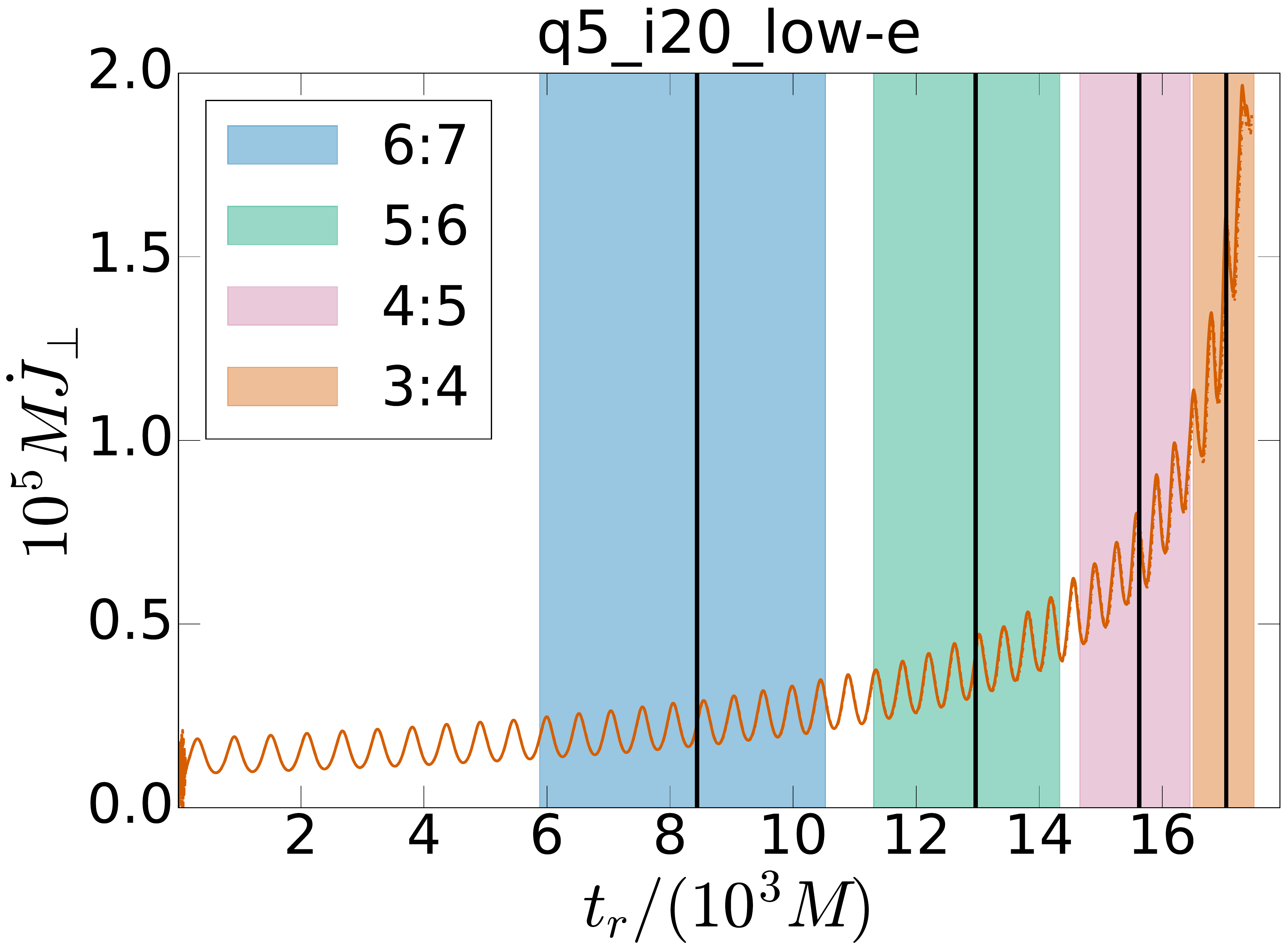}
\includegraphics[width=0.33\columnwidth]{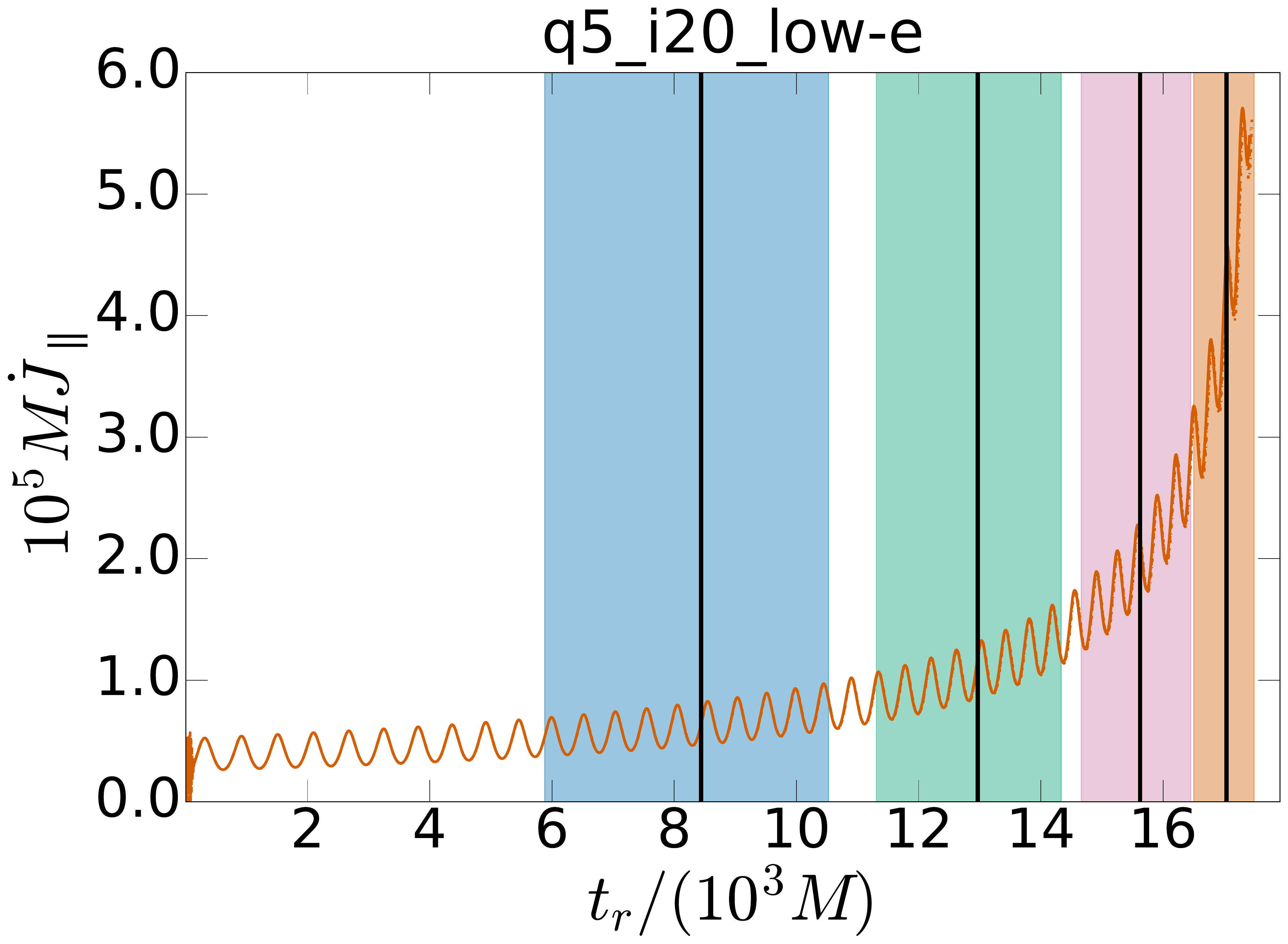}
\includegraphics[width=0.33\columnwidth]{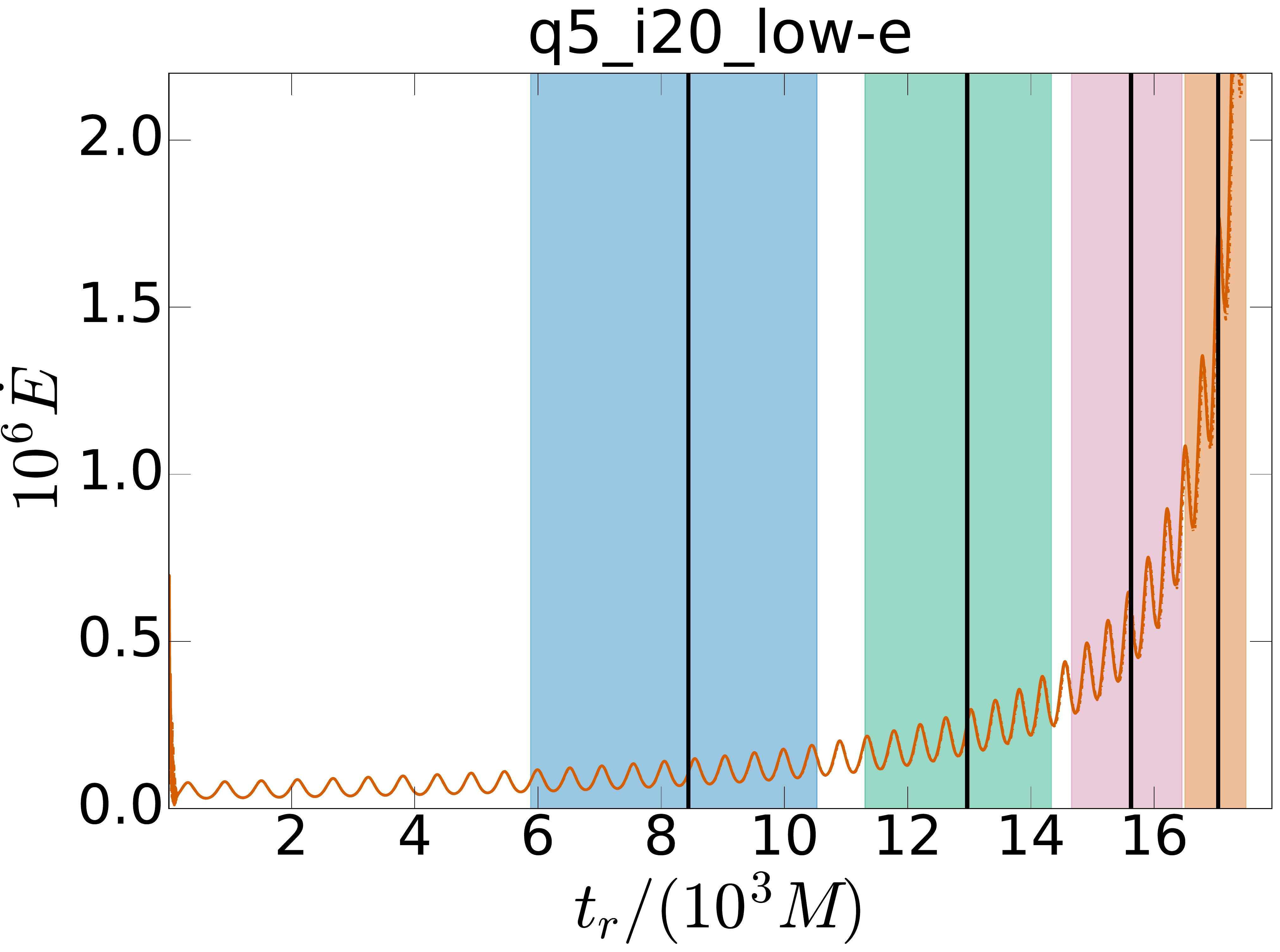} \\
\includegraphics[width=0.33\columnwidth]{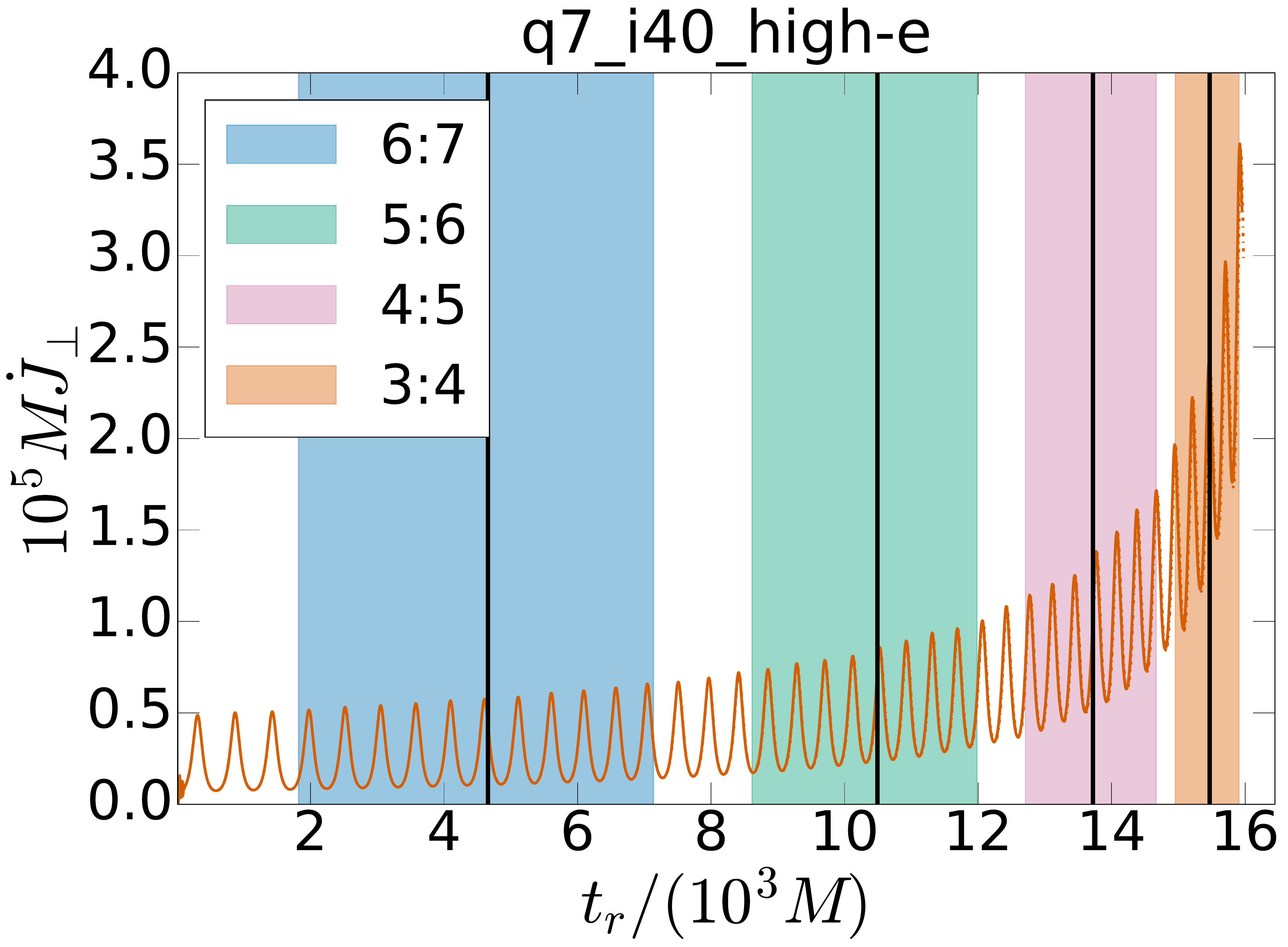}
\includegraphics[width=0.33\columnwidth]{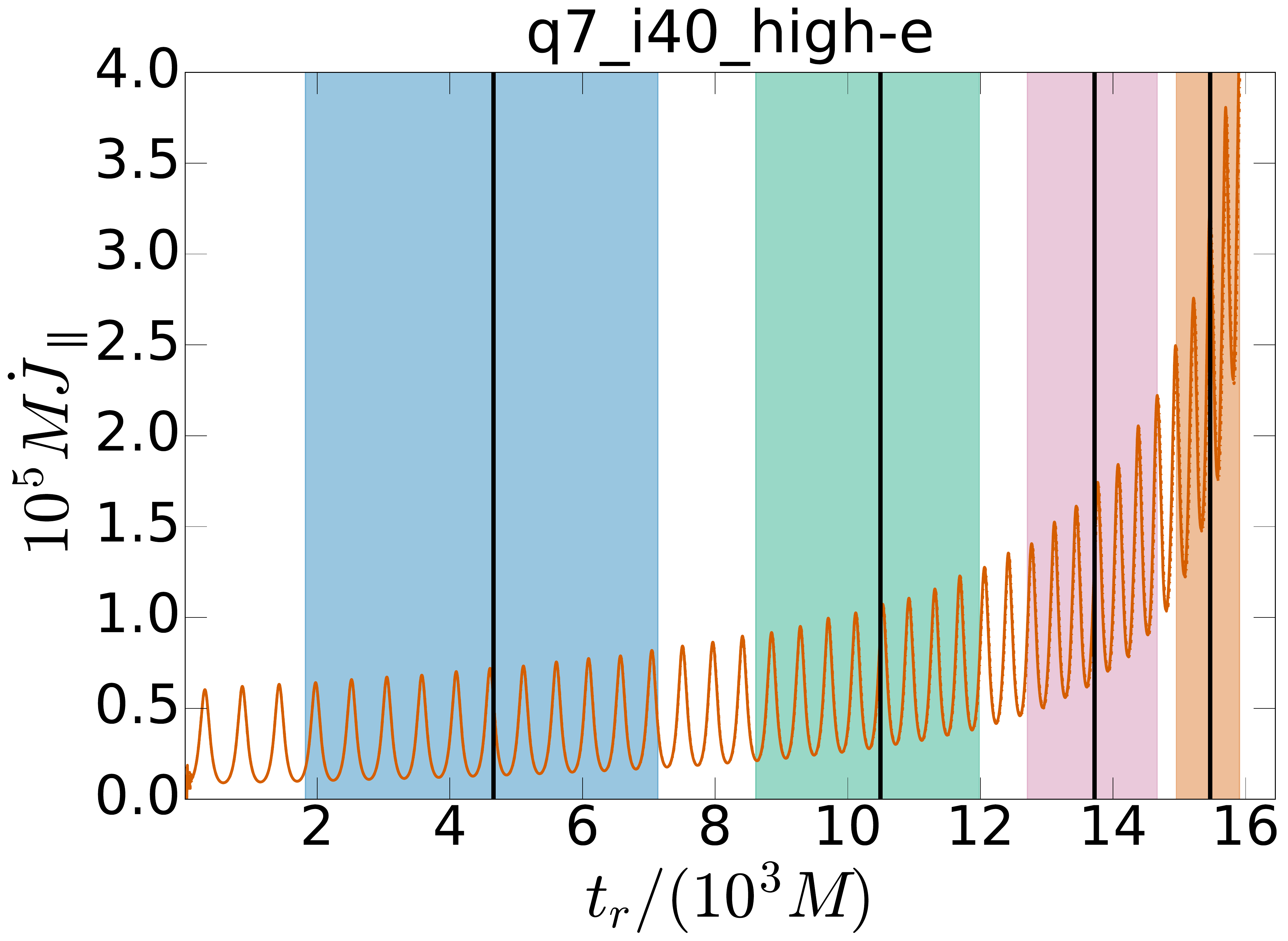}
\includegraphics[width=0.33\columnwidth]{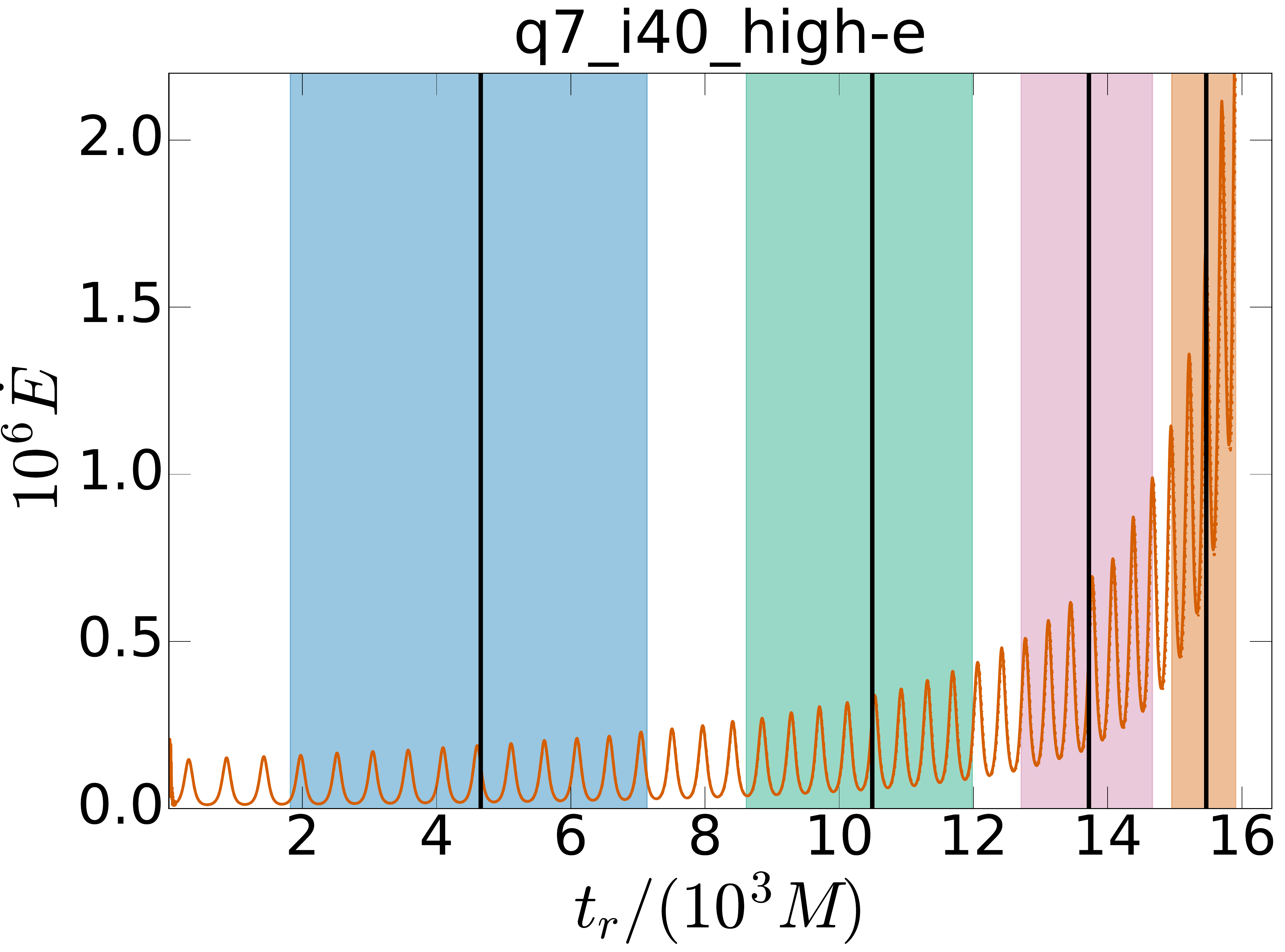} \\
\includegraphics[width=0.33\columnwidth]{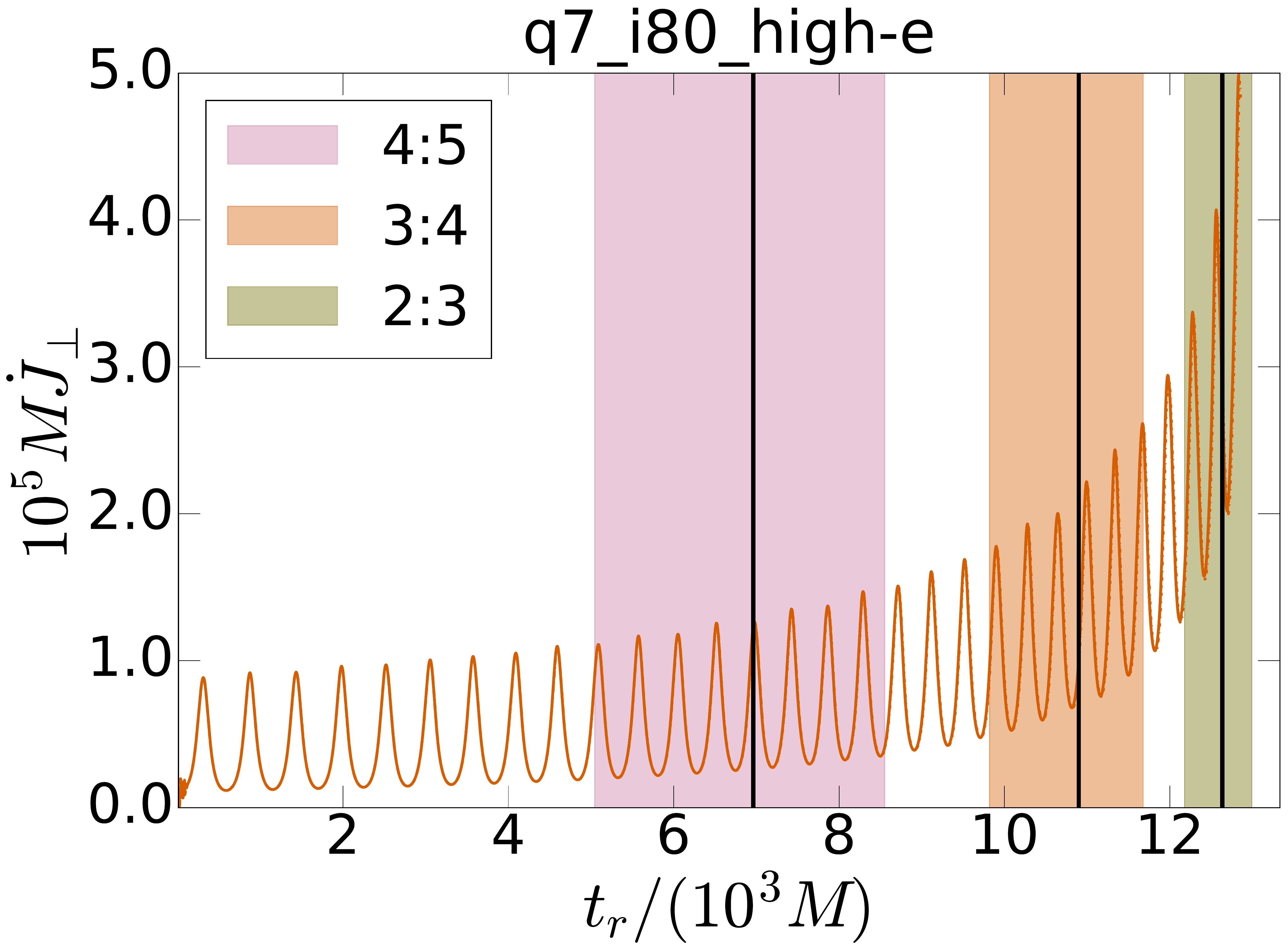}
\includegraphics[width=0.33\columnwidth]{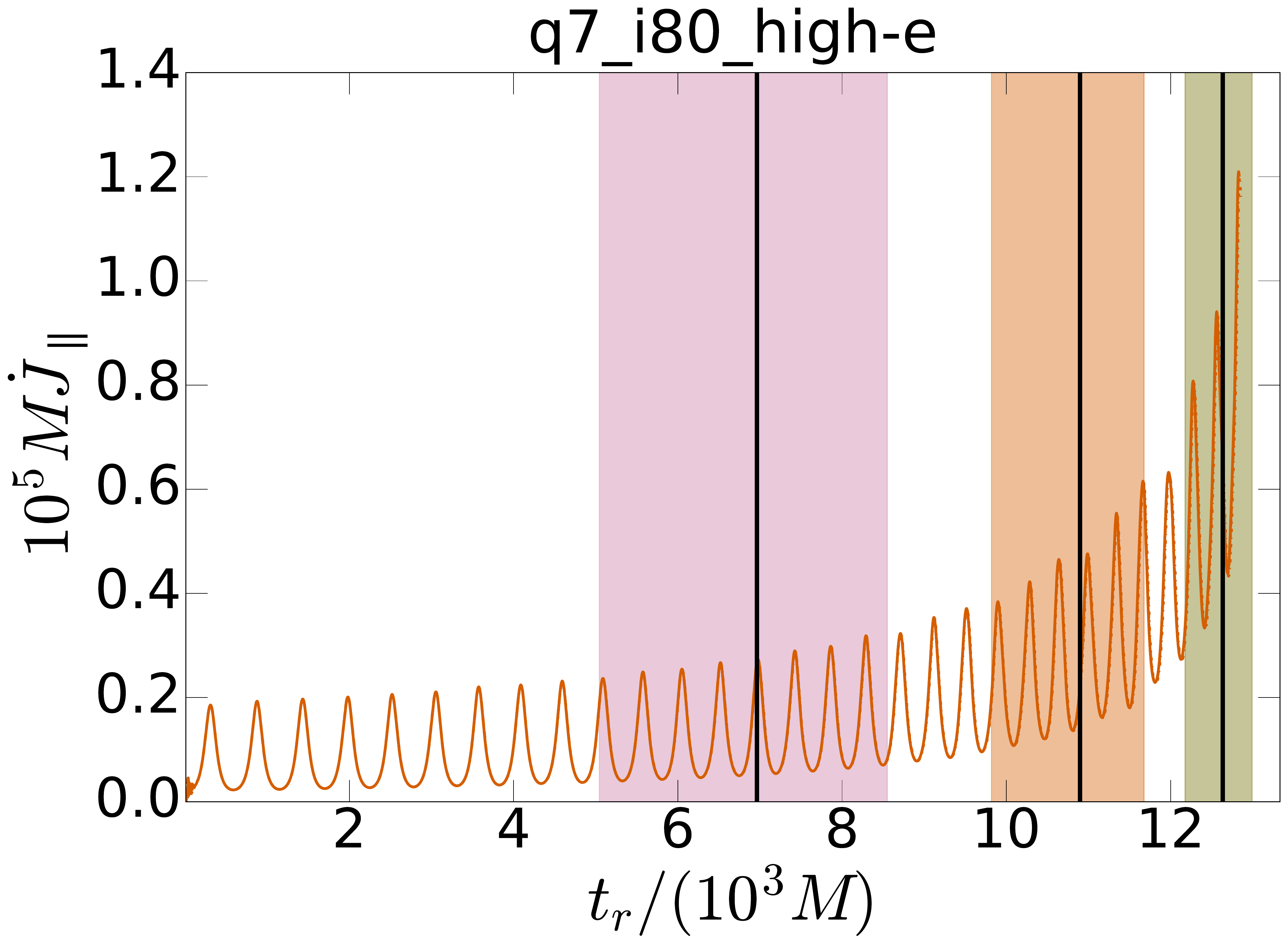}
\includegraphics[width=0.33\columnwidth]{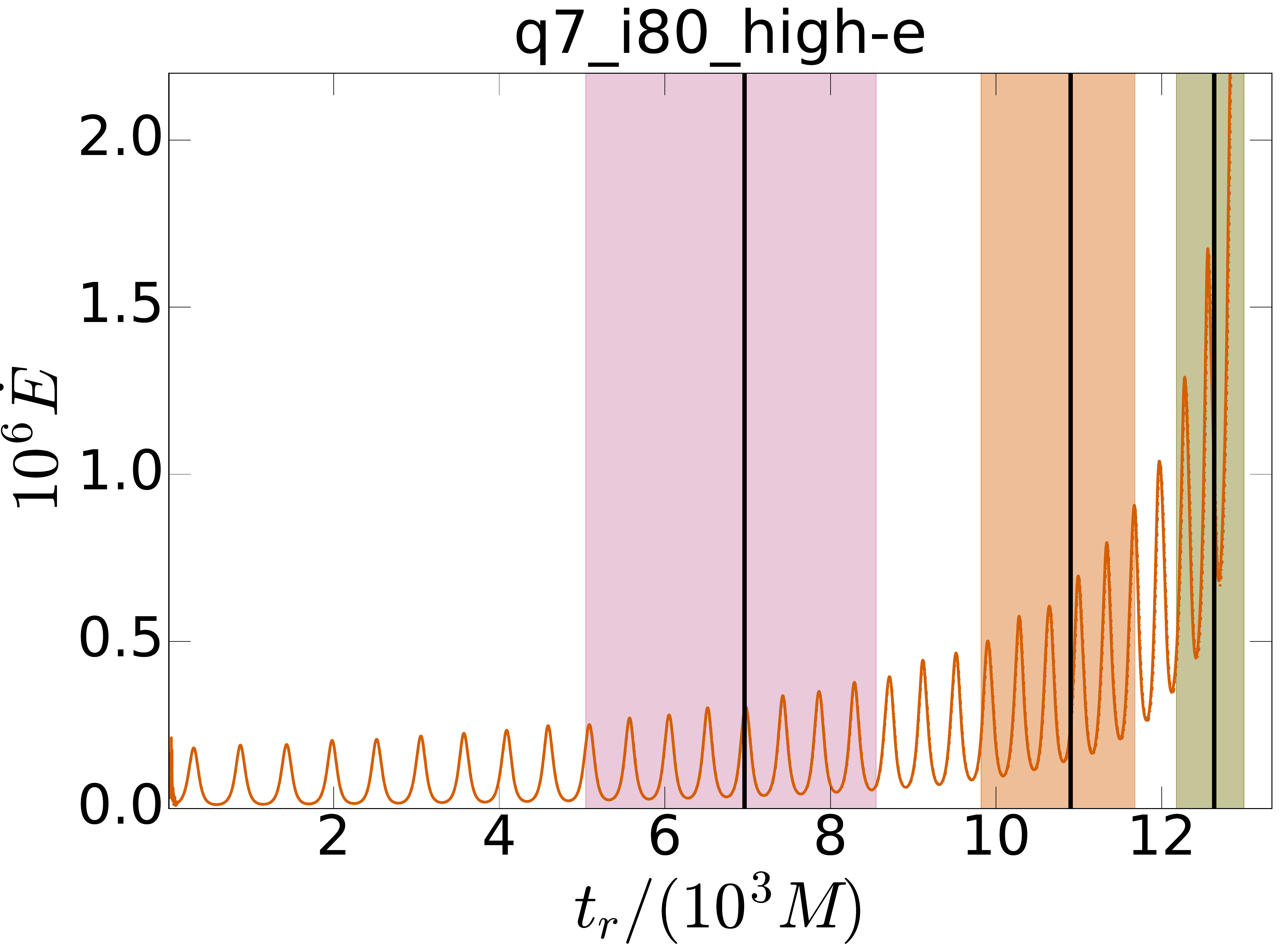}
\caption{
\label{fig:fluxplots}
Gravitational wave fluxes during resonances.  Each row of plots corresponds to one numerical simulation as indicated in the title of each panel.  The columns correspond to three gravitational wave fluxes.  \emph{Left:}   Angular momentum flux orthogonal to primary spin direction $\hat{\bm\chi}(t)$.   \emph{Middle:}  Angular momentum flux tangential to $\hat{\bm\chi}(t)$.  \emph{Right:}  Energy flux. Resonances are indicated by the shaded areas. The fluxes
  are shown at three resolutions, with differing linestyles; on the scale of the plot, however, these are indistinguishable in the plots except close to merger.
 }
\end{figure}

As table~\ref{tab:Resonances} and figure~\ref{fig:resphases} indicate,
our inspirals remain on resonance for one to two resonant cycles.
This may lend some hope that secular
accumulations from terms like those in \eqref{eq:selfforce} might be
directly visible, leading to a noticeable
change in the evolution of constants of motion ($\mathcal E$,
$\mathcal L_z$, and $\mathcal Q$) during the inspiral.
Returning to equation \eqref{eq:selfforce}, we see that the impact of passage through a resonance is that a subset of the oscillatory contributions to $d\mathcal C/dt$ momentarily ``freeze out'', with $\dot \Phi_{kn}$ passing through zero at the resonance for pairs $(k,n)$ commensurate with the resonance.
Near the resonance, the phases of these contributions can be approximated by Taylor expanding $\Phi_{kn}$ around the time $t_{\rm res}$ when resonance is achieved (see e.g.~\cite{Ruangsri:2013hra})
\begin{equation}
\label{eq:ResonantPhase}
\Phi_{kn} \approx \Phi_{kn} (t_{\rm res}) + \frac12 (k \dot \Omega^\theta - n \dot \Omega^r)|_{t_{\rm res}} (t-t_{\rm res})^2 \,.
\end{equation}
While the amplitude and sign of these resonant contributions to $d\mathcal C/dt$ depends on the coefficients $\dot \mathcal{C}_{kn}$ and the phase on resonance $\Phi_{kn} (t_{\rm res})$, these terms have a distinct time evolution: their contribution is approximately symmetric about $t_{\rm res}$, and they contribute over a time scale of approximately the resonant width $\Delta t_{\rm res}$.
This characteristic time evolution is clearly seen in the plots of $\cos \Phi_{kn}$ in figure~\ref{fig:resphases} around each resonant passage.
Thus the simplest way to diagnose whether a resonant passage impacts the dynamics is to search for modulations of $d \mathcal C/dt$ with this behavior.
Meanwhile, the time integral of these terms results in a jump in $\mathcal C$  which accumulates over a period approximately equal to $\Delta t_{\rm res}$.
The detailed time evolution of these contributions involves Fresnel integrals (e.g.~\cite{Berry:2016bit}), and is antisymmetric across the resonance.

To investigate the possibility of resonances affecting the orbital evolution, we compute the instantaneous gravitational wave energy flux $\dot E(t_r)$ and the gravitational wave angular momentum fluxes $\dot{\bm J}(t_r)$ as suitable products of spherical harmonic modes of the gravitational waveform~\cite{Boyle:2008} for all modes $l\le 8$. 
These fluxes are given in terms of the retarded time 
\begin{equation}
t_r = t - R - 2M \ln\left(\frac{R}{2M}-1\right),
\end{equation}
where $R$ is the gravitational wave extraction radius of each simulation ($\sim 500M$).  For the angular momentum flux, we project parallel and orthogonal to the primary BH-spin direction $\hat{\bm\chi}(t)$,
\begin{equation}
\dot{J}_{||}(t_r) = \dot{\bm J}(t_r) \cdot \hat{\bm\chi}(t),
\end{equation}
\begin{equation}
\dot{J}_\perp(t_r) = \left| \dot{\bm J}(t_r) - J_{||}(t_r)\hat{\bm \chi}(t)\right|.
\end{equation}
We use $\dot J_{||}$  as a proxy for the azimuthal angular momentum flux of the test body (corresponding to the constant of motion 
  $\mu \mathcal L_z$),  and $\dot{J}_\perp$ 
as a very rough proxy for the evolution of the square root of the Carter constant
$\mu \sqrt{\mathcal Q}$. 

In figure~\ref{fig:fluxplots} we plot the energy- and angular momentum-fluxes for three simulations with the time-intervals on resonance indicated.
Figure~\ref{fig:fluxplots} shows short-period modulations due to eccentricity, 
with sharp spikes near each periastron passage. 
However, no additional modulations over the
resonant timescales are discernible. We have also examined the
angular momentum ${\bm J}$ and total energy $E$ by integrating the
fluxes, and see no excess build-up or deficit over the resonant widths.

The amplitude of any changes that occur during resonance depends on the resonant phase 
and it is possible that all of our simulations have phases that make resonant effects undetectable.
Unfortunately, the resonant phase cannot be easily changed
in a numerical simulation while keeping all other parameters constant.
Since it is unlikely that all resonances are encountered with unfavorable phases, we interpret the absence of detectable resonant modulations 
as evidence that the mass-ratios and eccentricities under consideration are not extreme enough to exhibit strong resonant effects.

In \ref{sec:ResEstimation} we estimate quantitatively the strength of resonant effects
by fitting a sufficiently-stiff function to peak-to-peak averages over the 
gravitational wave fluxes averaged roughly over the resonant timescale. Since
the chosen function is too stiff to capture resonant-timescale effects, 
resonant accumulations should be visible as jumps in the residuals over that
timescale. Using this method on the most extreme cases of 
the {\tt q7\_i40\_high-e} and {\tt q7\_i80\_high-e} runs we bound resonant effects upon these runs from above at less than the 0.4\% level.

\section{Conclusions}
\label{sec:Conclusions}

We have presented a suite of binary black hole simulations
  performed with the \spec~code that aim to explore the relation
  between BBH systems and generic Kerr geodesics.
Our simulations are eccentric and explore both precessing and nonprecessing 
configurations.  The simulations cover 30--60 radial passages and multiple precession 
cycles, and are performed at comparatively high mass-ratios of $q=5$ and $q=7$.

As a first step to understanding these fully generic binary inspirals,
we developed methods for extracting the instantaneous, fundamental
frequencies of motion while partially accounting for dissipative
effects.  In the case of equatorial inspirals, these are the
frequencies of radial motion $\Omega^r$ and azimuthal motion
$\Omega^\phi$ averaged over a radial orbit.  Here the frequencies can
be extracted most cleanly, 
since the interdependence of the polar and the radial
motion does not come into play and does not produce frequency modulations. The ratio of these
frequencies provides the rate of periastron advance and is a
physically measurable quantity.  We have compared the precession rate
of our eccentric binaries to geodesic orbits in Kerr, and extracted
the leading self-force correction to this rate.  Our results show that
second-order SF corrections to these rates is small, and that much of
the dependence on eccentricity and black hole spin can be absorbed into
the leading Kerr behavior. 

In the case of our generic, precessing binaries, we can cleanly
extract the fundamental frequencies of radial and polar motion, 
as well as the average orbital frequency $\left<\Omega\right>$.  
The fundamental azimuthal frequency $\Omega^\phi$ differs from $\left<\Omega\right>$ 
because motion in the orbital plane is a mixture of 
azimuthal and polar motions for precessing binaries.  Unfortunately, it turns 
out that $\Omega^\phi$ is affected by interactions of the polar and radial 
motions to a large degree, cf.~\ref{sec:FreqValidation}.
Nevertheless, we have extracted
the ratios of radial, polar, and average orbital frequencies from our simulations, cf. figure~\ref{fig:kij_inc}.
The ratio $K^{r\theta}= \Omega^r/\Omega^\theta$ can be predicted for
Kerr geodesics, and in this case we have compared to analytic theory
and extracted the SF corrections to our precession rates.  

Finally, with our fully generic orbits we were able to identify low-order 
orbital resonances in the radial-polar motion, and show that our
systems pass through resonances with $\Omega^r/\Omega^\theta=$ 5:6, 4:5, 3:4 and 2:3.  
Our simulations remain on resonance usually for more than one resonant cycle, 
and sometimes up to two resonant cycles.
The $r$--$\theta$ resonances have been
shown to have an impact on the adiabatic inspirals of extreme mass
ratio systems\cite{Hinderer:2008dm,Flanagan:2010cd}, and our study is the first to identify them in a
numerical spacetime.  
However, at the mass ratios we have achieved we find
no clear impact on the fluxes of energy and angular momentum of our
systems (cf.~figure \ref{fig:fluxplots}).

 In two of our low-eccentricity
simulations, strong residual oscillations in the extracted frequencies
dominate the comparison.  These modulations arise from small but
visible variations in the radial separation of the binary, persist at
several resolutions, and will be the subject of future investigation.

In this study we have focused on the dynamics of the black holes
themselves.  In future studies we will examine the gravitational
waveforms produced by these systems, and explore the extraction of the
fundamental frequencies directly from the waveform.  Waveforms from
our equatorial and fully generic system will serve to test how
detectable these binaries are using current quasi-circular and
eccentric waveform models, as well as help develop those models
further.  Waveforms from these binaries can also elucidate what
constraints can be placed on eccentricities in future gravitational
wave detections.

Future work will include extending our suite of simulations to higher
mass ratios, and to cover a larger range of eccentricities and binary
configurations.  We will extract the redshift factor
\cite{Zimmerman:2016ajr} from generic black hole binaries, which is
analogous to the Lorentz factor of the black holes.  With three
frequencies and the Lorentz factor, we can imagine forming an explicit
map between high mass ratio binaries and perturbed orbits in Kerr.
Comparisons to post-Newtonian predictions for the frequencies of
motion would supplement our SF-inspired comparisons to Kerr geodesics.
In addition, SF predictions in for Kerr orbits are developing rapidly
(e.~g.~\cite{vandeMeent:2016pee,vandeMeent:2016hel}), which will allow
for a quantitative comparison to analytic approximations, as well as
an extraction of higher-order SF effects.

In previous studies of circular orbits, it was suggested that
re-expanding geodesic predictions about the symmetric mass ratio
$\nu = m_1 m_2/M^2$ and as functions of $M \Omega^\phi$ would provide
for faster convergence to the numerical results
\cite{LeTiec-Mroue:2011,Tiec:2013twa}.  With a SF prediction for the
precession rates in our equatorial and generic binaries, we will be
able test this promising idea in a new regime.

\ack

We thank Tanja Hinderer and Serguei Ossokine for valuable discussions.
We also thank Serguei Ossokine for assistance in computing
gravitational wave fluxes of angular momentum and energy from our simulations.  
We also thank the anonymous referee for valuable comments which led us to discover 
an error in our original manuscript. 
Calculations were performed with the Spectral Einstein
Code ({\tt SpEC})~\cite{SpECwebsite}.  We gratefully acknowledge
support for this research from NSERC of Canada, the Canada Research
Chairs Program, the Canadian Institute for Advanced Research, and the
Vincent and Beatrice Tremaine Postdoctoral Fellowship (A.Z.).  We
acknowledge hospitality of the Albert-Einstein Instiute at Potsdam,
where part of this work was completed.  Computations were performed on
the GPC supercomputer at the SciNet HPC Consortium~\cite{SciNet}.
SciNet is funded by the Canada Foundation for Innovation (CFI) under the
auspices of Compute Canada, the Government of Ontario, Ontario
Research Fund - Research Excellence and the University of Toronto.
Further computations were performed on the Briar{\'e}e supercomputer
from the Universit{\'e} de Montr{\'e}al, managed by Calcul Qu{\'e}bec
and Compute Canada, and funded by CFI, NanoQu{\'e}bec, RMGA and the Fonds de recherche du Qu{\'e}bec - Nature et Technologie (FRQ-NT). 

\appendix

\section{The Kerr metric}
\label{sec:KerrDetails}

The Kerr metric expressed in Boyer-Lindquist coordinates $x^\mu = (t,r,\theta,\phi)$  is given by
\begin{eqnarray}
ds^2 &=&  - \left( 1-\frac{2Mr}{\rho^2}\right) dt^2 - \frac{4 M a r \sin^2 \theta}{\rho^2} dt d\phi + \frac{\rho^2}{\Delta} dr^2 + \rho^2 d\theta^2 
\nonumber \\ & &
+ \frac{\sin^2\theta}{\rho^2} \left[(r^2+a^2)^2 - a^2 \Delta \sin^2 \theta \right]d\phi^2 \,.
\end{eqnarray}
where $\rho^2 = r^2 +a^2 \cos \theta$, $\Delta = r^2 - 2 M r +a^2$, and $M = M_{\rm Kerr}$ in this appendix only.
 
The equations of motion for a test particle are given by \eqref{eq:RadialEoM}--\eqref{eq:PhiEoM}, and the explicit forms of the potentials are (see e.g.~\cite{Fujita:2009bp})
\begin{eqnarray}
\label{eq:RPotential}
R(r) & = & \left[ \mathcal E (r^2 +a^2) - a \mathcal L_z\right]^2 - \Delta \left[(a \mathcal E - \mathcal L_z)^2 +r^2 + \mathcal Q\right] \,, \\
\label{eq:ThetaPotential}
\Theta(\theta) & = & \mathcal Q - \cos^2 \theta \left[ a^2 (1-\mathcal E^2) + \mathcal L_z^2 \csc^2 \theta \right] \,, 
\end{eqnarray}
and
\begin{eqnarray}
\label{eq:TPotential}
T_r(r) = \frac{r^2+a^2}{\Delta}\left[ \mathcal E (r^2 +a^2) - a \mathcal L_z\right] \,, \qquad
& T_\theta(\theta)  = - a^2 \mathcal E \sin^2 \theta \,, \\
\label{eq:PhiPotential}
\Phi_r(r)  =  \frac{a}{\Delta} \left[ \mathcal E(r^2 +a^2) - a \mathcal L_z \right] \,, 
&\Phi_\theta(\theta) = \mathcal L_z \csc^2 \theta\,.
\end{eqnarray}
The radial equation can be converted into one for the radial phase $\chi^r$, and the polar equation into one for the phase $\chi^\theta$ in a straightforward manner using \eqref{eq:RadEllipse} and \eqref{eq:theta_phase}. It is these equations that we integrate to generate Kerr orbits of fixed $(p,e,i)$ to test our frequency extraction methods.

 \section{Envelope subtraction method in the slow time approximation}
 \label{sec:EnvSubtract}
 
In this appendix we demonstrate the utility of the envelope subtraction method discussed in section \ref{sec:Simulations}.
The goal of this method is to locate the successive maxima $t^+_i$ or minima $t^-_i$ of the underlying motion of an oscillator with slowly varying parameters, and use them to calculate the frequency of oscillation.
The variation of the amplitude and midline of the underlying oscillation shifts the position of the extrema from where they would be in the absence of the slow changes.
We quantify the utility of the envelope subtraction method in removing these shifts by analyzing a toy model which captures some of the behavior of $\Omega(t)$ in our equatorial inspirals.

To model the orbital frequency, consider a function $f(\tilde t,t)$ of a fast time $t$ and a slow time $\tilde t = \epsilon t$, with $\epsilon \ll 1$ giving the ratio of inspiral to orbital time scales.
In order to represent an underlying oscillation at a single fixed frequency $\omega$, we assume that $f$ is periodic with period $P = 2 \pi/\omega$ in the fast time,
\begin{equation}
f(\tilde t,t+ P) = f(\tilde t,t).
\end{equation}
Since $\tilde t = \epsilon t$, after a period $f$ does not return to precisely the same value, but is modified by the advance of the slow time variable.

The maxima of the underlying oscillation are the extrema of $f$ in the limit $\epsilon \to 0$.  For simplicity we phrase the discussion in terms of  maxima; the procedure is identical for minima.
For nonzero $\epsilon$, we find the maxima $t_i$ by solving
\begin{equation}
\label{eq:ExtremaTwoTime}
\frac{df}{dt}= \partial_t f + \epsilon \partial_{\tilde t} f = 0
\end{equation}
order by order in $\epsilon$, under the ansatz
\begin{equation}\label{eq:B3}
t_i = t^{(0)}_i + \epsilon t^{(1)}_i +  O(\epsilon^2)\,.
\end{equation}
At leading order, we have as expected
\begin{equation}
\left. \partial_t f \right|_{t^{(0)}_i} = 0 \,,
\end{equation}
which shows that $t^{(0)}_i$ are the maxima of the underlying oscillation.
We seek $t^{(0)}_i$, but we have only access to $t_i$ when analyzing $f$.  From equation~(\ref{eq:B3}) we see that the naive maxima $t_i$ incur an error $O(\epsilon)$.
 
Returning to equation \eqref{eq:ExtremaTwoTime}, we Taylor expand all quantities around $t = t_i^{(0)}$ to find at the next order that
\begin{equation}
t_i^{(1)} = - \left. \frac{\partial_{\tilde t} f}{\partial_t^2 f}\right|_{t_i^{(0)}} \,.
\end{equation} 
This provides the leading error on our estimation of the maxima. 
We can iterate this procedure to compute higher order errors, but it turns out we can instead identify a different function, the envelope subtracted $f$, whose maxima are closer to $t_i^{(0)}$.

We first fit a function through the maxima $t_i$ (in practice a spline of at least cubic order).
This envelope $f^+(\tilde t)$ describes the slow time variation passing from maximum to maximum, meaning that $f^+(\tilde t) = f(\tilde t, t_i^+)$.  By Taylor expansion about $t_i^{(0)}$ we have
\begin{equation}
f^+ (\tilde t)  = f(\tilde t, t_i^{(0)}) -\epsilon^2  \left.  \frac{(\partial_{\tilde t} f)^2}{2 \partial_t^2f}\right|_{t_i^{(0)},\tilde t} + O(\epsilon^3) \,,
\end{equation}
where we have simplified this function by recalling that $f(\tilde t,t_i^{(0)}+nP) = f(\tilde t,t^{(0)}_i)$.

Next, we define the ``envelope-subtracted'' function
\begin{equation}
\label{eq:ESf}
\hat f  = f - f^+ \,,
\end{equation}
and seek its maxima.
To motivate this definition, imagine we could seek the maxima of
\begin{equation}
f(\tilde t,t) - f(\tilde t,t^{(0)}_i)\,,
\end{equation}
which are given by the condition
\begin{equation}
\partial_t f(\tilde t, t) - \epsilon[\partial_{\tilde t} f(\tilde t, t) - \partial_{\tilde t} f(\tilde t,t^{(0)}_i) ] = 0\,.
\end{equation}
The solutions are in fact the desired maxima, $t = t_i^{(0)}$, but we cannot formulate this ideal subtraction because we cannot identify $f(\tilde t,t^{(0)}_i)$. 
Fortunately, $\hat f$ is is equal to $f(\tilde t,t^{(0)}_i)$ up to $O(\epsilon^2)$, and so we search for the maxima of \eqref{eq:ESf}.

We find that the maxima of $\hat f$, which we denote $\hat t_i$, are
\begin{eqnarray}
\hat t_i   =  t_i^{(0)} + \epsilon^3 \left. \delta t_i  \right|_{t_i^{(0)}} + O(\epsilon^4) \,, \\
 \delta t_i \equiv - \frac{(\partial_{\tilde t} f)^2}{\partial_t^2 f} \partial_{\tilde t} \left(\ln \partial_{\tilde t} f - \frac 12 \ln \partial_t^2 f \right)  \,.
\end{eqnarray}
The error in $\hat t_i$ (relative to the desired maxima $t_i^{(0)}$) is $O(\epsilon^3)$, i.e.~envelope substraction has resulted in an improvement by two powers of $\epsilon$.

When we actually compute the frequencies, the error is even lower.
We have 
\begin{eqnarray}
\hat t_{i+1} - \hat t_i &=&  t^{(0)}_{i+1} - t^{(0)}_{i}  + \epsilon^3 (\delta t_{i+1} - \delta t_i) +O(\epsilon^4) \nonumber \\
& = & P + \epsilon^3[ \delta t(t_i^{(0)}+P, \epsilon t_i^{(0)} + \epsilon P) - \delta t_i] + O(\epsilon^4)\nonumber\\
& = & P + \epsilon^4 P \partial_{\tilde t} (\delta t) |_{t^{(0)}_i} + O(\epsilon^5)
\end{eqnarray}
where we have used the periodicity of $f$ in the variable $t$ and expanded the slow time function at $t_{i+1}$ about $t_i$. 
The above expressions are understood to be evaluated at $t_i^{(0)}$.
The differencing at the extrema leads to an error $O(\epsilon^4)$ due to another cancellation of the leading order correction to the period, due to slow evolution of the function and its envelope from maximum to maximum.
Without the envelope subtraction procedure, the differencing at successive peaks results in an $O(\epsilon^2)$ error in the extraction of the period.

\section{Evolution of eccentricity and semi-major axis}
\label{sec:PetersEvo}

In this appendix we collect the relevant results of \cite{PetersMathews1963,Peters1964} for the evolution of the eccentricity of a binary at lowest PN order.
In this case, we consider a binary in an eccentric, Newtonian orbit, which sources gravitational wave emission through the quadrupole formula.
The resulting orbit-averaged emission of energy and angular momentum changes the Newtonian parameters over the course of the slow inspiral.
Peters \cite{Peters1964} provides differential equations for orbit-averaged changes to the semi-major axis and eccentricity, $da/dt$ and $de/dt$, in terms of $(a,e)$.
These can be solved numerically given initial conditions $(a_0,e_0)$, or combined and integrated to give an analytic expression for $a(e)$,
\begin{equation}
a  = c_0 \frac{e^{12/19}}{1-e^2}\left(1 + \frac{121}{304}e^2 \right)^{870/2299} \,,
\end{equation}
where the integration constant $c_0$ is determined by $(a_0,e_0)$.
In order to re-express this equation as the evolution of eccentricity with averaged orbital frequency $\langle\Omega\rangle$, we recall Kepler's law for Newtonian orbits, $m  = \langle \Omega \rangle^2 a^3$.
We can then write
\begin{equation}
\label{eq:Omega_vs_e}
\frac{\langle \Omega \rangle}{\langle \Omega \rangle_0} = \left(\frac{e}{e_0}\right)^{-18/19} \left(\frac{1-e_0^2}{1-e^2} \right)^{-3/2} \left( \frac{1 + \frac{121}{304}e^2}{1 + \frac{121}{304}e_0^2}\right)^{-1305/2299}.
\end{equation}
Given and initial eccentricity and starting frequency, this formula provides us with an curve describing $e(\langle \Omega \rangle)$. 
We compare this leading post-Newtonian result to our numerical simulations in figure~\ref{fig:eccentricities}.

\section{Validation of frequency extraction methods}
\label{sec:FreqValidation}

\begin{figure}
\centering
\includegraphics[scale=0.19]{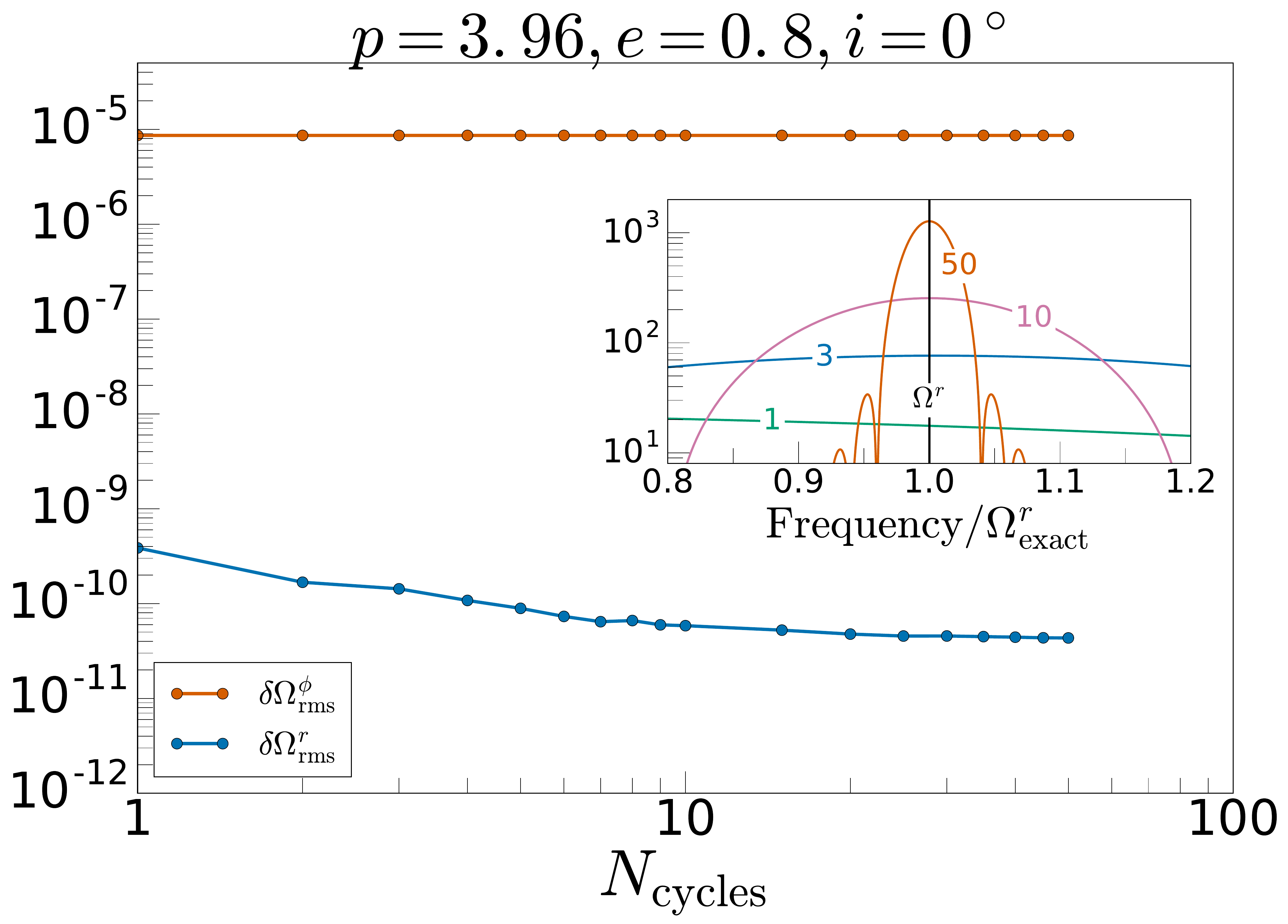}
\includegraphics[scale=0.19]{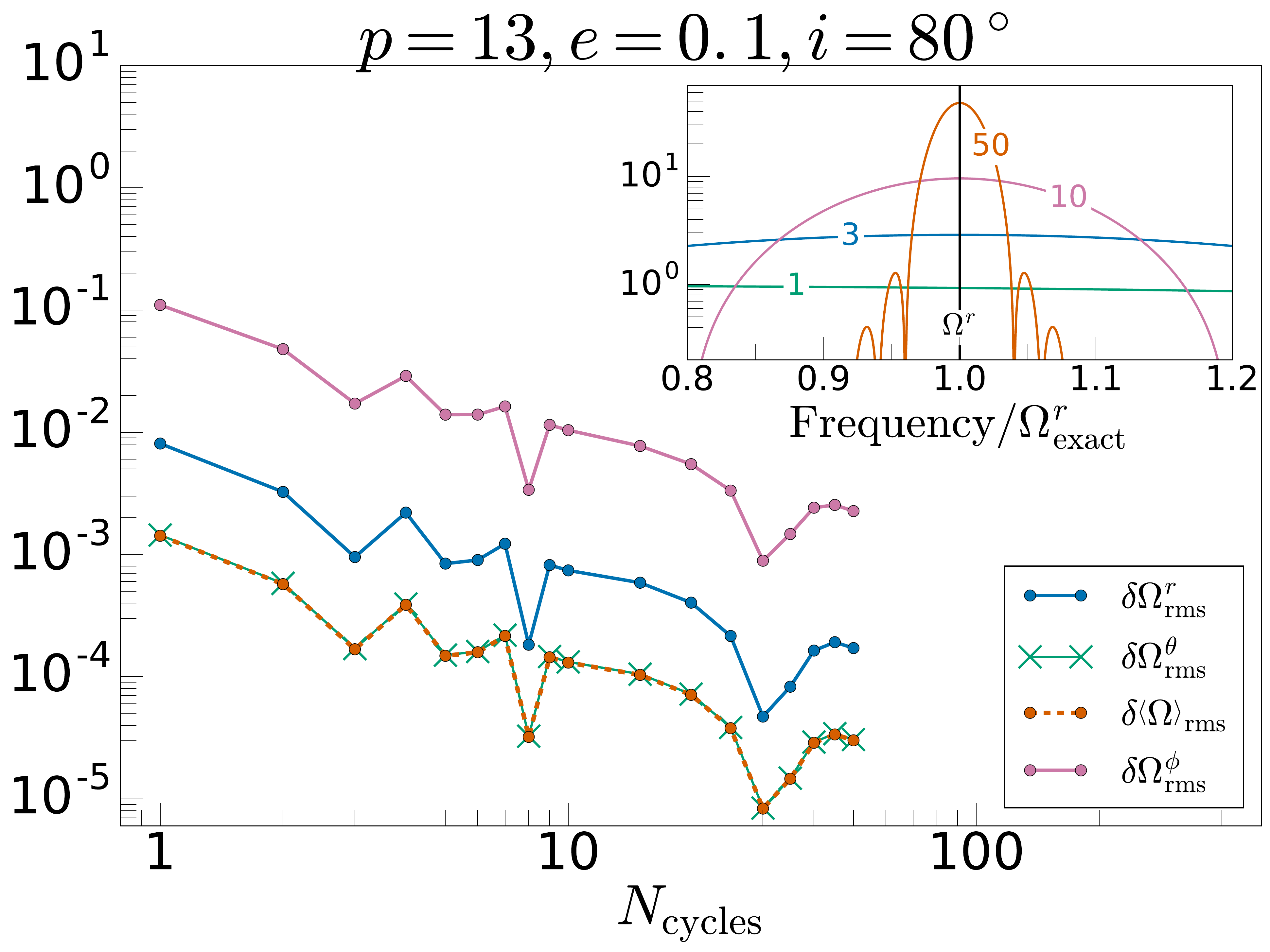}
\caption{\label{fig:kerrconverge} Performance of frequency extraction for Kerr-geodesics.  Shown are the RMS differences 
  $\delta\Omega^a_{\rm rms} = ||\Omega^a -
  \Omega^a_\mathrm{exact}||/\Omega^a_\mathrm{exact}$,
  where $\Omega^a$ is computed from the geodesic trajectory with the same techniques we employ for our BBH analysis, and $\Omega^a_\mathrm{exact}$ is the exact frequency of the Kerr geodesic,
for which we also show   $\Omega^\theta$ and $\left<\Omega\right>$.  We repeat the frequency-extraction using intervals covering $N_{\rm cycles}$ radial oscillation periods, and plot the resulting $\delta\Omega^a_{\mathrm{rms}}$ as a function of $N_{\rm cycles}$  
The insets show absolute values of FFTs on $\Omega(t)$ computed over widths of $1, 3, 10$ and $50$ radial cycles, with the solid black line highlighting $\Omega^r_\mathrm{exact}$.
  Even at 50 cycles the peaks remain unacceptably broad.}
\end{figure}

This appendix assesses the accuracy of our 
frequency extraction techniques by application to Kerr geodesics, whose fundamental 
frequencies are known analytically.
 We compute one equatorial Kerr geodesic (eccentricity $e=0.8$,
  semilatus rectum $p=3.96M$), and one inclined Kerr geodesic
  (inclination $i=80^\circ$, $e=0.1$, $p=13M$). The geodesics are integrated with high precision by an ODE integrator, but we output only a discrete time series with spacing similar to that of the BBH simulations of our main analysis.  We then apply the
  frequency extraction techniques described in
  sections~\ref{sec:EquatorialFrequencies} and~\ref{sec:Inclined} to the discrete time series.  In
  addition to computing frequencies over $N_{\rm cycles}=1$ radial
  cycle, we also compute these frequencies over $N_{\rm cycles}\ge 2$
  radial cycles by changing ``$i+1$'' to ``$i+N_{\rm cycles}$'' in
  equations~(\ref{eq:omphiequatorial}) and (\ref{eq:omrequatorial}).
 For each value of $N_{\rm cycles}$ we compute the relative difference
  $\delta\Omega^a_i = (\Omega^a_i -
  \Omega^a_{\mathrm{exact}})/\Omega^a_\mathrm{exact}$, and 
  and its root-mean-square
  \begin{equation}
  \label{eq:rms}
  \delta\Omega^a_{\rm rms} \equiv \left(\frac{1}{N}\sum_i^N(\delta\Omega^a_i)^2\right)^{1/2}.
\end{equation}
Here, $N$ is the number of elements $\delta\Omega^a_i$ for the
  current value of $N_{\rm cycles}$.

  The left panel of figure~\ref{fig:kerrconverge} shows the
  results for the equatorial geodesic.  Equatorial geodesics are
  strictly periodic and equations~(\ref{eq:omphiequatorial}) and
  (\ref{eq:omrequatorial}) should recover the exact frequencies, up to
  the numerical accuracy of our extraction techniques.  Indeed,
  $\delta\Omega^\phi_{\rm rms}$ in the left panel is dominated by
  our numerical procedure to compute $\Omega$, which involves a
  fourth-order spline interpolant in the construction of
  $\dot{\bm r}$, and $\delta\Omega^\phi_{\rm rms}$ decreases as the
  fourth power of the time-sampling of the geodesic data.  The
  accuracy of the extraction of $\Omega^r$ is limited by the accuracy
  $\sim 10^{-6}M_{\rm Kerr}$ with which our procedure determines the times $t_i^+$ of the maxima of
  $\Omega(t)$.  Our extraction errors for the equatorial orbits are
  several orders of magnitude smaller than the systematic errors that
  arise due to genuine multi-periodicity for inclined orbits. 

  We now turn to the analysis of the inclined geodesic, which is
  presented in the right panel of figure~\ref{fig:kerrconverge}.  The
  radial frequency $\Omega^r$ is recovered to an accuracy better than
  $1\%$ when $N_{\rm cycles}=1$.  The polar frequency $\Omega^{\theta}$ and the average orbital
  frequency $\left<\Omega\right>$ are recovered to nearly $0.1\%$. We
  also illustrate extraction of the azimuthal frequency $\Omega^\phi$.
  To extract $\Omega^\phi$, we begin by defining an ``in-plane'' separation vector
  ${\bm \rho}={\bm r}-({\bm r}\cdot\hat{\bm\chi})\hat{\bm\chi}$, where
  $\hat{\bm\chi}$ is the direction of the black hole angular momentum.
  We compute an instantaneous azimuthal frequency by
\begin{equation}
  \label{eq:omrho}
  \Omega^\rho(t) = \frac{{\bm\rho} \times \dot{\bm\rho}}{\bm\rho^2}
  \end{equation}
and obtain the averaged azimuthal frequency $\Omega^\phi$ by evaluating the right-hand-side of~(\ref{eq:omphiequatorial}), 
substituting $\Omega^\rho(t)$ for $\Omega(t)$.
We recover $\Omega^\phi$ with a fractional accuracy of about $10\%$ (for $N_{\rm cycles}=1$).

   The errors $\delta\Omega^a_{\rm rms}$ for the inclined
   geodesic stem from the interaction of the radial and polar motion.
   Simply put, subsequent periastron passages occur at different
   values of $\theta$.  As the extraction interval is lengthened
   (larger $N_{\rm cycles}$), this dependence is averaged out over a
   longer time-intervals, and $\delta\Omega^a_{\rm rms}$ fall
   $\propto 1/N_{\rm cycles}$.  The magnitudes of
   $\delta\Omega^a_{\rm rms}$ for each frequency depend on how
   important the interactions between radial and polar motion are for
   that particular frequency.  The chosen geodesic has high
   inclination $i=80^\circ$ and fairly small eccentricity $e=0.1$;
   while the radial motion provides the dominant modulations (and our
   choice to average over radial periods
   $[t_i^+, t_{i+N_{\rm cycles}}^+]$ is appropriate), the considered
   geodesic emphasizes the impact of the polar motion.  As this
   example demonstrates, $\Omega^\phi$ is most susceptible to
   modulations, a finding that we confirm for the BBH
   systems, and which
   provides the basis for our preference of $\left<\Omega\right>$ over
   $\Omega^\phi$.

\begin{figure}
\centering
\includegraphics[scale=0.164]{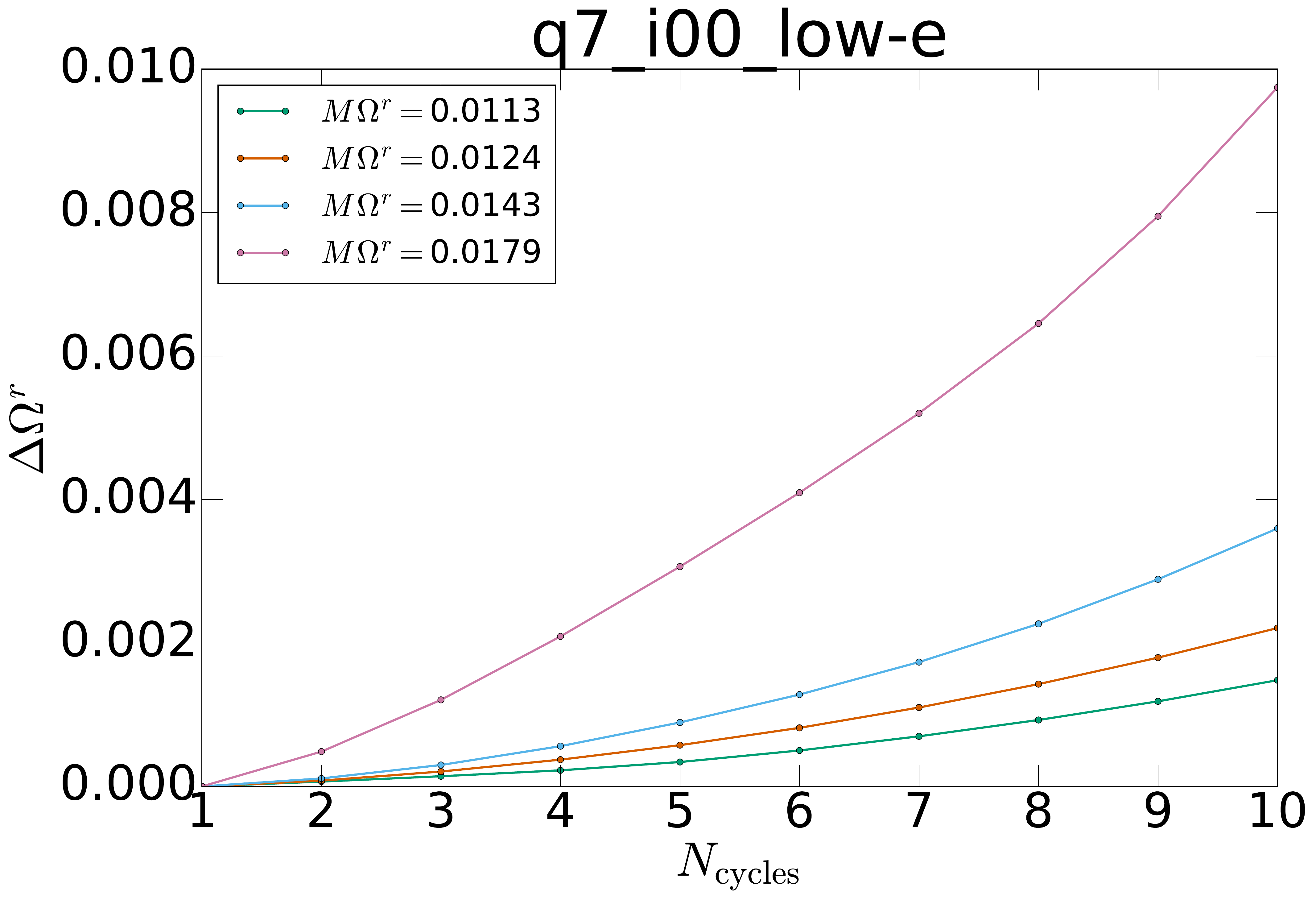}
\includegraphics[scale=0.164]{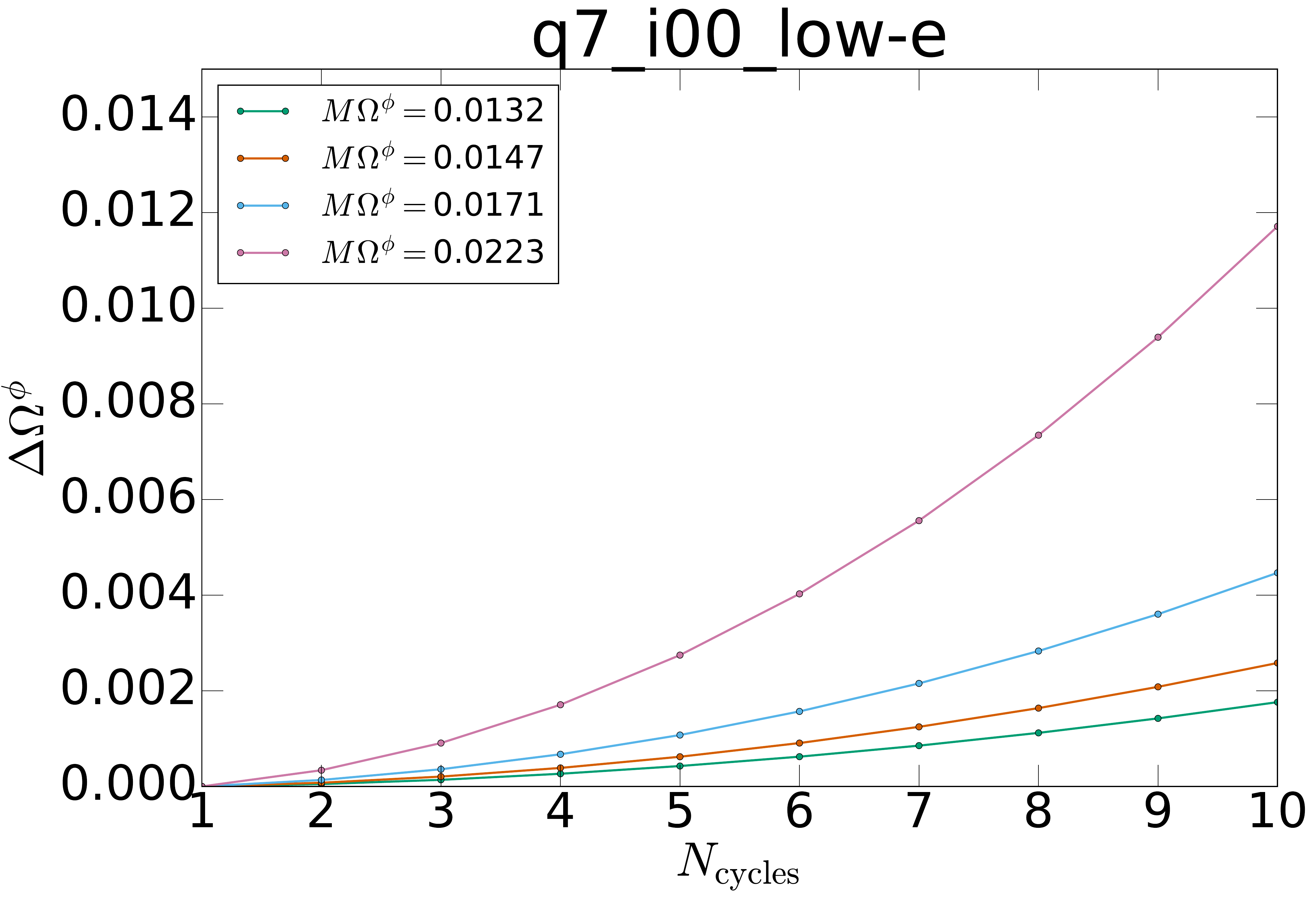} 
\\
\includegraphics[scale=0.164]{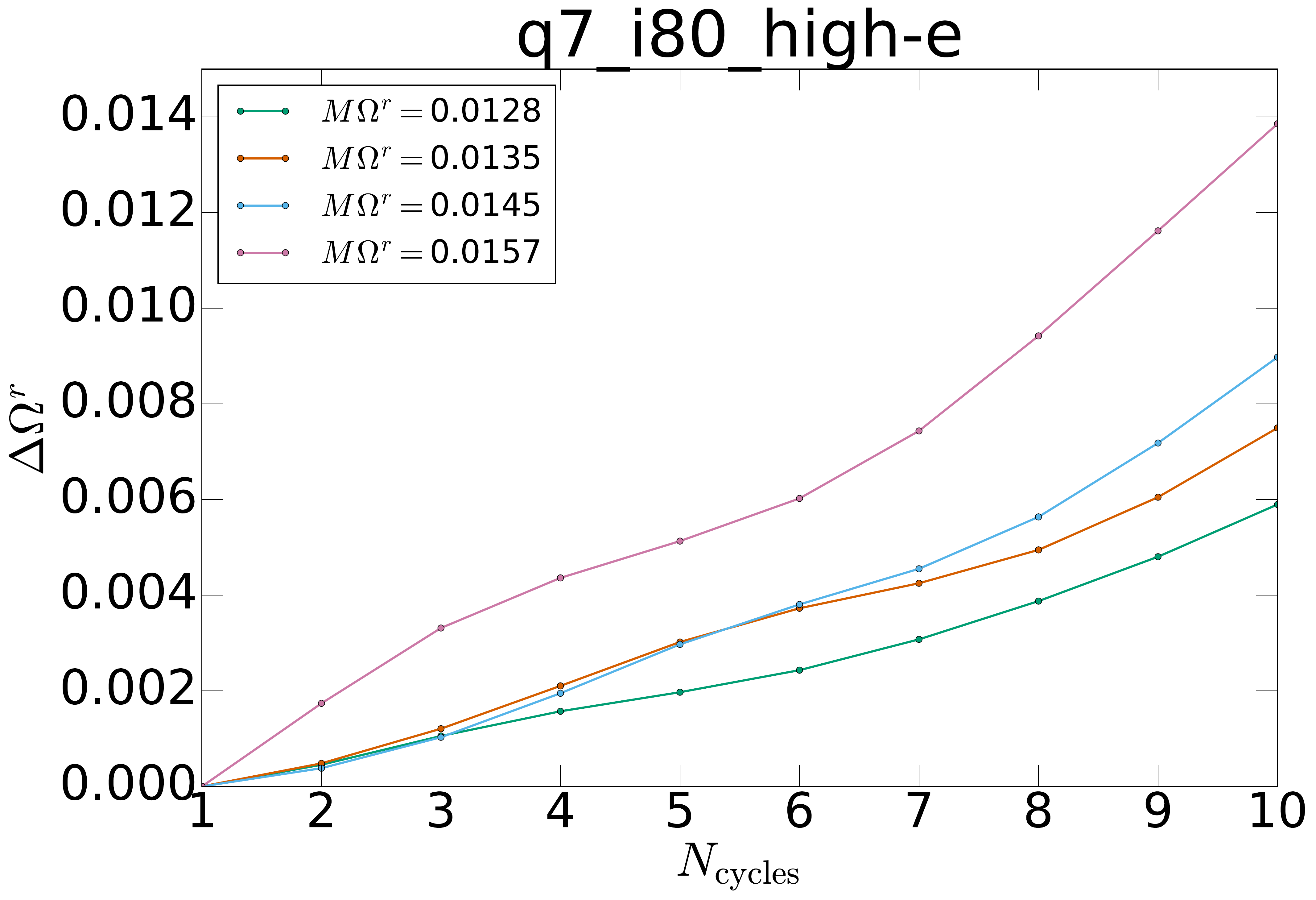}
\includegraphics[scale=0.164]{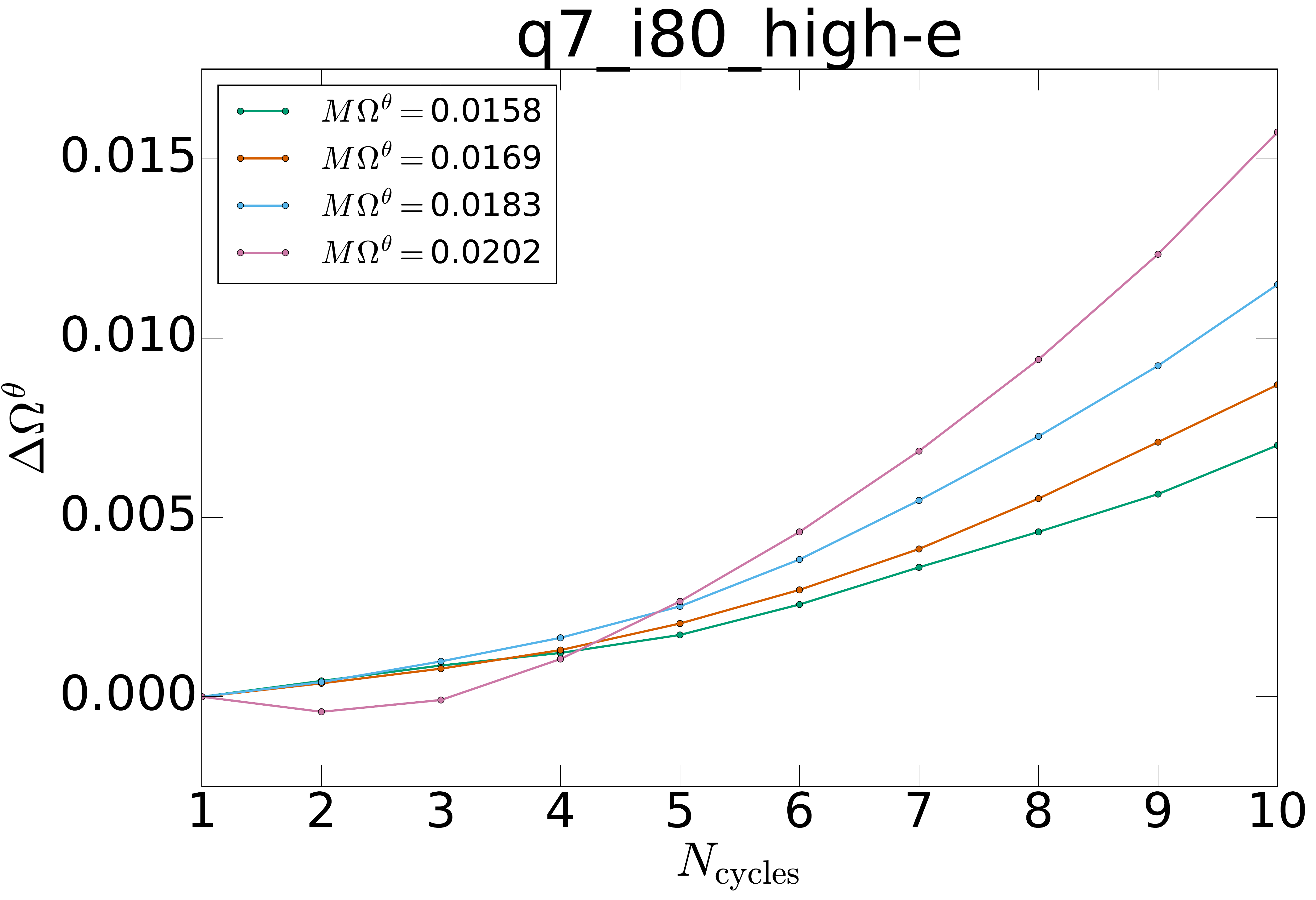} 
\\
\includegraphics[scale=0.164]{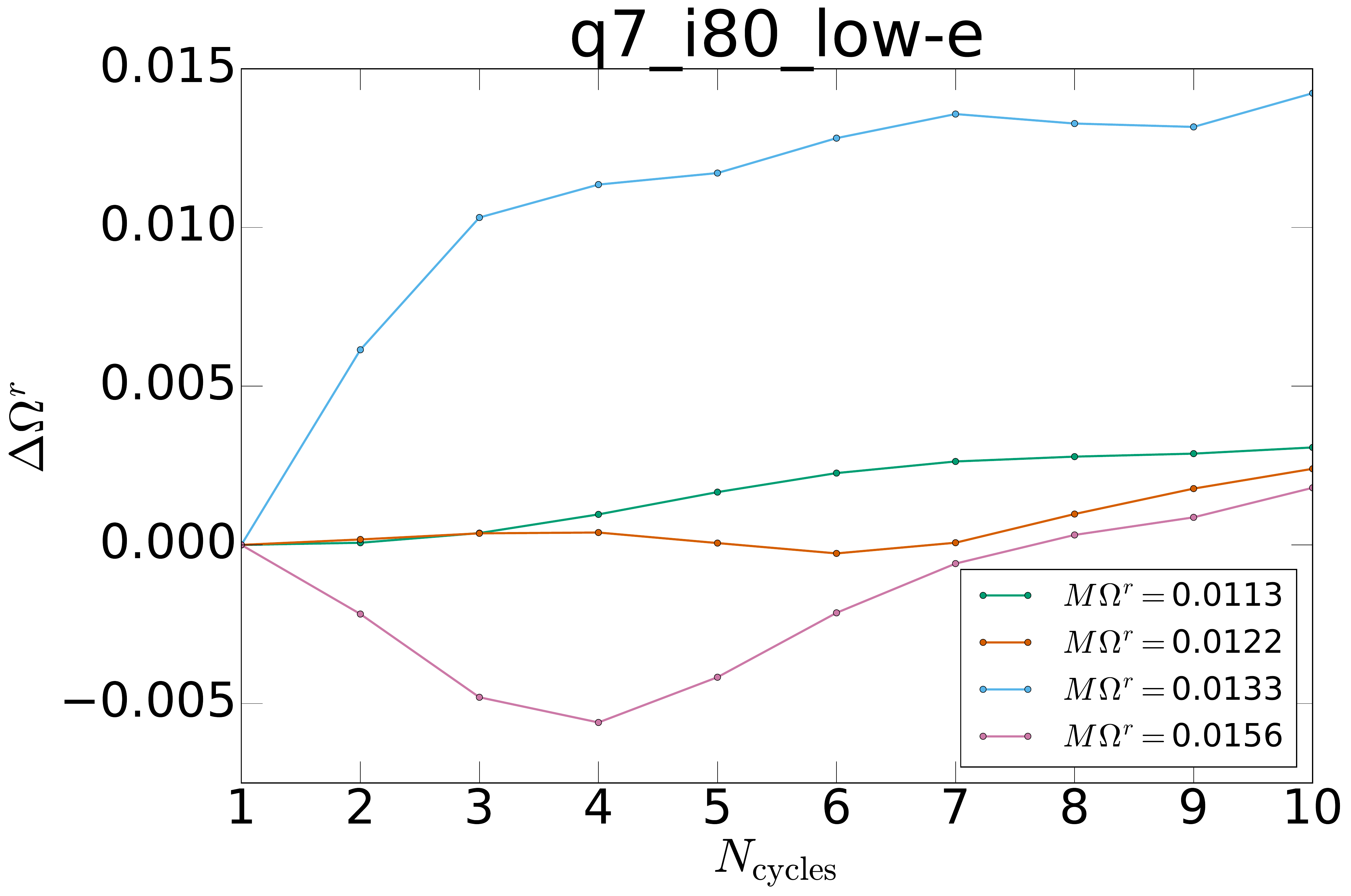}
\includegraphics[scale=0.164]{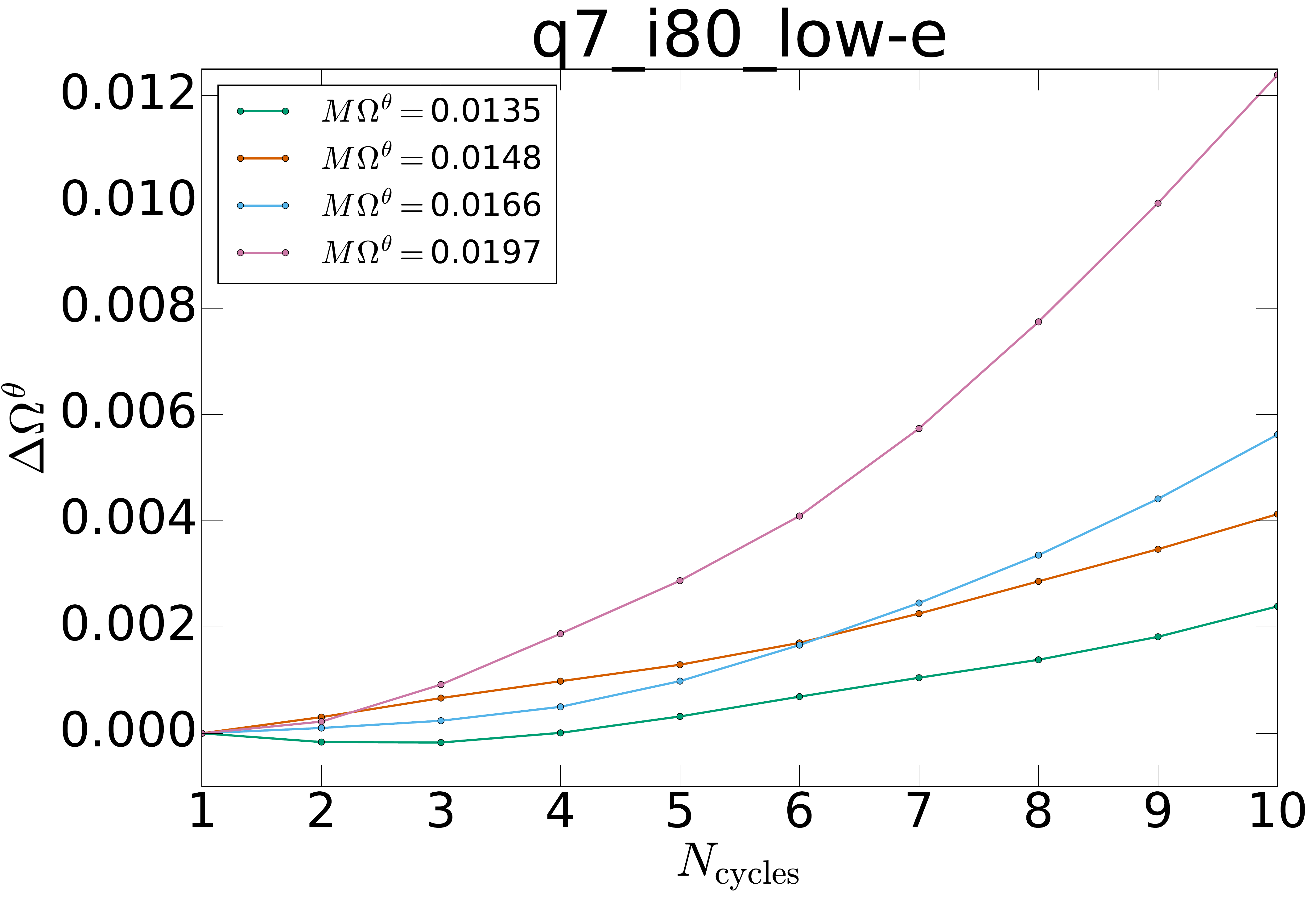} \\
\caption{\label{fig:bbhconverge}  Robustness of frequency-extraction
under change of width of extraction window.  We extract frequencies using windows
of $N_{\rm cycles}$ radial cycles, and plot the relative deviation from 
the same frequencies extracted at $N_{\rm cycles}=1$.  Different panels represent different simulations, and the curves
within each panel correspond to different times $\tilde t^+_i$.
For the equatorial simulations (top panels), the radial and azimuthal frequencies are shown, whereas for the inclined simulations (middle and bottom panels), the radial and polar frequencies are plotted.}
\end{figure}

   The insets in figure~\ref{fig:kerrconverge} illustrate
     frequency-extraction with more conventional Fourier techniques,
     in the form of
     periodograms computed from Fourier transforms of Hann-windowed samples of $\Omega(t)$
     over window-sizes of $N_\mathrm{cycles}=1,\,3,\ 10$, and $50$ radial
     passages.  For sufficiently long window-sizes, the periodograms
     do converge to the exact Kerr-frequency.  However, at window
     sizes practical for the inspiral rate of our BBH simulations
     ($N_{\rm cycles}=1, 3$), periodograms do not yield any useful
     information, and even when using 50 radial cycles, the achieved
     accuracy is only $\sim 1\%$.

Figure~\ref{fig:bbhconverge} examines the impact of the
     length of the extraction intervals (parameterized by
     $N_{\rm cycles}$) for our BBH simulations. 
We extract frequencies $\Omega^a_N$ computed over intervals of length $N=N_{\rm cycles}$, which we 
associate with the midpoint times $\tilde{t}^+_{i, N}=(t_i^++t_{i+N}^+)/2$.  We interpolate the time series 
$\Omega^a_N$ onto the mid-times $\tilde t^+_i$, and compute the relative difference from the 
$N_\mathrm{cycles}=1$ case, 
\begin{equation}
\Delta\Omega_N^a = \frac{\Omega^a_N- \Omega^a_1}{\Omega^a_1},
\end{equation}
where the subscript denotes the value of $N_{\rm cycles}$ over which the frequency is computed. 
    Because the characteristic
  frequencies increase during the inspiral, we report
  $\Delta\Omega_N^a$ at select values of $\Omega^a$, rather than
the RMS error across the whole series.

In all cases we find that $\Delta\Omega^a$ increases with $N_{\rm cycles}$, with the largest
  $\Delta\Omega^a$ closer to merger (i.e.~at larger $\tilde t^+_i$). 
  For equatorial orbits our frequency extraction is very precise, and
  so $N_{\rm cycles}=1$ will ensure the least impact by errors arising
  from dissipative inspiral.  For precessing orbits, there are two
  competing effects: First, with increasing $N_{\rm cycles}$, the
  impact of cycle-to-cycle modulations diminishes, leading to
  extracted sequences $\Omega^a_i$ that show less variations between
  neighboring extrema~$i$.  Second, increasing $N_{\rm cycles}$ leads
  to larger systematic biases of the extracted frequencies
  $\Omega^a_i$, cf. figure~\ref{fig:bbhconverge}.  To avoid
  contamination by such systematic biases, we choose
  $N_{\rm cycles}=1$ throughout, accepting possibly larger variations
  of the extracted frequencies.

\begin{figure}
\centering
\includegraphics[width=0.495\columnwidth]{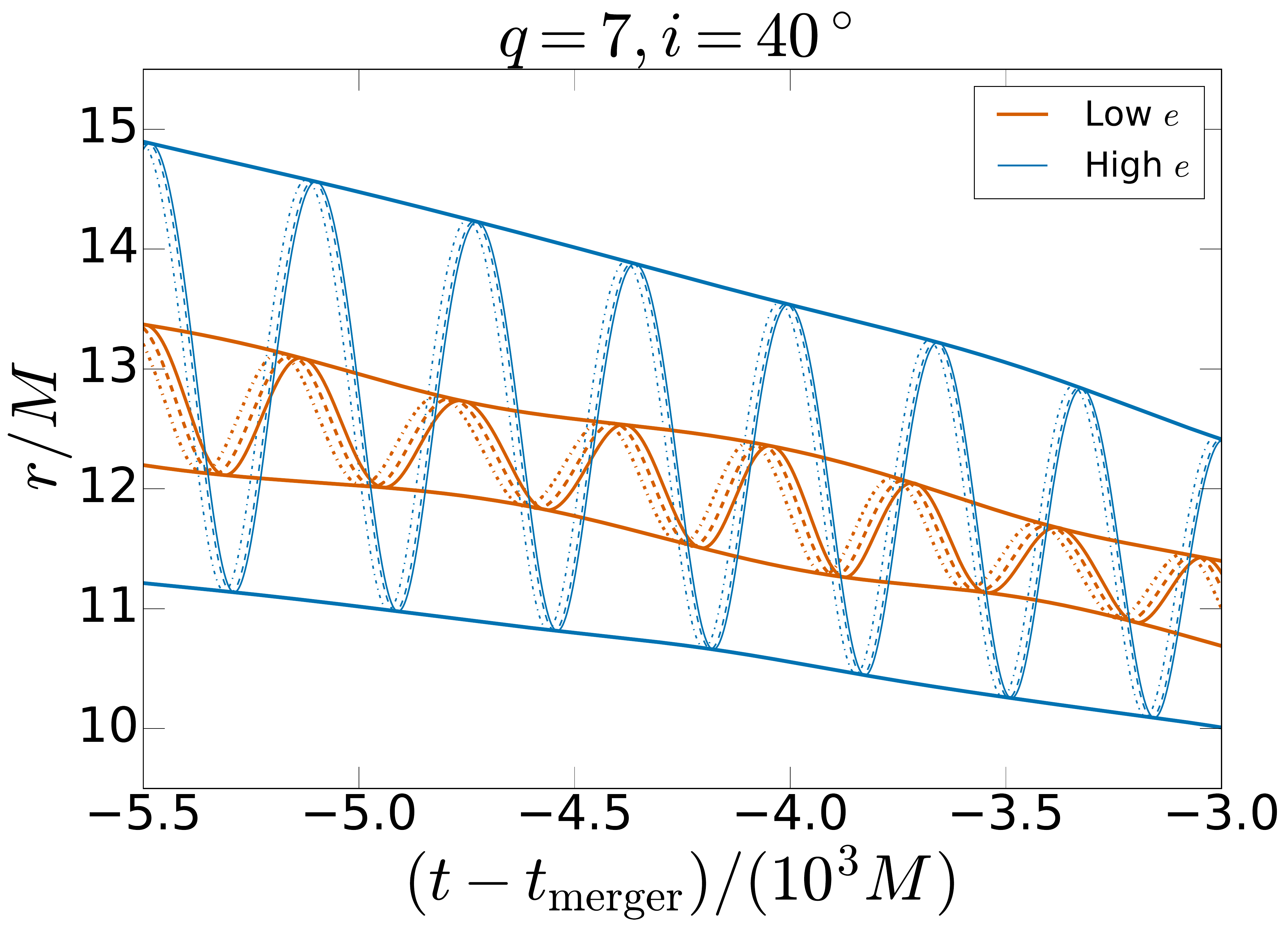}
\includegraphics[width=0.495\columnwidth]{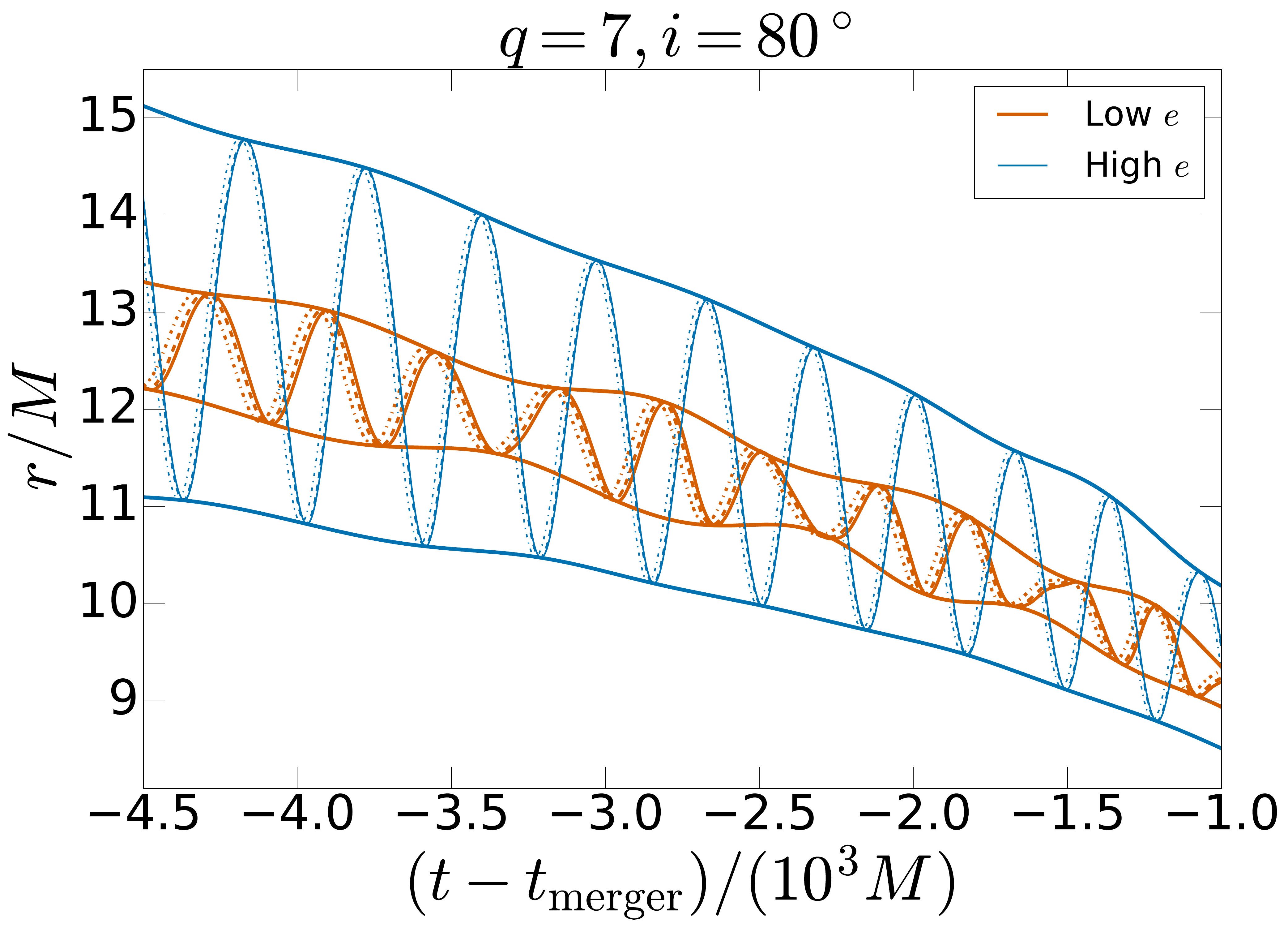}
\caption{\label{fig:bad_r_fig} Separation $r(t)$ for the two ``ill-behaved'' low-eccentricity simulations,
zoomed in to show the aperiodic behaviour.   Also plotted are the envelopes of $r(t)$ to more clearly show the strong modulations in the low-eccentricity simulations.
Three different numerical resolutions are plotted as different linestyles; the behaviour is 
convergent and mostly identical across resolutions.}
\end{figure}

Some of our simulations result in extracted frequencies that are noticably more erratic than the rest of the runs.  This effect is most striking
for the low-eccentricity highly-inclined orbits, cf. lower left panel of figure~\ref{fig:kij_inc}. These simulations are ill-behaved 
throughout our study, yielding far less smooth precession rate curves 
than their counterparts (cf.~figure \ref{fig:kij_inc}). 
Study of the separation $r(t)$ for these simulations (cf.~figure \ref{fig:BBHsepvst}) 
reveals unusually strong
aperiodic features in the trajectories, which persists
across our different numerical resolutions, which we show in figure \ref{fig:bad_r_fig}. 
Curiously, the 
aperiodicity is far less dramatic at higher eccentricities, and indeed worsens
during approach to merger, where the eccentricity is lower. This may be a sign
that for these particular parameter choices quasi-periodic effects are comparable
to dissipative ones, and that therefore $N_\mathrm{cycles}=1$ (or any constant
$N_\mathrm{orbits}$) is inappropriate here. Alternatively, this may be an
effect of the damped-harmonic \spec~simulation gauge \cite{Lindblom2009c, Choptuik:2009ww, Szilagyi:2009qz}, 
whose behaviour is largely untested for
dynamically generic orbits.

\section{Estimation of resonant accretion}
\label{sec:ResEstimation}
To increase the sensitivity to potential resonant effects, we
can  average the fluxes $\dot J_\perp$, $\dot J_\parallel$ and $\dot E$
  over $N$ radial oscillation periods, where we slide the averaging
  window to start at each maximum of the respective flux.  From~(\ref{eq:selfforce}), 
  we expect that non-resonant terms are
  averaged out by this procedure.  Therefore, off-resonance, the
  averaged fluxes 
  \begin{equation}
    \langle \dot{\mathcal C} \rangle \approx \dot{\mathcal C}_{00}
  \end{equation}
correspond to the $\dot{\mathcal C}_{00}$ term.  
When averaging over a time window smaller than the time on resonance, and when the window is centered on the resonance,
  we can approximate $\Phi_{kn}(t)\approx
  \Phi_{kn}(t_{\rm res})$ over the averaging window.
  Then \eqref{eq:selfforce} yields
  \begin{equation}\label{eq:Cavg-resonant}
    \langle \dot{\mathcal C} \rangle \approx\dot{\mathcal C}_{00}+\dot{\mathcal C}_{kn}e^{-i\Phi_{kn}(t_{\rm res})}.
  \end{equation}
Therefore, during resonance, the averaged fluxes $\langle \dot{\mathcal C}\rangle$ should
  show a relative deviation with peak amplitude $\delta\dot C\sim \dot{\mathcal C}_{kn}e^{-\Phi_{kn}(t_{\rm res})}/\dot{\mathcal C}_{00}$ compared to orbits off-resonance.

\begin{figure}
\centerline{
  \includegraphics[width=0.495\columnwidth]{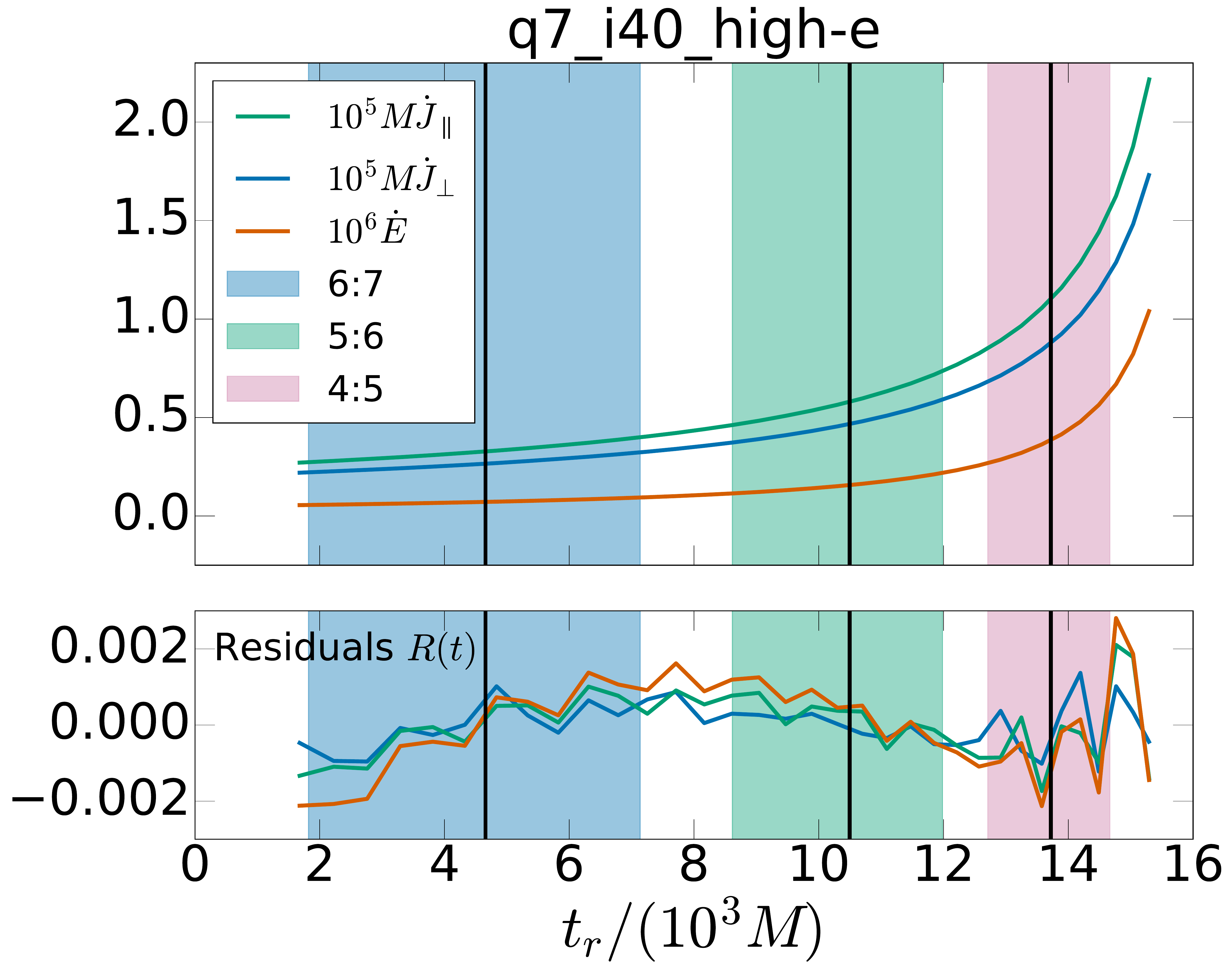} 
  \includegraphics[width=0.495\columnwidth]{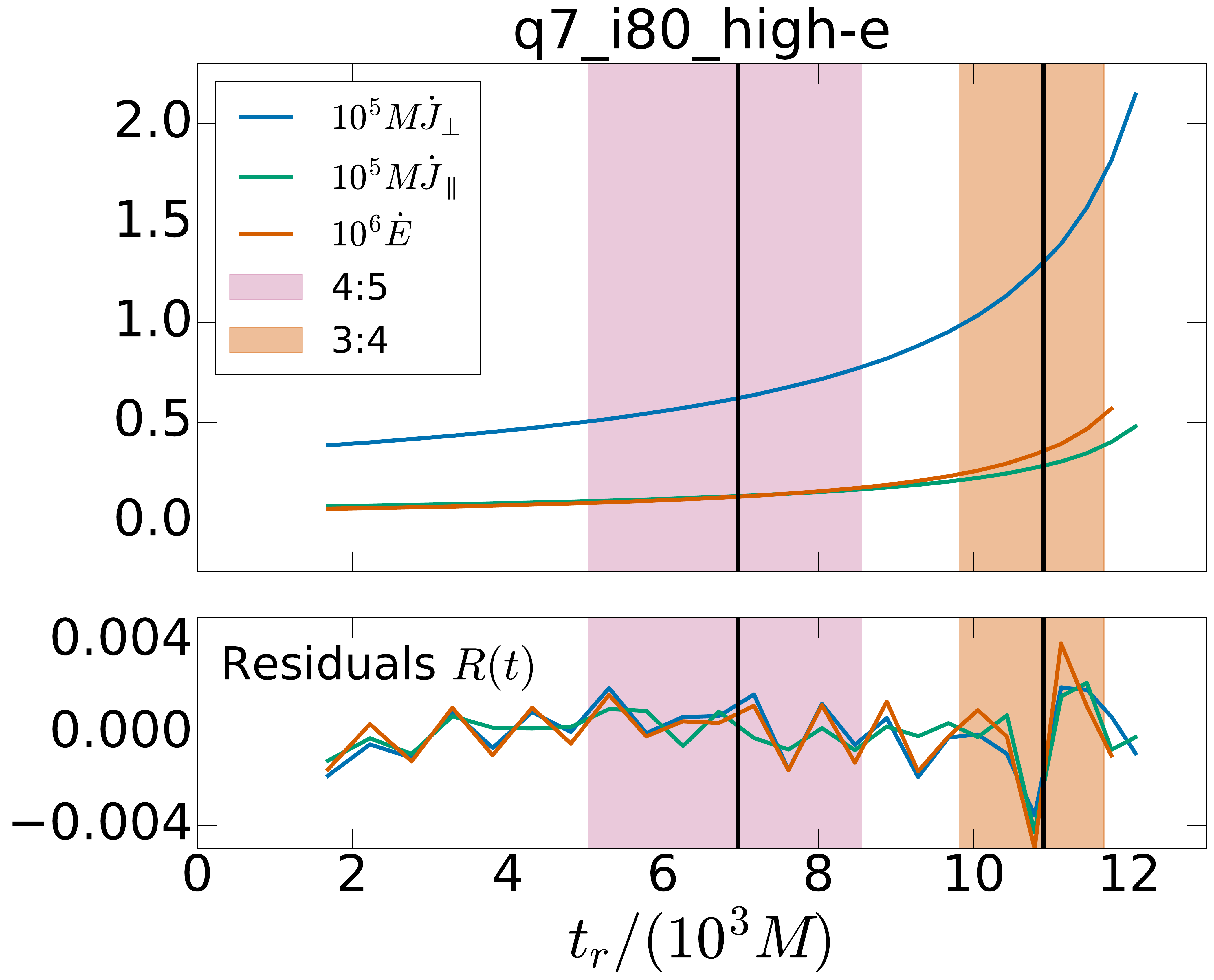}}
\caption{
\label{fig:fluxAvg}
Gravitational wave fluxes averaged over $N=5$ radial oscillation periods for the runs {\tt q7\_i40\_high-e} and  {\tt q7\_i80\_high-e}.
The top panels show the actual data, and the lower panels the normalized residuals with respect to a fit that removes
the overall inspiral trend, cf.~\eqref{eq:FluxResidual}.}
\end{figure}

The resulting averaged fluxes are shown in the top panel of
  figure~\ref{fig:fluxAvg} for two of our most extreme simulations, 
  \texttt{q7\_i40\_high-e} and \texttt{q7\_i80\_high-e}.  
  On the vertical scale of these plots, no variations of
  the fluxes are discernible between on- and off-resonance.  To
  sharpen our bound,
  we remove the overall inspiral component of the fluxes through a fit
  to a function $f_{\rm fit}(t)=A_0(A_1-t)^{A_2}+A_3$.  The functional form
  is chosen to capture the overall inspiral behavior, while not being able to
  capture intermediate variations that are expected on resonance.  Therefore,
  we expect $f_{\rm fit}(t)\approx \dot{\mathcal C}_{00}$.
  The normalized residual
\begin{equation}\label{eq:FluxResidual}
    R(t) = \frac{\langle \dot{\mathcal C}\rangle - f_{\rm fit}}{\langle\dot{\mathcal C}\rangle}
  \end{equation}
 should therefore vanish off-resonance.  
On-resonance,~(\ref{eq:Cavg-resonant}) suggests that 
(dropping terms of order unity)
$R\sim \dot{\mathcal C}_{kn}e^{-i\Phi_{kn}(t_{\rm res}) }/\dot{\mathcal C}_{00}$.  
As can be seen from the lower panels of figure~\ref{fig:fluxAvg}, we
  find numerically $R\lesssim 0.004$.  The short-period variations in $R$
  in this figure are caused by the numerical accuracy with which we
  can extract periastron passages and perform the averaging.  The
  overall smooth trend ($R$ is slightly negative at small $t$)
  indicates the quality with which our fitting function $f_{\rm fit}$
  can capture the overall inspiral dynamics.  Besides these two
  properties of $R(t)$, figure~\ref{fig:fluxAvg} does not show any
  systematic deviations at the highlighted resonances.
  Therefore, we conclude for these resonances that
\begin{equation}
    \frac{\dot{\mathcal C}_{kn}e^{-i\Phi_{kn}(t_{\rm res})}}{\dot{\mathcal C}_{00}}\lesssim 0.004.
  \end{equation}
  
\section*{References}
\bibliographystyle{iopart-num}
\bibliography{References/References}

\providecommand{\newblock}{}
\begin{thebibliography}{100}
\expandafter\ifx\csname url\endcsname\relax
  \def\url#1{{\tt #1}}\fi
\expandafter\ifx\csname urlprefix\endcsname\relax\def\urlprefix{URL }\fi
\providecommand{\eprint}[2][]{\url{#2}}

\bibitem{LIGOVirgo2016a}
Abbott B~P {\em et~al.\/} (LIGO Scientific Collaboration, Virgo Collaboration)
  2016 {\em Phys.\ Rev.\ Lett.\/} {\bf 116}(6) 061102 (\textit{Preprint}
  \eprint{1602.03837})
  \urlprefix\url{http://link.aps.org/doi/10.1103/PhysRevLett.116.061102}

\bibitem{TheLIGOScientific:2016qqj}
{Abbott} B~P {\em et~al.\/} (LIGO Scientific Collaboration, Virgo
  Collaboration) 2016 {\em Phys.\ Rev.\ D\/} {\bf 93} 122003 (\textit{Preprint}
  \eprint{1602.03839})

\bibitem{Abbott:2016nmj}
Abbott B~P {\em et~al.\/} (LIGO Scientific Collaboration, Virgo Collaboration)
  2016 {\em Phys. Rev. Lett.\/} {\bf 116} 241103 (\textit{Preprint}
  \eprint{1606.04855})

\bibitem{LIGO-DataAnalysis-Whitepaper:2015}
Cadonati L {\em et~al.\/} (LIGO Scientific Collaboration, Virgo Collaboration)
  2015 The {LSC}-{V}irgo white paper on gravitational wave searches and
  astrophysics URL \url{https://dcc.ligo.org/LIGO-T1500055/public}

\bibitem{TheLIGOScientific:2016pea}
Abbott B~P {\em et~al.\/} (Virgo, LIGO Scientific) 2016 {\em Phys. Rev.\/} {\bf
  X6} 041015 (\textit{Preprint} \eprint{1606.04856})

\bibitem{PetersMathews1963}
Peters P~C and Mathews J 1963 {\em Phys.\ Rev.\/} {\bf 131} 435
  \urlprefix\url{http://link.aps.org/abstract/PR/v131/p435}

\bibitem{Khan:2015jqa}
Khan S, Husa S, Hannam M, Ohme F, Pürrer M, Jiménez~Forteza X and Bohé A
  2016 {\em Phys. Rev.\/} {\bf D93} 044007 (\textit{Preprint}
  \eprint{1508.07253})

\bibitem{Taracchini:2013rva}
Taracchini A, Buonanno A, Pan Y, Hinderer T, Boyle M, Hemberger D~A, Kidder
  L~E, Lovelace G, Mroue A~H, Pfeiffer H~P, Scheel M~A, Szil{\'a}gyi B, Taylor
  N~W and Zenginoglu A 2014 {\em Phys.\ Rev.\ D\/} {\bf 89 (R)} 061502
  (\textit{Preprint} \eprint{1311.2544})

\bibitem{Hannam:2013oca}
Hannam M, Schmidt P, Boh{\' e} A, Haegel L, Husa S {\em et~al.\/} 2014 {\em
  Phys.\ Rev.\ Lett.\/} {\bf 113} 151101 (\textit{Preprint} \eprint{1308.3271})

\bibitem{Pan:2013rra}
{Pan} Y, {Buonanno} A, {Taracchini} A, {Kidder} L~E, {Mrou{\'e}} A~H,
  {Pfeiffer} H~P, {Scheel} M~A and {Szil{\'a}gyi} B 2013 {\em Phys.\ Rev.\ D\/}
  {\bf 89} 084006 (\textit{Preprint} \eprint{1307.6232})

\bibitem{Babak:2016tgq}
Babak S, Taracchini A and Buonanno A 2016  (\textit{Preprint}
  \eprint{1607.05661})

\bibitem{baumgarteShapiroBook}
Baumgarte T~W and Shapiro S~L 2010 {\em Numerical Relativity: Solving
  Einstein's Equations on the Computer\/} (New York: Cambridge University
  Press)

\bibitem{Ajith:2012az}
Ajith P, Boyle M, Brown D~A, Brugmann B, Buchman L~T {\em et~al.\/} 2012 {\em
  Class.\ Quantum Grav.\/} {\bf 29} 124001

\bibitem{Hinder:2013oqa}
Hinder I {\em et~al.\/} (The NRAR Collaboration) 2014 {\em Class.\ Quantum
  Grav.\/} {\bf 31} 025012 (\textit{Preprint} \eprint{1307.5307})

\bibitem{Mroue:2013PRL}
{Mrou{\'e}} A~H, {Scheel} M~A, {Szil{\'a}gyi} B, {Pfeiffer} H~P, {Boyle} M,
  {Hemberger} D~A, {Kidder} L~E, {Lovelace} G, {Ossokine} S, {Taylor} N~W,
  {Zengino{\u g}lu} A, {Buchman} L~T, {Chu} T, {Foley} E, {Giesler} M, {Owen} R
  and {Teukolsky} S~A 2013 {\em Phys.\ Rev.\ Lett.\/} {\bf 111} 241104
  (\textit{Preprint} \eprint{1304.6077})

\bibitem{Chu:2015kft}
{Chu} T, {Fong} H, {Kumar} P, {Pfeiffer} H~P, {Boyle} M, {Hemberger} D~A,
  {Kidder} L~E, {Scheel} M~A and {Szil{\'a}gyi} B 2016 {\em Class.\ Quantum
  Grav.\/} {\bf 33} 165001 (\textit{Preprint} \eprint{1512.06800})

\bibitem{Jani:2016wkt}
Jani K, Healy J, Clark J~A, London L, Laguna P and Shoemaker D 2016 {\em Class.
  Quant. Grav.\/} {\bf 33} 204001 (\textit{Preprint} \eprint{1605.03204})

\bibitem{Husa:2015iqa}
Husa S, Khan S, Hannam M, Pürrer M, Ohme F, Jim\'{e}nez~Forteza X and Boh\'{e}
  A 2016 {\em Phys. Rev.\/} {\bf D93} 044006 (\textit{Preprint}
  \eprint{1508.07250})

\bibitem{SperhakeEtAl:2008}
Sperhake U, Berti E, Cardoso V, Gonz\'alez J~A, Br\"ugmann B and Ansorg M 2008
  {\em Phys. Rev. D\/} {\bf 78}(6) 064069
  \urlprefix\url{http://link.aps.org/doi/10.1103/PhysRevD.78.064069}

\bibitem{Hinder2008}
Hinder I, Vaishnav B, Herrmann F, Shoemaker D~M and Laguna P 2008 {\em Phys.\
  Rev.\ D\/} {\bf 77} 081502 (pages~5)
  \urlprefix\url{http://link.aps.org/abstract/PRD/v77/e081502}

\bibitem{Hinder:2008kv}
Hinder I, Herrmann F, Laguna P and Shoemaker D 2010 {\em Phys.\ Rev.\ D\/} {\bf
  82} 024033 (\textit{Preprint} \eprint{0806.1037})

\bibitem{Mroue2010}
Mrou\'{e} A~H, Pfeiffer H~P, Kidder L~E and Teukolsky S~A 2010 {\em Phys.\
  Rev.\ D\/} {\bf 82} 124016 (\textit{Preprint} \eprint{arXiv:1004.4697
  [gr-qc]})

\bibitem{East:2015yea}
{East} W~E, {Paschalidis} V and {Pretorius} F 2015 {\em ApJ Letters\/} {\bf
  807} L3 (\textit{Preprint} \eprint{1503.07171})

\bibitem{Gold:2012tk}
Gold R and Bruegmann B 2013 {\em Phys.\ Rev.\ D\/} {\bf 88} 064051
  (\textit{Preprint} \eprint{1209.4085})

\bibitem{Huerta:2016rwp}
{Huerta} E~A, {Kumar} P, {Agarwal} B, {George} D, {Schive} H~Y, {Pfeiffer} H~P,
  {Chu} T, {Boyle} M, {Hemberger} D~A, {Kidder} L~E, {Scheel} M~A and
  {Szil{\'a}gyi} B 2016 {\em Submitted to Phys.~Rev.~D.; arXiv:1609.05933\/}
  (\textit{Preprint} \eprint{1609.05933})

\bibitem{Morscher:2014doa}
Morscher M, Pattabiraman B, Rodriguez C, Rasio F~A and Umbreit S 2015 {\em
  Astrophys. J.\/} {\bf 800} 9 (\textit{Preprint} \eprint{1409.0866})

\bibitem{2008ApJ...676.1162S}
{Sadowski} A, {Belczynski} K, {Bulik} T, {Ivanova} N, {Rasio} F~A and
  {O'Shaughnessy} R 2008 {\em Astrophys.\ J.\/} {\bf 676} 1162--1169
  (\textit{Preprint} \eprint{astro-ph/0710.0878})

\bibitem{Rodriguez:2015oxa}
Rodriguez C~L, Morscher M, Pattabiraman B, Chatterjee S, Haster C~J and Rasio
  F~A 2015 {\em Phys.\ Rev.\ Lett.\/} {\bf 115} 051101 (\textit{Preprint}
  \eprint{1505.00792})

\bibitem{Rodriguez:2016}
Rodriguez C~L, Chatterjee S and Rasio F~A 2016 {\em Phys. Rev. D\/} {\bf 93}(8)
  084029 \urlprefix\url{http://link.aps.org/doi/10.1103/PhysRevD.93.084029}

\bibitem{Hopman:2006}
Hopman C and Alexander T 2006 {\em The Astrophysical Journal Letters\/} {\bf
  645} L133 \urlprefix\url{http://stacks.iop.org/1538-4357/645/i=2/a=L133}

\bibitem{OLeary2009}
{R M O'Leary, B Kocsis and A Loeb} 2009 {\em Mon.\ Not.\ Roy.\ Astr.\ Soc.\/}
  {\bf 395} 2127--2146 (\textit{Preprint} \eprint{arXiv:astro-ph/0807.2638})

\bibitem{Kocsis:2012}
Kocsis B and Levin J 2012 {\em Phys.\ Rev.\ D\/} {\bf 85} 123005

\bibitem{Tsang:2013}
Tsang D 2013 {\em The Astrophysical Journal\/} {\bf 777} 103
  \urlprefix\url{http://stacks.iop.org/0004-637X/777/i=2/a=103}

\bibitem{SamsingEtAl:2014}
{Samsing} J, {MacLeod} M and {Ramirez-Ruiz} E 2014 {\em Astrophys.\ J.\/} {\bf
  784} 71 (\textit{Preprint} \eprint{1308.2964})

\bibitem{AntoniniEtAl:2016}
Antonini F, Chatterjee S, Rodriguez C~L, Morscher M, Pattabiraman B, Kalogera V
  and Rasio F~A 2016 {\em The Astrophysical Journal\/} {\bf 816} 65
  \urlprefix\url{http://stacks.iop.org/0004-637X/816/i=2/a=65}

\bibitem{Lidov:1962}
{Lidov} M~L 1962 {\em Planetary and Space Science\/} {\bf 9} 719--759

\bibitem{Kozai:1962}
{Kozai} Y 1962 {\em Astronom.\ J.\/} {\bf 67} 591

\bibitem{Ford-Kozinsky-Rasio2000}
Ford E~B, Kozinsky B and Rasio F~A 2004 {\em Astrophys. J.\/} {\bf 605} 966
  (\textit{Preprint} \eprint{astro-ph/9905348})

\bibitem{Ford-Kozinsky-Rasio2004}
Ford E~B, Kozinsky B and Rasio F~A 2004 {\em Astrophys.\ J.\/} {\bf 605}
  966--966

\bibitem{FabryckyAndTremaine:2007}
{Fabrycky} D and {Tremaine} S 2007 {\em Astrophys.\ J.\/} {\bf 669} 1298--1315
  (\textit{Preprint} \eprint{0705.4285})

\bibitem{NaozEtAl:2013}
{Naoz} S, {Farr} W~M, {Lithwick} Y, {Rasio} F~A and {Teyssandier} J 2013 {\em
  Mon.\ Not.\ Roy.\ Soc.\/} {\bf 431} 2155--2171 (\textit{Preprint}
  \eprint{1107.2414})

\bibitem{MerritBook:2013}
{Merritt} D 2013 {\em Dynamics and Evolution of Galactic Nuclei\/} (Princeton:
  Princeton University Press)

\bibitem{AntoniniAndPerets:2012}
{Antonini} F and {Perets} H~B 2012 {\em Astrophys.\ J.\/} {\bf 757} 27
  (\textit{Preprint} \eprint{1203.2938})

\bibitem{KatzAndDong:2012}
{Katz} B and {Dong} S 2012 {\em ArXiv e-prints\/} (\textit{Preprint}
  \eprint{1211.4584})

\bibitem{AntoniniEtAl:2014}
{Antonini} F, {Murray} N and {Mikkola} S 2014 {\em Astrophysical Journal\/}
  {\bf 781} 45 (\textit{Preprint} \eprint{1308.3674})

\bibitem{BodeAndWegg:2014}
{Bode} J~N and {Wegg} C 2014 {\em Mon.\ Not.\ Roy.\ Ast.\ Soc.\/} {\bf 438}
  573--589

\bibitem{Seto:2013wwa}
Seto N 2013 {\em Phys.\ Rev.\ Lett.\/} {\bf 111} 061106 (\textit{Preprint}
  \eprint{1304.5151})

\bibitem{Wen2003}
Wen L 2003 {\em Astrophys.\ J.\/} {\bf 598} 419--430

\bibitem{Stephan:2016}
Stephan A~P, Naoz S, Ghez A~M, Witzel G, Sitarski B~N, Do T and Kocsis B 2016
  {\em Mon. Not. Roy. Astron. Soc.\/} {\bf 460} 3494--3504 (\textit{Preprint}
  \eprint{1603.02709})

\bibitem{Antognini:2014}
Antognini J~M, Shappee B~J, Thompson T~A and Amaro-Seoane P 2014 {\em Mon. Not.
  Roy. Astron. Soc.\/} {\bf 439} 1079--1091 (\textit{Preprint}
  \eprint{1308.5682})

\bibitem{AmaroSeoane2016}
Amaro-Seoane P and Chen X 2016 {\em Mon. Not. Roy. Astron. Soc.\/} {\bf 458}
  3075--3082 (\textit{Preprint} \eprint{1512.04897})

\bibitem{Abbott:2016ymx}
Abbott B~P {\em et~al.\/} (Virgo, LIGO Scientific) 2016  (\textit{Preprint}
  \eprint{1607.07456})

\bibitem{BrownZimmerman2009}
{D Brown and P Zimmerman} 2010 {\em Phys.\ Rev.\ D\/} {\bf 81} 024007
  (\textit{Preprint} \eprint{arXiv:gr-qc/0909.0066})

\bibitem{Huerta:2013qb}
Huerta E and Brown D~A 2013 {\em Phys.\ Rev.\ D\/} {\bf 87} 127501
  (\textit{Preprint} \eprint{1301.1895})

\bibitem{AmaroSeoane:2012je}
Amaro-Seoane P, Aoudia S, Babak S, Binetruy P, Berti E {\em et~al.\/} 2012 {\em
  Class.\ Quantum Grav.\/} {\bf 29} 124016 (\textit{Preprint}
  \eprint{1202.0839})

\bibitem{Seoane:2013qna}
Seoane P~A {\em et~al.\/} (eLISA) 2013  (\textit{Preprint} \eprint{1305.5720})

\bibitem{AmaroSeoane:2009ui}
Amaro-Seoane P and Santamaria L 2010 {\em Astrophys.J.\/} {\bf 722} 1197--1206
  (\textit{Preprint} \eprint{0910.0254})

\bibitem{Tanay:2016zog}
Tanay S, Haney M and Gopakumar A 2016 {\em Phys. Rev.\/} {\bf D93} 064031
  (\textit{Preprint} \eprint{1602.03081})

\bibitem{AckayEtAl:2015}
{Akcay} S, {Le Tiec} A, {Barack} L, {Sago} N and {Warburton} N 2015 {\em Phys.\
  Rev.\ D\/} {\bf 91} 124014 (\textit{Preprint} \eprint{1503.01374})

\bibitem{Tiec:2015cxa}
Le~Tiec A 2015 {\em Phys. Rev.\/} {\bf D92} 084021 (\textit{Preprint}
  \eprint{1506.05648})

\bibitem{LoutrelEtAl:2016}
{Loutrel} N, {Yunes} N and {Pretorius} F 2014 {\em Phys.\ Rev.\ D\/} {\bf 90}
  104010 (\textit{Preprint} \eprint{1404.0092})

\bibitem{1972ApJ...178..347B}
Bardeen J~M, Press W~H and Teukolsky S~A 1972 {\em Astrophys.\ J.\/} {\bf 178}
  347

\bibitem{Flanagan:2010cd}
Flanagan E~E and Hinderer T 2012 {\em Phys.\ Rev.\ Lett.\/} {\bf 109} 071102
  (\textit{Preprint} \eprint{1009.4923})

\bibitem{Flanagan:2012kg}
Flanagan E~E, Hughes S~A and Ruangsri U 2014 {\em Phys. Rev.\/} {\bf D89}
  084028 (\textit{Preprint} \eprint{1208.3906})

\bibitem{SXSWebsite}
Simulating e{X}treme {S}pacetimes \url{http://www.black-holes.org/}

\bibitem{Poisson2011}
E~Poisson A~P and Vega I 2011 {\em Living Rev.\ Rel.\/} {\bf 14} 7

\bibitem{Shah:2012gu}
Shah A~G, Friedman J~L and Keidl T~S 2012 {\em Phys. Rev.\/} {\bf D86} 084059
  (\textit{Preprint} \eprint{1207.5595})

\bibitem{vandeMeent:2015lxa}
van~de Meent M and Shah A~G 2015 {\em Phys.\ Rev.\ D\/} {\bf 92} 064025
  (\textit{Preprint} \eprint{1506.04755})

\bibitem{vandeMeent:2016pee}
van~de Meent M 2016 {\em Phys. Rev.\/} {\bf D94} 044034 (\textit{Preprint}
  \eprint{1606.06297})

\bibitem{Merlin:2016boc}
Merlin C, Ori A, Barack L, Pound A and van~de Meent M 2016  (\textit{Preprint}
  \eprint{1609.01227})

\bibitem{LeTiec-Mroue:2011}
Le~Tiec A, Mrou\'e A~H, Barack L, Buonanno A, Pfeiffer H~P, Sago N and
  Taracchini A 2011 {\em Phys.\ Rev.\ Lett.\/} {\bf 107} 141101
  (\textit{Preprint} \eprint{1106.3278})

\bibitem{LeTiec:2011dp}
Le~Tiec A, Barausse E and Buonanno A 2012 {\em Phys. Rev. Lett.\/} {\bf 108}
  131103 (\textit{Preprint} \eprint{1111.5609})

\bibitem{Tiec:2013twa}
Le~Tiec A, Buonanno A, Mrou\'e A~H, Hemberger D~A, Lovelace G, Pfeiffer H~P,
  Kidder L~E, Scheel M~A, Szilagy B, Taylor N~W and Teukolsky S~A 2013 {\em
  Phys.\ Rev.\ D\/} {\bf 88}(12) 124027 (\textit{Preprint} \eprint{1309.0541})

\bibitem{Warburton:2013yj}
Warburton N, Barack L and Sago N 2013 {\em Phys. Rev.\/} {\bf D87} 084012
  (\textit{Preprint} \eprint{1301.3918})

\bibitem{Schmidt:2002qk}
Schmidt W 2002 {\em Class.\ Quantum Grav.\/} {\bf 19} 2743 (\textit{Preprint}
  \eprint{gr-qc/0202090})

\bibitem{LichtenbergLieberman}
{Lichtenberg} A~J and {Lieberman} M~A 1983 {\em {Regular and stochastic
  motion}\/} (Applied Mathematical Sciences, New York: Springer)

\bibitem{Mino:2003yg}
Mino Y 2003 {\em Phys. Rev.\/} {\bf D67} 084027 (\textit{Preprint}
  \eprint{gr-qc/0302075})

\bibitem{Hinderer:2008dm}
Hinderer T and Flanagan E~E 2008 {\em Phys. Rev.\/} {\bf D78} 064028
  (\textit{Preprint} \eprint{0805.3337})

\bibitem{Pound:2015wva}
Pound A 2015 {\em Phys. Rev.\/} {\bf D92} 104047 (\textit{Preprint}
  \eprint{1510.05172})

\bibitem{Isoyama:2014mja}
Isoyama S, Barack L, Dolan S~R, Le~Tiec A, Nakano H, Shah A~G, Tanaka T and
  Warburton N 2014 {\em Phys. Rev. Lett.\/} {\bf 113} 161101 (\textit{Preprint}
  \eprint{1404.6133})

\bibitem{Vines:2015efa}
Vines J and Flanagan E~E 2015 {\em Phys. Rev.\/} {\bf D92} 064039
  (\textit{Preprint} \eprint{1503.04727})

\bibitem{Carter:1968rr}
Carter B 1968 {\em Phys. Rev.\/} {\bf 174} 1559--1571

\bibitem{Drasco:2003ky}
Drasco S and Hughes S~A 2004 {\em Phys. Rev.\/} {\bf D69} 044015
  (\textit{Preprint} \eprint{astro-ph/0308479})

\bibitem{Fujita:2009bp}
Fujita R and Hikida W 2009 {\em Class.\ Quantum Grav.\/} {\bf 26} 135002
  (\textit{Preprint} \eprint{0906.1420})

\bibitem{Hughes:2001jr}
Hughes S~A 2001 {\em Phys. Rev.\/} {\bf D64} 064004 [Erratum: Phys.
  Rev.D88,no.10,109902(2013)] (\textit{Preprint} \eprint{gr-qc/0104041})

\bibitem{LukesGerakopoulos:2010rc}
Lukes-Gerakopoulos G, Apostolatos T~A and Contopoulos G 2010 {\em Phys. Rev.\/}
  {\bf D81} 124005 (\textit{Preprint} \eprint{1003.3120})

\bibitem{Henrard1982}
{Henrard} J 1982 {\em Celestial Mechanics\/} {\bf 27} 3--22

\bibitem{vandeMeent:2013sza}
van~de Meent M 2014 {\em Phys. Rev.\/} {\bf D89} 084033 (\textit{Preprint}
  \eprint{1311.4457})

\bibitem{Teukolsky}
Teukolsky S 1973 {\em Astrophys.\ J.\/} {\bf 185} 635

\bibitem{Hirata:2010xn}
Hirata C~M 2011 {\em Phys. Rev.\/} {\bf D83} 104024 (\textit{Preprint}
  \eprint{1011.4987})

\bibitem{vandeMeent:2014raa}
van~de Meent M 2014 {\em Phys. Rev.\/} {\bf D90} 044027 (\textit{Preprint}
  \eprint{1406.2594})

\bibitem{Berry:2016bit}
Berry C~P~L, Cole R~H, Ca\~{n}izares P and Gair J~R 2016  (\textit{Preprint}
  \eprint{1608.08951})

\bibitem{SpECwebsite}
\url{http://www.black-holes.org/SpEC.html}

\bibitem{Friedrich1985}
Friedrich H 1985 {\em Commun.\ Math.\ Phys.\/} {\bf 100} 525--543
  \urlprefix\url{http://www.springerlink.com/content/w602g633428x8365}

\bibitem{Garfinkle2002}
Garfinkle D 2002 {\em Phys.\ Rev.\ D\/} {\bf 65} 044029

\bibitem{Pretorius2005c}
{Pretorius} F 2005 {\em Class.\ Quantum Grav.\/} {\bf 22} 425
  (\textit{Preprint} \eprint{gr-qc/0407110})

\bibitem{Lindblom2006}
Lindblom L, Scheel M~A, Kidder L~E, Owen R and Rinne O 2006 {\em Class.\
  Quantum Grav.\/} {\bf 23} S447 (\textit{Preprint} \eprint{gr-qc/0512093})

\bibitem{Lindblom2009c}
Lindblom L and Szil\'agyi B 2009 {\em Phys.\ Rev.\ D\/} {\bf 80} 084019
  (\textit{Preprint} \eprint{arXiv:0904.4873})

\bibitem{Choptuik:2009ww}
Choptuik M~W and Pretorius F 2010 {\em Phys.\ Rev.\ Lett.\/} {\bf 104} 111101
  (\textit{Preprint} \eprint{0908.1780})

\bibitem{Szilagyi:2009qz}
Szil{\' a}gyi B, Lindblom L and Scheel M~A 2009 {\em Phys.\ Rev.\ D\/} {\bf 80}
  124010 (\textit{Preprint} \eprint{0909.3557})

\bibitem{Lovelace:2011nu}
Lovelace G, Boyle M, Scheel M~A and Szil{\'a}gyi B 2012 {\em Class.\ Quantum
  Grav.\/} {\bf 29} 045003 (\textit{Preprint} \eprint{arXiv:1110.2229 [gr-qc]})

\bibitem{Scheel2009}
{M A Scheel, M Boyle, T Chu, L E Kidder, K D Matthews and H P Pfeiffer} 2009
  {\em Phys.\ Rev.\ D\/} {\bf 79} 024003 (\textit{Preprint}
  \eprint{arXiv:gr-qc/0810.1767})

\bibitem{Hemberger:2013hsa}
{Hemberger} D~A, {Lovelace} G, {Loredo} T~J, {Kidder} L~E, {Scheel} M~A,
  {Szil{\'a}gyi} B, {Taylor} N~W and {Teukolsky} S~A 2013 {\em Phys.\ Rev.\
  D\/} {\bf 88} 064014 (\textit{Preprint} \eprint{1305.5991})

\bibitem{Ossokine:2013zga}
Ossokine S, Kidder L~E and Pfeiffer H~P 2013 {\em Phys.\ Rev.\ D\/} {\bf 88}
  084031 (\textit{Preprint} \eprint{1304.3067})

\bibitem{Rinne2006}
Rinne O 2006 {\em Class.\ Quantum Grav.\/} {\bf 23} 6275--6300
  \urlprefix\url{http://stacks.iop.org/0264-9381/23/6275}

\bibitem{Rinne2007}
Rinne O, Lindblom L and Scheel M~A 2007 {\em Class.\ Quantum Grav.\/} {\bf 24}
  4053--4078 \urlprefix\url{http://stacks.iop.org/0264-9381/24/4053}

\bibitem{Pfeiffer2003}
Pfeiffer H~P, Kidder L~E, Scheel M~A and Teukolsky S~A 2003 {\em Comput.\
  Phys.\ Commun.\/} {\bf 152} 253--273 (\textit{Preprint}
  \eprint{gr-qc/0202096})

\bibitem{Caudill-etal:2006}
{Caudill} M, {Cook} G~B, {Grigsby} J~D and {Pfeiffer} H~P 2006 {\em Phys.\
  Rev.\ D\/} {\bf 74} 064011 (\textit{Preprint} \eprint{gr-qc/0605053})

\bibitem{Lovelace2008}
Lovelace G, Owen R, Pfeiffer H~P and Chu T 2008 {\em Phys.\ Rev.\ D\/} {\bf 78}
  084017

\bibitem{York1999}
York J~W 1999 {\em Phys.\ Rev.\ Lett.\/} {\bf 82} 1350--1353

\bibitem{OwenThesis}
Owen R 2007 {\em Topics in Numerical Relativity: {T}he periodic standing-wave
  approximation, the stability of constraints in free evolution, and the spin
  of dynamical black holes\/} Ph.D. thesis California Institute of Technology
  \urlprefix\url{http://resolver.caltech.edu/CaltechETD:etd-05252007-143511}

\bibitem{Buonanno:2010yk}
Buonanno A, Kidder L~E, Mrou\'{e} A~H, Pfeiffer H~P and Taracchini A 2011 {\em
  Phys.\ Rev.\ D\/} {\bf 83} 104034 (\textit{Preprint} \eprint{1012.1549})

\bibitem{SavitzkyGolay1964}
Savitzky A and Golay M~J~E 1964 {\em Analytical Chemistry\/} {\bf 36}
  1627--1639 (\textit{Preprint} \eprint{http://dx.doi.org/10.1021/ac60214a047})
  \urlprefix\url{http://dx.doi.org/10.1021/ac60214a047}

\bibitem{Boyle2007}
Boyle M, Brown D~A, Kidder L~E, Mrou{\'e} A~H, Pfeiffer H~P, Scheel M~A, Cook
  G~B and Teukolsky S~A 2007 {\em Phys.\ Rev.\ D\/} {\bf 76} 124038
  (\textit{Preprint} \eprint{0710.0158})

\bibitem{Peters1964}
Peters P~C 1964 {\em Phys.\ Rev. B\/} {\bf 136} 1224
  \urlprefix\url{http://link.aps.org/abstract/PR/v136/pB1224}

\bibitem{Ruangsri:2013hra}
Ruangsri U and Hughes S~A 2014 {\em Phys. Rev.\/} {\bf D89} 084036
  (\textit{Preprint} \eprint{1307.6483})

\bibitem{Boyle:2008}
{Boyle} M, {Buonanno} A, {Kidder} L~E, {Mrou{\'e}} A~H, {Pan} Y, {Pfeiffer} H~P
  and {Scheel} M~A 2008 {\em Phys.\ Rev.\ D\/} {\bf 78} 104020
  (\textit{Preprint} \eprint{0804.4184})

\bibitem{Zimmerman:2016ajr}
Zimmerman A, Lewis A~G~M and Pfeiffer H~P 2016 {\em Phys. Rev. Lett.\/} {\bf
  117} 191101 (\textit{Preprint} \eprint{1606.08056})

\bibitem{vandeMeent:2016hel}
van~de Meent M 2016  (\textit{Preprint} \eprint{1610.03497})

\bibitem{SciNet}
Loken C, Gruner D, Groer L, Peltier R, Bunn N, Craig M, Henriques T, Dempsey J,
  Yu C~H, Chen J, Dursi L~J, Chong J, Northrup S, Pinto J, Knecht N and Zon R~V
  2010 {\em J. Phys.: Conf. Ser.\/} {\bf 256} 012026

\end{thebibliography}

\end{document}